\documentclass[useAMS,fleqn,usenatbib]{mnras}
\usepackage{newtxtext,newtxmath}
\usepackage[T1]{fontenc}

\usepackage{upgreek}
\usepackage{graphicx}
\usepackage{amsmath}
\usepackage{amssymb}
\usepackage{lineno}
\usepackage{hyperref}
\usepackage{academicons}
\usepackage{xcolor}
\usepackage{soul}
\usepackage{orcidlink}

\newcommand{\pd}{\partial}                        
\renewcommand{\vec}[1]{\ensuremath{\bmath{#1}}}   
\newcommand{\hv}[1]{\ensuremath{\hat{\bmath{#1}}}}
\newcommand{\ham}{\mbox{\ensuremath{\hat{H}}}}    
\newcommand{\td}{\mbox{\ensuremath{\tilde{\cal E}}}} 

\title[3D Maser polarization simulation]{3D Maser polarization simulation for J=1-0 SiO masers in the circumstellar envelope of an AGB star}
\author[M. Phetra et al.]{
	M. Phetra\orcidlink{0009-0008-1973-2662},$^{1,2}$
	M. D. Gray\orcidlink{0000-0002-2094-846X},$^{2}$\thanks{E-mail: malcolm@narit.or.th}
	K. Asanok\orcidlink{0000-0002-4125-0941},$^{2}$
	S. Etoka\orcidlink{0000-0003-3483-6212},$^{3}$
	B. H. Kramer\orcidlink{0000-0001-8168-5141},$^{2,4}$
	K. Sugiyama\orcidlink{0000-0002-6033-5000},$^{2}$
	\newauthor
	and W. Nuntiyakul\orcidlink{0000-0002-1664-5845}$^{5}$
	\\
	$^{1}$Graduate School, Chiang Mai University, Chiang Mai 50200\\
	$^{2}$National Astronomical Research Institute of Thailand, Chiang Mai 50180, Thailand\\
	$^{3}$Jodrell Bank Centre for Astrophysics, School of Physics and Astronomy, University of Manchester, M13 9PL, UK\\
	$^{4}$Max Planck Institute for Radio Astronomy, Auf dem H\"{u}gel 69, Bonn 53121, Germany\\
	$^{5}$Department of Physics and Materials Science, Faculty of Science, Chiang Mai University, Chiang Mai 50200, Thailand\\
}

\date{Accepted 2025 April 17. Received 2025 March 24; in original form 2024 September 10}

\begin{document}
\pagerange{\pageref{firstpage}--\pageref{lastpage}} \pubyear{2025}
\maketitle
\label{firstpage}

\begin{abstract}
SiO masers from AGB stars exhibit variability in intensity and polarization during a 
pulsation period. This variability is explained by radiative transfer and magnetic 
properties of the molecule. To investigate this phenomenon, a 3D maser simulation is employed 
to study the SiO masers based on Zeeman splitting. We demonstrate that the magnetic field 
direction affects maser polarization within small tubular domains with isotropic pumping, and yields 
results that are similar to those obtained from 1D modelling. This work also studies larger clouds 
with different shapes.
We use finite-element domains with internal node distributions to 
represent the maser-supporting clouds. We calculate solutions for the population
inversions in all transitions and at every node. These solutions show that saturation 
begins near the middle of a domain, moving towards the edges and particularly the ends of long axes, as 
saturation progresses, influencing polarization. When the observer's view of the domain changes, the 
plane of linear polarization responds to the projected shape and the projected magnetic field axis. 
The angle between the observer's line of sight and the magnetic field may cause jumps in the plane of polarization. 
Therefore, we can conclude that polarization is influenced by both the cloud's major axis orientation 
and magnetic field direction. We have investigated the possibility of explaining observed polarization plane 
rotations, apparently within a single cloud, by the mechanism of line-of-sight overlap of two magnetized maser clouds.
\end{abstract}


\section{Introduction}
\label{s:intro}
Maser (microwave amplification by stimulated emission of radiation) emission from 
many rotational transitions of the SiO molecule is frequently observed
towards asymptotic giant branch (AGB) stars with oxygen-rich circumstellar envelopes (CSEs).
The range of frequencies covered is from about 43\,GHz ($J=1-0$ transitions) to at least 
345\,GHz ($J=8-7$). Maser emission arises from vibrational states $v=0-4$, mostly from the most
common isotoplogue, $^{28}$SiO, but also from the rarer isotopic substitutions $^{29}$SiO and
$^{30}$SiO. For a recent summary, see \citet{2021ApJS..253...44R}.
Masers from the $v=0$ state of $^{28}$SiO are typically weak, and accompanied by
thermal emission, appearing as narrow maser spikes superimposed on a broader thermal
base. The two most common maser transitions are the $v=1$ and $v=2$, $J=1-0$
transitions of $^{28}$SiO, at the respective frequencies of 43.1220 and 42.820\,GHz,
which have been found together in 89 per cent of 9727 sources detected
by the BAaDE survey \citep{2024IAUS..380..314L}. The maser emission from these two transitions
typically follows the optical light curve of the stellar continuum, with a delay of roughly 0.1 - 0.2 of the period \citep{Pardo2004}, but is more nearly in phase with the IR light curve. This has been taken to show that there is a strong radiative element to the pumping scheme \citep{Cernicharo1997}. The spatial
structure of SiO masers at 43 and 86\,GHz has been shown to consist of broken rings of emission
in the sky plane through VLBI (Very-long-baseline interferometry) observations, for example
\citet{1994ApJ...430L..61D}. These rings show proper motion that is related to the stellar
pulsation cycle \citep{2003ApJ...599.1372D}. More recent observations, for example of
WX~Psc with the Korean-Japanese {\it KaVA} interferometer \citep{2016ApJ...822....3Y}, have
45\,microarcsec relative alignment of the $J=1-0,v=1$ and $v=2$ rings, and demonstrate that the $v=2$
ring is generally the smaller. Some modern instruments
also have the ability to register maser positions accurately against other structures: for example the stellar continuum with {\it ALMA} \citep{2018iss..confE..13H}, and dust shells local to
and outside the SiO maser zone using the {\it VLTI} and Keck interferometers \citep{2015A&A...576A..70P}.

Long-term monitoring observations of SiO masers in full polarization with VLBI arrays show variability not only in intensity but also in linear and circular polarization, for example \citet{Kemball2009,Assaf2013}. In
observations of the $v=1, J=1-0$ masers towards TX~Cam, typical polarization levels were 15-30 per cent
linear and 3-5 per cent circular, but circular polarization levels of tens of per cent have
been observed in some individual features \citep{Kemball2009}. 
Roughly consistent values (tens of per cent linear
and a few per cent circular) were detected towards R~Cas \citep{Assaf2013}. In R~Cas, features with higher
linear polarization correlated with weaker Stokes-$I$ intensity. The 
EVPA (electric vector position angle) distribution in R~Cas was
distinctly bimodal, favouring either tangential or radial alignment \citep{Assaf2013}. Near
the optical minimum (phase 0.452-0.675) of one cycle, alignment was mostly tangential, changing to
radial through maximum (phase 0.744-1.158). However, this behaviour was not consistently followed
in the next cycle, where the tangentially-dominated range (phase 1.243-1.432) did not encompass the optical minimum.
In other objects, tangential orientation of the EVPA was found towards IK~Tau, but a rather disordered
distribution, in R~Cnc \citep{2009ApJS..185..574C}. In addition to global EVPA variation, both TX~Cam and R~Cas exhibit individual maser features (meaning a single VLBI emission patch) in which the EVPA rotates through an
angle of approximately $\upi/2$ \citep{Kemball1997,Kemball2011,Assaf2013}. The effect persists
over several frequency channels, and
appears not to be particularly rare, as there is evidence for similar rotations in some
maser structures in the CSEs of the Miras R~Aqr and o~Cet \citep{2006A&A...456..339C}. Over about 2
pulsation cycles of R~Cas there is a tendency for the EVPA to align perpendicular to the direction
of proper motion, which is dominated by a slow (0.4\,km\,s$^{-1}$, phase averaged) radial expansion
\citep{Assaf2018ApJ}.

Polarization can be used to interpret the magnetic field morphology in a region, given some 
assumptions about the mechanism through which the magnetic field influences polarization. In the analysis
results discussed in this paragraph, and later, the assumed mechanism in AGB star
CSEs unless otherwise stated is `saturation polarization' \citep{Lankhaar2024A&A}, where the molecular
symmetry axis is defined by the Zeeman effect, but populations of the Zeeman
sublevels are controlled by radiation transfer. We will often, loosely, refer to this as
the Zeeman mechanism. In a little more detail, `Zeeman mechanism' here refers to the 
classic range of saturation where the magnetic precession rate is much larger
than the rate of stimulated emission, which is in turn much greater than the decay rate, or
$g\Omega \gg R \gg \Gamma$, and the precession rate is small
compared to the Doppler width ($ g\Omega \ll \Delta\omega$). This regime corresponds to Case~(2a)
in the seminal work of \citet{Goldreich1973}, hereafter
GKK73, but also incorporates Case~(2b) ($R \ll \Gamma$), and is typical of closed-shell
molecules like SiO in which the Zeeman splitting depends on the nuclear magneton. In Case~(2a), 
GKK73 demonstrated that the EVPA of linear polarization is a function of the angle, $\theta$, between the magnetic field direction and the ray to the observer.  The solution predicts an EVPA flip of 90 degrees at $\theta = \arcsin(\sqrt{2/3}) \simeq 54.7$ degrees, or the Van Vleck angle ($\theta_{VV}$). At $\theta > \theta_{VV}$,
levels of linear polarization up to $\simeq 1/3$ can be readily generated at modest degrees of
saturation ($R\sim 10$ times the saturation intensity, \citealt{Western1984Apj285}). 
Larger fractions of linear polarization can
be obtained when $\theta < \theta_{VV}$, but only for significantly greater $R$, and \citet{Western1984Apj285}
question whether the magnetic field magnitudes, perhaps $>100$\,G, required to keep $R$ in the classic range
in this case are acceptable on the grounds of dynamical influence on the CSE. Such fields are
also marginal for the condition $g\Omega \ll \Delta\omega$. It is
therefore important to note that several alternatives, or modifications, to the
Zeeman model have been suggested, including
asymmetric pumping of the magnetic substates of the maser levels by the infrared 
pumping radiation \citep{Western1983ApJ275},
Wiebe and Watson's non-Zeeman mechanism for generation of circular polarization (based on line-of-sight magnetic
field variations, \citealt{1998ApJ...503L..71W}),
extreme saturation (GKK73 Case~3, $R > g\Omega$, for example \citealt{Deguchi1990ApJ354,Nedoluha1990ApJ354}), and
anisotropic resonant scattering (ARS, \citealt{2022MNRAS.511..295H}). Circular polarization has
often been ignored in modelling, since in the classic Zeeman regime Stokes-$V$ is zero at line
centre; it is not considered in detail by GKK73. Circular (and linear) polarization are modelled
by \citet{Watson2001}, with predictions of the behaviour of both types with $\theta$. Usefully,
Stokes-$V$ is proportional to $B$ in the Case(2a) regime (within a function of $\theta$), whilst
the fractional linear polarization is largely insensitive to $B$ if $g\Omega$ is much smaller
than the Doppler width.

Maser polarization has been a controversial subject in the weak-splitting case. To understand the competing theories and their predictions completely requires a detailed reading of the following works: \citet{Elitzur1993ApJ,Elitzur1996ApJ} and \citet{Western1984Apj285}, \citet{Nedoluha1990ApJ354}, \citet{Watson2001}. Briefly, the limiting solutions of GKK73 are achieved only at significantly higher levels of saturation in the models by Watson and co-workers than in those by Elitzur. Moreover, \citet{Elitzur1996ApJ} predicts that linear polarization is only possible for $\sin(\theta) > \sqrt{1/3}$, and that Stokes V should depend on $1/\cos(\theta)$. However \cite{Gray2003MNRAS} confirmed that maser polarization should follow the Watson models, and \cite{Dinh-v-Trung2009MNRAS} demonstrated that linear polarization is
generated for $\sin(\theta) < \sqrt{1/3}$. More recent years have seen the development of two one-dimensional (1D) maser polarization simulation codes that broadly agree, described in LV19 \citep{Lankhaar2019} and TGK23 \citep{Tobin2023}.

In more detail, \citet{Lankhaar2019,Lankhaar2024A&A} have presented a modelling program called CHAracterizing Maser Polarization ({\sc champ}) that can compute solutions in the case where magnetic and hyperfine splittings are both present and have similar frequency spacing. Higher rotational systems of arbitrary $J$ can also be analysed (systems
up to $J=3-2$ have previously been considered by \citealt{Deguchi1990ApJ354} and \citealt{Nedoluha1990ApJ354}).
The results regarding SiO masers confirm that the 90-degree flip of the EVPA and the drop of fractional linear polarization at the Van Vleck angle occurs in the regime where ($g\Omega > R \gg \Gamma$), but disappears for a very high stimulated emission rate, such that $R > g\Omega$. TGK23 \citep{Tobin2023} used the {\sc prism}, (Polarized Radiation Intensity from Saturated Masers) code which included in-source Faraday rotation in their modelling. The simulation is also in 1D but divided into one- and two-direction radiation propagation, and investigates the effect of optical depth and magnetic field strength on the polarization. The results show that strong Faraday rotation affects the linear polarization, especially at magnetic field strengths of order $\gtrsim 10$\,G, introducing additional EVPA flips and fluctuations in addition to the flip at $\theta_{VV}$.

Historically, attempts to recover the magnetic field from polarization-sensitive
observations have been hampered by many uncertainties including unknown intrinsic
beam angles, the importance of anisotropic pumping and rival theories (see above).
A classic Zeeman interpretation and the \citet{Elitzur1996ApJ} theory were used
to recover a field of 5-10\,G from VLBI observations of TX~Cam \citep{Kemball1997}.
Single-dish observations of a sample of late-type stars,
including 43 Miras (a majority) were used to calculate
a mean magnetic field of 3.5\,G in the SiO zone, with
a range of 0-20\,G \citep{2006A&A...450..667H}. This
calculation also used Elitzur's theory, and assumed the classic Zeeman range.
Recovery of the direction of $B$ on the plane of the sky suffers from the
ambiguity inherent in the $\upi /2$ flip through $\theta_{VV}$. If $\theta < \theta_{VV}$,
the sky component of $B$ is parallel to the EVPA; otherwise, it is perpendicular
(for example \citealt{2006A&A...450..667H}). Variation of the angle of $B$ to the
line of sight, passing through $\theta_{VV}$ was used to explain the probable detection of 
such an EVPA flip of a $v=1, J=1-0$ SiO maser in a single cloud from TX Cam \citep{Kemball2011}.
\citet{Lankhaar2024A&A} use an excitation analysis to obtain pumping anisotropy parameters for SiO masers, followed by polarization-sensitive radiative transfer modelling to compute the polarization. They conclude that SiO masers in the vicinity of a late-type star are probably pumped very anisotropically.

In this work, we aim to extend existing numerical work related to polarization in astrophysical
masers by investigating some effects that are very difficult, or even impossible to simulate
in one dimension. These include effects on polarization due to modest departures from 
axial symmetry, effects due to distortion of a basic pseudo-spherical shape into oblate
and prolate variants, and the overlap of two magnetized clouds with differently oriented
magnetic fields. We also aim to demonstrate convergence with 1D results through the study of
a narrow tube domain. 

In section \ref{sec:Theory}, the equations governing maser polarization are introduced. Section \ref{sec:Simul} describes the specifics of setting up the model. We report the simulation results in Section~\ref{sec:Results}, then discuss and conclude in Section~\ref{sec:Dis_and_Con}. 

\section{Theory} \label{sec:Theory}
Two- and three-dimensional models of polarized maser
radiation transfer have been attempted previously, for
example \citet{Western1983Apj268}. However, the present
authors believe that this is the first 3D model that
can solve the problem for saturated masers in arbitrary cloud shapes with
full spectral coverage, and with the ability to vary the
magnetic field within the cloud. Here, we introduce the
main equations governing the electric field components of all rays and the response of the molecules, subject
to an applied magnetic field. 

\subsection{General equations}
Our model is semi-classical, with classical electric fields interacting with
a molecular response represented through a quantum-mechanical density matrix (DM). The
transfer of the electric field through the molecular medium is derived ultimately
from Maxwell's equations, whilst the evolution equations for the molecular DM
are derived from the time-dependent Schr$\ddot{\text{o}}$dinger equation. In this main text
we write down only final results of the analysis; detailed derivations are presented in Appendix~\ref{a:deriv}.
The equations presented below are at a stage where the broad-band electric field has been
defined, and applied to interaction terms between molecular dipoles and electric fields. The
rotating wave approximation (RWA), for example \citet{Western1983Apj268}, has been applied
to eliminate non-resonant parts of the interaction terms, and a discrete-frequency Fourier
transform \citep{1978PhRvA..17..701M} has been performed on the DM and electric field equations, converting them from the time domain to (angular) frequency. A diagonal element of the DM, $\varrho_{p,n}$, 
for energy level $p$ and Fourier component $n$, is related to off-diagonal elements
and electric fields at a general position by
\begin{equation}\label{eq:diag_DM}
	\begin{split}
		\varrho_{p,n}=& 2\upi\tilde{\mathcal{L}}_{p,n} P_{p,n}
        -\frac{\upi\tilde{\mathcal{L}}_{p,n}}{2\hbar } \sum_{\xi=1}^J w_{\Omega,\xi} \sum_{m=-\infty}^{\infty} \\
                        & \Biggl\{ \sum_{j=1}^{p-1} \left[
         S_{pj,m+n}^{(\xi)} \hv{d}_{pj}^{\ast} \cdot \vec{\td}_{\xi,m}^{\ast}
        + S_{pj,m-n}^{\ast (\xi)} \hv{d}_{pj} \cdot \vec{\td}_{\xi,m} \right] \\
		&-\sum_{j=p+1}^{N} \left[ S_{jp,m-n}^{\ast (\xi)} \hv{d}_{jp} \cdot \vec{\td}_{\xi,m} 
        + S_{jp,m+n}^{(\xi)} \hv{d}_{jp}^{\ast} \cdot \vec{\td}_{\xi,m}^{\ast} \right]
        \Biggr\} ,
	\end{split}
\end{equation}
which is derived in detail as equation~(\ref{eq:diag_vi}).
At this stage, the DM elements are functions of position $\boldsymbol{r}$ and frequency channel, since we have
integrated out the molecular velocity $\boldsymbol{v}$ (see Appendix~\ref{ss:velint}). 
The factor $\tilde{\mathcal{L}}_{p,n}$ represents the complex Lorentzian function defined
in equation~(\ref{eq:diaglor}), while $P_{p,n}$ is the rate of increase of population in
level $p$, channel $n$, due to all non-maser processes from all other levels. The diagonal element
is influenced by $J$ rays, each with solid-angle weighting $w_{\Omega,\xi}=\sqrt{\delta \Omega_j/(4\upi)}$,
where $\delta \Omega_j$ is the actual solid angle associated with ray $j$. Within each ray is a sum
over Fourier components, or channels, index $m$. Electric field amplitude vectors, $\vec{\td}_{\xi,m}$, are 
indexed by ray and by channel and exert their influence on the DM through dipole operators,
$\hv{d}_{pj}$, attached to transitions between upper level $p$ and lower level $j$. This
effect is mediated through the slow parts of off-diagonal DM elements, $S_{pj,m+n}^{(\xi)}$,
which are indexed by ray and channel and also have two energy-level indices, labelled $p$ (upper) and $j$
(lower), which are unequal in off-diagonal elements that represent coherences between pairs of levels. 

Slowly-varying parts of off-diagonal DM elements of the general form $S_{pq,n}^{(\xi)}$ are coupled to 
(complex) electric field amplitudes, off-diagonal elements in other channels and transitions, and the
populations of levels $p$ and $q$ by,
\begin{equation}\label{eq:off_DM}
	\begin{split}
		S_{pq,n}^{(\xi)} \! = \! &-\frac{i\upi w_{\Omega,\xi}}{\hbar} 
             \!\!\!\!\! \sum_{m=-\infty}^{\infty} \!\!\Biggl\{
        2i\hv{d}_{pq} \! \cdot \! \vec{\td}_{\xi,m} (\Xi_{p,n}^{pq}\varrho_{p,n-m}-\Xi_{q,n}^{pq}\varrho_{q,n-m}) \\
&\!\!\!\!\!\! + \!\! \sum_{j=1}^{q-1} \!\! \left[ 
    \Xi_{pj,n}^{pq} S_{pj,m+n}^{(\xi)} \hv{d}_{qj}^{\ast} \!\cdot\! \vec{\td}_{\xi,m}^{\ast} 
  + \Xi_{qj,n}^{pq} S_{qj,m-n}^{\ast (\xi)} \hv{d}_{pj} \!\cdot\! \vec{\td}_{\xi,m} 
                     \!  \right]\\
&\!\!\!\!\!\! + \!\!\!\! \sum_{\!\!j=q+1}^{p-1} \!\!\! \left[ 
     \Xi_{pj,n}^{pq} S_{pj,n-m}^{(\xi)} \hv{d}_{jq} \!\cdot\! \vec{\td}_{\xi,m} 
  - \Xi_{jq,n}^{pq} S_{jq,n-m}^{(\xi)} \hv{d}_{pj} \!\cdot\! \vec{\td}_{\xi,m}
                     \! \right]\\
&\!\!\!\!\!\!\! - \!\!\!\!\! \sum_{\!\!j=p+1}^{N} \!\!\! \left[ 
    \Xi_{jp,n}^{pq} S_{jp,m-n}^{\ast (\xi)} \hv{d}_{jq} \!\cdot\! \vec{\td}_{\xi,m} 
    \! + \Xi_{jq,n}^{pq} S_{jq,m+n}^{(\xi)} \hv{d}_{jp}^{\ast} \!\cdot\! \vec{\td}_{\xi,m}^{\ast}
                     \! \right]
	\Biggr\},
	\end{split}
\end{equation}
where the symbol $\Xi_{jq,n}^{pq}$, defined in equation~(\ref{eq:bigxi}), represents the 
convolution of a complex Lorentzian function for off-diagonal DM elements, which is dependent on velocity (see
equation~\ref{eq:off1lor}) with an unknown 3D distribution of molecular velocities in the transition $pj$.
The existence of non-zero terms in the lower three lines of equation~(\ref{eq:off_DM}) is determined by
the presence of allowed dipoles. If an off-diagonal DM element has level indices that correspond
to a `type~2' transition (see Appendix~\ref{ss:rwa} and Fig.~\ref{fig:trans_diagram}), we do not 
at this stage discard that term.

The complex amplitude of the electric field of ray $\xi$ in channel $n$ evolves according to
the radiative transfer equation,
\begin{equation}\label{eq:RT_eq}
\frac{d \vec{\td}_{\xi,n}}{d s_\xi} + \kappa(\vec{r}) \vec{\td}_{\xi,n} = 
\frac{D}{w_{\Omega,\xi}} \sum_{p=2}^{N_G} \sum_{j=1}^{p-1} \hv{d}_{pj}^* S_{pj,n}^{(\xi)} ,
\end{equation}
where the parameter $D$ is defined in equation~(\ref{eq:defdconst}), $\kappa$ is a
complex attenuation coefficient, and $ds_\xi$ is an element of distance
along the path of ray $\xi$.

Equations (\ref{eq:diag_DM}) to (\ref{eq:RT_eq}) are very general; they are also rather difficult
to solve. They may be applied, in principle,
to any $J$ to $(J-1)$ rotational system with Zeeman splitting due to an external magnetic field $\boldsymbol{B}$.
Much previous work has concentrated on the $J=1-0$ system, so we reduce our equations suitably.
The level $J=1$ splits into sub-levels with $M_J=-1,0,1$. We define the level index $p$ as 1 for the undivided level $J$=0, and 2, 3, and 4, respectively for the sublevels of $J=1$ with magnetic quantum number, $M_J=-1,0,1$. We assume that transitions between the split levels ('type~2' transitions that
change $M_J$, but not $J$) have vastly smaller frequencies than transitions with $\Delta J = 1$. There are three transitions with indices $pq$ equal to 21, 31, and 41, which are called the $\sigma^{+}$, $\pi$, and 
$\sigma^{-}$ helical transitions, respectively. See Fig.~\ref{fig:trans_diagram} for a diagram of the $J=1-0$
system. Further details of the $J=1-0$ reduction appear in Appendix~\ref{ss:j1-0pattern}, including a pairing
of the diagonal elements into inversions, such as 
$\Delta_{p1}=\varrho_p - \varrho_1$, for upper level $p$.

\begin{figure}
	\centering
	\includegraphics[width=0.30\textwidth]{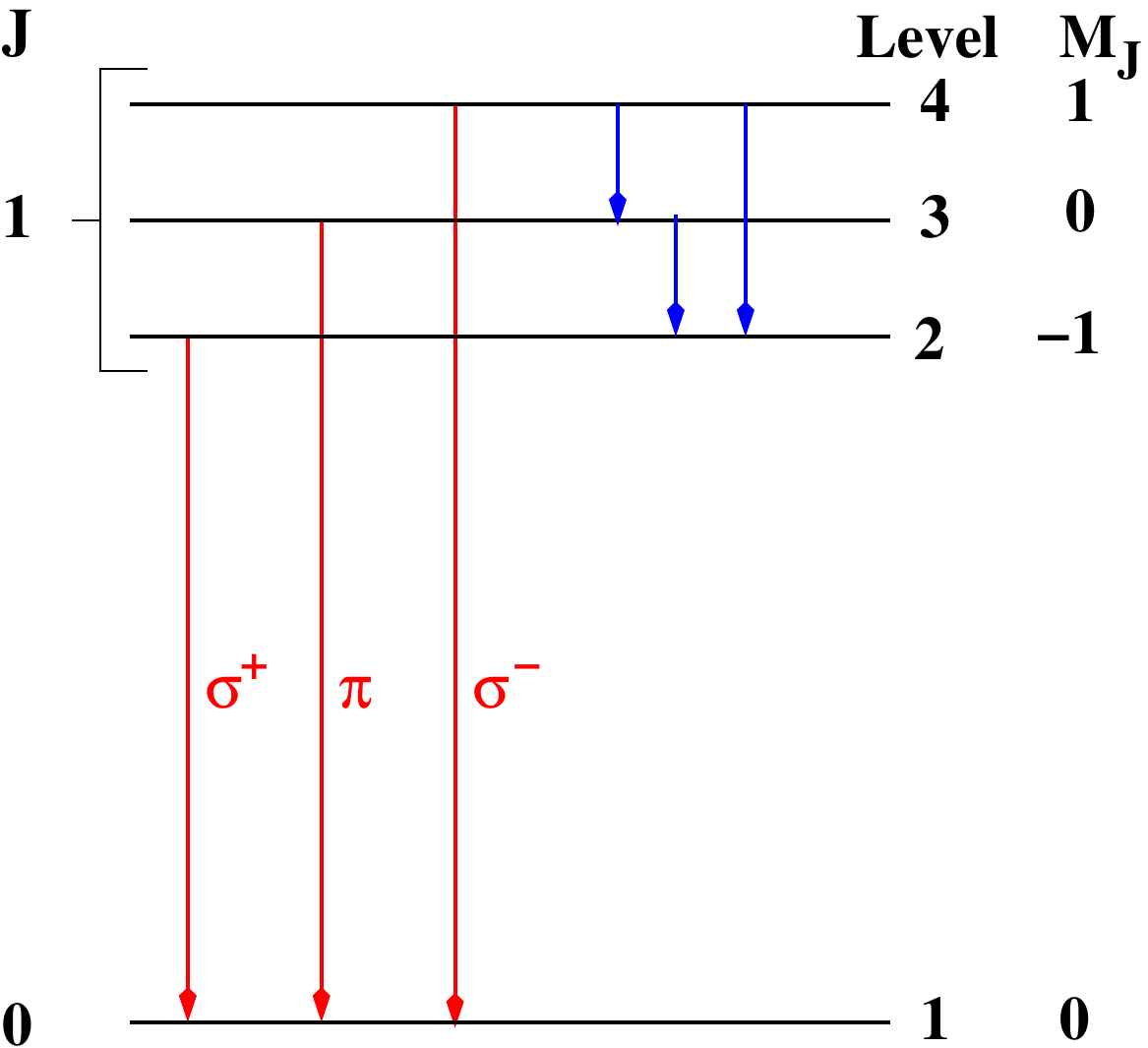}
	\caption{A transition diagram of the Zeeman effect for a $J=1-0$ system. The three dipole-allowed 
    transitions (`type~1') are shown in red ($\sigma^{+}$, $\pi$, and $\sigma^{-}$, as marked). 
    The type~2 transitions (not present when
    their rotational level is degenerate)
    are shown in blue.}
	\label{fig:trans_diagram}
\end{figure}

\subsection{Approximations and Scaling}
\label{ss:proxandscal}
Following a reduction to the $J=1-0$ system, we apply a number of approximations to further
simplify the equation set. We begin with an assumption of complete velocity redistribution (CVR).
This allows us to approximate the molecular velocity distributions as 3D Gaussians. Details in
Appendix~\ref{ss:cvr_approx} show that, under CVR, we may reduce the symbols $\Xi_{pj,n}^{pq}$ in
equation~(\ref{eq:off_DM}) to the simpler $\Xi_n^{pq}$, which are complex Voigt profiles \citep{Olver2010nist}
in general.

The next approximation we make is that there are no
population pulsations, that is Fourier components of
$\varrho_{p,n}$ are empty for $n \neq 0$. This is similar
to the extreme ($\bar{m}=0$) case of spectral limiting in \citet{Wyenberg2021MNRAS}, and we
attempt to make a rough estimate of the fidelity of this approximation in
Section~\ref{ss:nodsols}.
The effect on diagonal DM elements in, for example,
equation~(\ref{eq:diag_DM}) is that all Fourier indices
inside the sums on the right-hand side reduce to $m$,
$\tilde{\mathcal{L}}_{p,n}$ becomes $1/(2\upi\Gamma)$,
and the whole expression becomes real. The second approximation is that we do not need to consider 
the sub-state mixing effect of the type-2 off-diagonal elements. This approximation restricts us to 
levels of saturation consistent with the
GKK73 Case (2) where $R \ll g\Omega$. Details of the implementation of these
approximations appears in Appendix~\ref{ss:classic}. A great advantage
of this simplified system is that equation~(\ref{eq:offred}) can be substituted into
equations~(\ref{eq:diagred}) and (\ref{eq:j10_eamp}), eliminating the off-diagonal DM
elements. The resulting equation for $\Delta_{p1}$, the inversion between upper level $p$
and level $1$ is
\begin{equation}\label{eq:deltafinal}
    \begin{split}
        \Delta_{p1} & = \frac{P_{p1}}{\Gamma} - \frac{\upi}{\hbar^2 \Gamma} \sum_{\xi=1}^J \frac{\delta \Omega_\xi}{4\upi}
                      \sum_{m=-\infty}^\infty \Re \left\{
                         2 \Xi_m^{p1} \Delta_{p1} | \hv{d}_{p1} \cdot \vec{\td}_{\xi,m} |^2 \right. \\
                    &\left. + \Xi_m^{q1} \Delta_{q1} | \hv{d}_{q1} \cdot \vec{\td}_{\xi,m} |^2
                            + \Xi_m^{q'1} \Delta_{q'1} | \hv{d}_{q'1} \cdot \vec{\td}_{\xi,m} |^2
                      \right\} ,
    \end{split}
\end{equation}
where $q,q'$ are the alternative upper levels $\neq p$ and $\Gamma$ is the loss rate.
The radiative transfer equation is re-cast into a vector-matrix form, with matrix
coefficients independent of the electric field amplitudes. Details appear in
Appendix~\ref{ss:formsol}.

Surviving variables were scaled according to the following scheme:
\begin{subequations}
	\begin{align}
		\hv{d}_{p1} & = d \vec{\delta}_{p1}, \\
		\vec{\td}_{\xi,n} & = \sqrt{\frac{I_{\text{sat}}\delta\omega}{2\upi c\epsilon_0}}\vec{\mathfrak{E}}_{\xi,n}, \\
		\text{and }\Delta_{p1}&=\frac{P_{p1}}{\Gamma}\mathfrak{D}_{p1},
	\end{align}
\end{subequations}
where $\vec{\delta}_{p1}$, $\vec{\mathfrak{E}}_{\xi,n}$ and $\mathfrak{D}_{p1}$ are the dimensionless forms, $\delta\omega$ is the angular frequency channel width, and  $I_{\text{sat}}$ is the saturation intensity \citep{Tobin2023}.
The Voigt profiles from equation~(\ref{eq:deltafinal}) were scaled by extracting a 
factor of $1/(2\sqrt{\pi} \Delta \omega_D)$,
where $\Delta\omega_D$ is the Doppler width.

The dimensionless inversion is now given by the scaled version of equation~(\ref{eq:deltafinal}):
\begin{equation}\label{eq:inversion}
      \mathfrak{D}_{p1} \! = \! 1 - \! \frac{\delta \omega}{\upi^{3/2} \! \Delta \omega_D} \!\! \sum_{\xi=1}^J 
             \!\! \frac{\delta\Omega_\xi}{4\upi} \!\!\!\! \sum_{m=-\infty}^{\infty} \!\!\!\!\! \Re \biggl\{
             \! \sum_{q=2}^4
             [2] \mathfrak{X}_m^{p1} \mathfrak{D}_{q1} | \vec{\delta}_{q1} \!\cdot \vec{\mathfrak{E}}_{\xi,m}|^2 \
             \!\! \biggr\},
\end{equation}
where $\Re$ implies take the real part, $\mathfrak{X}$ is the dimensionless Voigt profile,
and the bracketed factor of 2 is applied only when $q=p$. The
scaled form of the radiative transfer equation, from equation~\ref{eq:gainmat} is
\begin{align}
   \frac{d}{d\tau_\xi}
   \left(
   \begin{array}{c}
   \mathfrak{E}_{\xi x_\xi,n} \\
   \mathfrak{E}_{\xi y_\xi,n}
   \end{array}
   \right)
=               \left[
   \begin{array}{cc}
      {\cal A}(\tau_\xi) & {\cal B}(\tau_\xi)  \\
      {\cal C}(\tau_\xi) & {\cal D}(\tau_\xi) 
   \end{array}
                \right]
                \left(
                   \begin{array}{c}
   \mathfrak{E}_{\xi x_\xi,n} \\
   \mathfrak{E}_{\xi y_\xi,n}
   \end{array}
                \right),
\label{eq:rtfinal}
\end{align}
where the matrix element ${\cal A}(\tau_\xi)$ is equal to
\begin{equation}
    {\cal A}(\tau_\xi) = \sum_{p=2}^4
                \mathfrak{X}_n^{p1}(\tau_\xi) \mathfrak{D}_{p1}(\tau_\xi) | \delta_{p1,x_\xi} |^2(\tau_\xi) .
\label{eq:dimlessA}
\end{equation}
Components of electric field amplitudes, and of dipoles, in equations~(\ref{eq:rtfinal}) and (\ref{eq:dimlessA})
are written explicitly in the ray frame (subscripts $x_\xi$ and $y_\xi$), where the ray
propagates along $z_\xi$. The scaled distance is the maser depth defined by $d\tau_\xi = \gamma_0 ds_\xi$, where the
gain coefficient, $\gamma_0$, is
\begin{equation}
	\gamma_0 = \frac{3 P_{21} \lambda_0^2 A_0}
                        {8 \upi^{1/2} \Gamma \Delta \omega_D}.
\label{eq:gainco0}                        
\end{equation}
The other three matrix coefficients may be calculated by making the following replacements
of $|\delta_{p1,x_\xi}|^2$ in equation~(\ref{eq:dimlessA}): for ${\cal B}$, the product $\delta_{p1,x_\xi}^*
\delta_{p1,y_\xi}$, for ${\cal C}$ the complex conjugate of the expression for ${\cal B}$, and
for ${\cal D}$, $|\delta_{p1,y_\xi}|^2$.

\section{Simulation set-up}\label{sec:Simul}
Simulations in the present work require the merging of a 3D geometrical model with
the polarization-sensitive equations developed in Section~\ref{sec:Theory}. We note that the
equations reduce to one and two-beam expressions in \citet{Tobin2023}, so that the main development
in the present work can be considered to be a generalization to many more rays. 
The 3D geometrical model used in the present work was introduced by \citet{Gray2018, Gray2019} and used to model processes involved in maser flaring. It included arbitrary saturation, but without polarization, of molecular populations
in a computational domain that represents condensations or clouds of arbitrary shape. 

\subsection{Domain set-up}\label{subsec:dom_set}
\begin{figure*}
	\centering
	\includegraphics[width=0.97\textwidth]{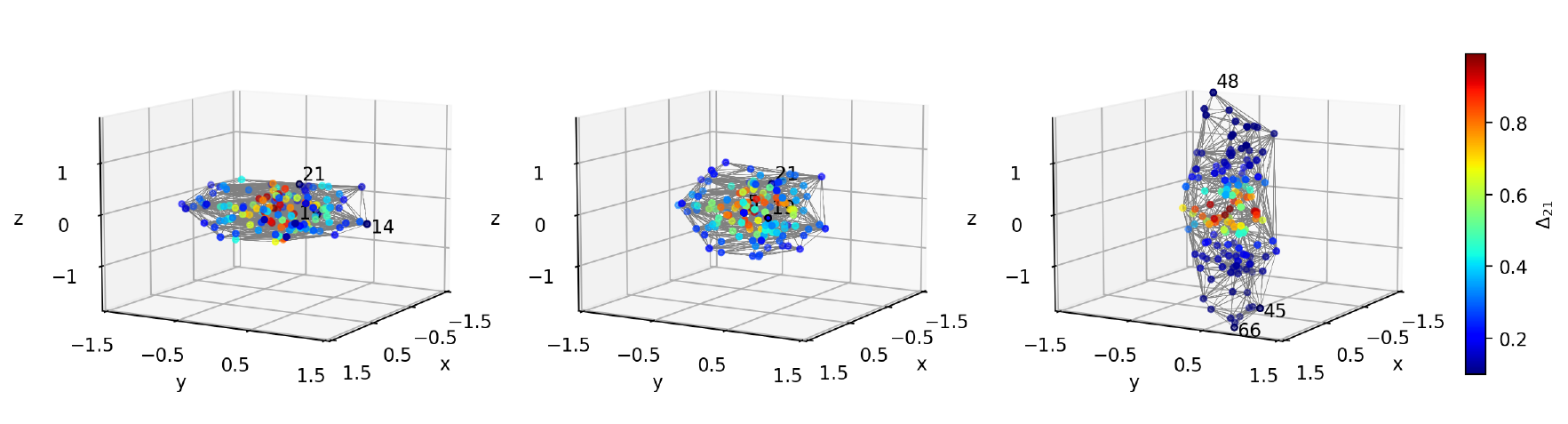}
	\caption{The node distribution, depicted as coloured dots, is shown with tetrahedral element shapes represented by gray lines, in the left, central, and right panels for the oblate, pseudo-spherical, and prolate shapes, respectively. The colours of the dots correspond to inversions in the $21$ transition, on a scale shown on the bar to the extreme
    right of the figure. The three nodes with the bluest colors are the most saturated in each panel, and 
    are marked with their node number to the top-right of each node.}
	\label{fig:3d_p137}
\end{figure*}

We start by creating a domain with a node distribution that represents a single cloud. The node positions 
are generated at random, in a manner that makes any position $(r,\theta,\phi)$, within an enclosing sphere
of radius 1.0, equally likely as the site of the next generated node. The volume occupied or dominated
by each node is not constant.
None of the nodes are constrained to coincide with the enclosing sphere, producing a domain surface
that is rough, and only approximately spherical. We calculated the radial distance from the domain origin
to the outermost point of contact with the domain for every ray source position: for
allocation of source positions, see Section~\ref{ss:solalg}. We call this quantity
the first-contact radius, and use it to quantify the departure of the domain from spherical symmetry.
We define the roughness index as the standard deviation of the first-contact radius divided by the mean
(over all ray source positions).
Much of the present work is based on a domain consisting of 137 nodes with a pseudo-spherical shape, 
named pointy137, which is shown in the central panel of Fig.~\ref{fig:3d_p137}. 
Each node is associated with physical parameters, including number density, temperature, 
velocity, and magnetic field, specified at its position.
Delaunay triangulation is 
employed to define tetrahedral simplex elements within the domain, each consisting of 4 nodes 
placed at their vertices. This domain consists of 757 tetrahedral elements in total, and the total volume 
of the domain is 
2.339 in domain units, calculated by summing the volumes of all elements. In domain
units, all nodes of a pseudo-spherical domain must lie within a radius of 1.0. The pointy137 domain
has a roughness index of 0.4032, measured over 1002 ray source positions. This is significantly
higher than the 249-node pseudo-spherical domain used in \citet{Gray2019} (roughness index of 0.0347
measured over 1692 rays).

Pseudo-spherical domains serve as the basis for constructing oblate and prolate versions, where the 
$z$-axis represents the axis of compression or stretching, respectively. 
If the basic pseudo-spherical domain is then deformed into a prolate or oblate structure, then we use
the method described in \cite{Sugano1998} to ensure volume conservation. 
The axial redistribution factors for achieving these shapes are 2:2:1 for oblate and 1:1:4 for prolate configurations in 
the $x$:$y$:$z$ directions. The modified domains in oblate and prolate shapes are 
illustrated in Fig.~\ref{fig:3d_p137}, depicted in the left and right panels, respectively.
We note that all domains used in the present work have rather small numbers of nodes and
elements, and that therefore the models computed below should be considered as
illustrative or test-of-concept examples. This issue is also mentioned briefly in Section~\ref{sec:Dis_and_Con}.

We seek a solution for the inversions of the three helical transitions at all nodes of
the domain using the finite element method (FEM). For the simplex elements utilized in this study, a scalar 
field $Q(x,y,z)$ at any position within an element is defined by the shape function coefficients 
of the element at that position as follows:
\begin{equation}\label{eq:FEMfund}
	Q(x,y,z)= \!\! \sum_{m=1}^{4}Q_m f_m(x,y,z)
                 = \!\! \sum_{m=1}^{4}Q_m (a_m + b_m x + c_m y + d_m z),
\end{equation}
where $Q_m$ is the field value at node $m$ and $f_m(x,y,z)$ is the shape function. Details on calculating these functions can be found in \cite{Liu2014}, Chapter 9 - FEM for 3D Solid Elements. Computed shape function coefficients,
$a_m$ to $d_m$, were subjected to a sum test to a tolerance set by the parameter \texttt{shtol}, equal to $7.5\times10^{-13}$. 

Starting with a ray that propagates through an element from an entrance point to an exit point, we parameterize distance along this straight-line route via the variable $t$, which is 0 on entry and 1 on exit.
In terms of $t$, the modified form of $f_m(x,y,z)$ becomes $A_m + B_m t$ where $A_m$ is the bracketed part of
the final expression in equation~(\ref{eq:FEMfund}), evaluated at the entry point, and
$B_m$ depends on the distance between the exit and entry positions. Since the solution of the RT
equation (see Appendix~\ref{ss:classic}, equation~\ref{eq:rtsoln}) depends upon repeated
applications of distance-integrated $2\times 2$ matrices of the type found in equation~(\ref{eq:rtfinal}),
it is sufficient to integrate matrix elements of the form we see in equation~(\ref{eq:dimlessA}).
For the sake of example, we will assume here that the profile function $\mathfrak{X}$ is independent
of position (implying constant kinetic temperature and magnetic field magnitude), and take $\mathfrak{G}$
to be an example of the elements ${\cal A}-{\cal D}$, so that a general form of equation~(\ref{eq:dimlessA}) is
\begin{equation}
    {\cal G} = \sum_{p=2}^4 \mathfrak{X}_n^{p1} \mathfrak{D}_{p1}(\tau_\xi) g(\tau_\xi),
\label{eq:coeffG}
\end{equation}
where $g(\tau_\xi)$ is one of the dipole product expressions described after equation~(\ref{eq:gainco0}).

We integrate equation~(\ref{eq:coeffG}) through the element, index $j$, from entry to exit point,
changing variable from $\tau_\xi$ to $t$, so that the range of integration is from 0 to 1. If both
distance-dependent quantities in equation~(\ref{eq:coeffG}) are represented through shape function
coefficients, the integral of ${\cal G}$ is
\begin{equation}\label{eq:[p1]^r}
	\begin{split}
I_j & = \tau_j \!\! \sum_{p=2}^4 \!\! \mathfrak{X}_n^{p1} \!\!\!\! \int_0^1 \!\left[ 
       \sum_{m=1}^M \!\!\mathfrak{D}_{p1,m} (A_m + B_m t) 
                                                       \right]
                                                       \!\!\left[
        \sum_{\mu=1}^M \! g_\mu (A_\mu + B_\mu t)
                                                       \right] dt
	\end{split}
\end{equation}
where subscripts $m,\mu$ refer to nodes from a total of $M$ per element and $\tau_j=\gamma_0 s_j$
where $s_j$ is the length of the ray path through element $j$. It is straightforward
to carry out the integral in equation~(\ref{eq:[p1]^r}). Moreover, since $g$ is a known function of
position, we can always suppress the $\mu$-sum by defining new parameters
\begin{subequations}
   \begin{align}
    X_j^{(\xi)} &=\sum_{\mu} g_\mu  (A_\mu + B_\mu / 2) \\
    Y_j^{(\xi)} &=\sum_{\mu} g_\mu  [(A_\mu / 2) + (B_\mu / 3)],
    \end{align}
\end{subequations}
such that the integrated form of equation~(\ref{eq:[p1]^r})
becomes
\begin{equation}\label{eq:p1int}
I_j = \tau_j \!\! \sum_{p=2}^4 \!\! \mathfrak{X}_n^{p1} \!\! \sum_{m=1}^M \!\! \mathfrak{D}_{p1,m} \!\left(
       A_m X_j^{(\xi)} + B_m Y_j^{(\xi)}
                                                                              \right) ,
\end{equation}
noting that $A_m,B_m,X_j$ and $Y_j$ are all ray-dependent. If we define a new saturation coefficient
$\Phi_{m,j}^{(\xi,g)}$ as everything inside the brackets in equation~(\ref{eq:p1int}) and sum over
all $N_\xi$ elements along the ray path, then the final integrated matrix element is
\begin{equation}\label{eq:finalmatco}
    I = \sum_{j=1}^{N_\xi} \tau_M \tau_j \sum_{p=2}^4 \mathfrak{X}_n^{p1} \sum_{m=1}^M
         \mathfrak{D}_{p1,m} \Phi_{m,j}^{(\xi,g)} ,
\end{equation}
where $\tau_M$ is the depth multiplier needed to increase the saturation state of the model from one job to the next.
The great advantage of the coefficients $\Phi_{m,j}^{(\xi,g)}$ is that they can be computed just once
at the initialization of the code. The number that need to be computed is less than $M \times J$
multiplied by the largest $N_\xi$, multiplied by the number of unique versions of $g$, of which there are $6$.

\subsection{Parameter set-up}
We take the Land\'{e} factor for the $J=1$ state to be 0.74048\,rad\,s$^{-1}$ per milligauss of applied magnetic field \citep{P-S&V2013,Lankhaar2019}. This factor provides a frequency shift for a sublevel with magnetic quantum number $M_J$ of
\begin{equation}\label{eq:DelOmz}
	\Delta\omega_{\text{Z}} = (1/2)g\Omega = 740.48 B M_J \;\;\mathrm{rad\,s^{-1} },
\end{equation}
where $B$ is the magnetic flux density at some domain node in units of Gauss. We define $N_{\text{Z}}$ to 
be the number of frequency channels per Zeeman shift, and we obtain the angular frequency width of $\delta\omega = \Delta\omega_{\text{Z}}/N_{\text{Z}}$ rad/s (which corresponds to an observational sampling time $t_{\text{p}} = 2\upi/\delta\omega$ s from a Fourier relation). The Doppler width in angular frequency is
\begin{equation}\label{eq:DelOmeD}
	\begin{split}
		\Delta\omega_{D}&=2\upi\Bigl(\frac{2k_BT_K\nu_0 ^2}{m_{\text{SiO}}c^2}\Bigr)^{1/2}\\
		&= 1.7504\times10^4\times\sqrt{T_K} \;\;\mathrm{rad\,s^{-1}},
	\end{split}
\end{equation}
where $\nu_0$ is the central frequency of the $\pi$ transition equal to 43.1220\,GHz, $m_{\text{SiO}}$ is the molecular mass of SiO which is 44$\times$1.67$\times10^{-27}$ kg, and $T_K$ is kinetic temperature. To reasonably simulate an AGB-star CSE, we adopt $B$=35\,G, which is on the large side, but has the following advantages. Firstly the large
magnetic field implies a smaller model, in the sense of fewer frequency channels, if $N_{\text{Z}}=1$. Secondly,
the condition $g\Omega \gg R$ is more secure. $B$ is usually aligned with the +z direction. We set $T_K$=1300 K, as suggested by \citet{Vlemmings2019}. The magnetic field should also not be made too large: combining 
equations \ref{eq:DelOmeD} and \ref{eq:DelOmz}, we find that $\Delta \omega_D/\Delta \omega_{\text{Z}}$ falls
below 10 unless $B < 74.8 \sqrt{T_{3}}$\,G, where $T_{3}$ is $T_K$ in units of 1000\,K.
With our value of $T_K$, we should never exceed 85\,G.

We define three new unit-less parameters for setting up the number of channels, which are $\texttt{gamdec}$, $\texttt{gausslor}$, and $\texttt{ndopw}$, defined respectively as the ratio of the loss rate and
channel width, $2\upi\Gamma/\delta\omega$, the ratio of the Doppler width and loss rate, $\Delta\omega_D/(2\upi\Gamma)$,
and the width of the full spectral profile in Doppler widths. In the present work, we use $\texttt{ndopw}=3.0$.
With parameter values from Table~\ref{tab:parms}, the baseline number of channels is 73, obtained by multiplying $\texttt{ndopw}\times\texttt{gamdec}\times\texttt{gausslor}$. To simplify calculations, we consider $N_{\text{Z}}$=1 channel. Two more channels result from the Zeeman shift of the line centre of the $\sigma$ transitions, with 1 channel each for left- and right-handed components. Therefore, the final number of channels is 75 and the central channels $m_{p1}$ 
are 37, 38, and 39 for the transitions $p1=21$, 31 and 41, respectively. All parameters of SiO and the frequency setup are summarized in Table~\ref{tab:parms}.

\begin{table*}
	\centering
	\caption{Parameters for a 3D simulation of $v$=1 and $J$=1-0 SiO maser transition in CSE of an AGB star.\label{tab:parms}}
	\begin{tabular}{c l l l}
		\hline
		Parameter & Description & Value & Unit \\
		\hline
		SiO parameters & & & \\
		$g\Omega/B$ & Zeeman splitting & 1480.96 & rad\,s$^{-1}$\,G$^{-1}$ \\
		$\Gamma$ & Loss rate & 5 & s$^{-1}$ \\
		$B$ & Magnetic field strength & 35 & G \\
		$\rho$ & Gas density & 10$^{8}$ & cm$^{-3}$ \\
		$T_K$ & Kinetic temperature & 1300 & K \\
		$I_{\text{bg}}/I_{\text{sat}}$ & Background intensity / Saturation intensity & 10$^{-6}$ & \\
		\hline
		Frequency parameter & & & \\
		$\Delta\omega_{\text{Z}}$ & Angular frequency of Zeeman splitting & 2.5917$\times10^{4}$ & rad/s\\
		$\Delta\omega_{\text{D}}$ & Angular frequency of Doppler width & 6.3110$\times10^{5}$ & rad/s\\
		$\delta\omega$ & Resolution of the angular frequency channels & 2.5917$\times10^{4}$ & rad/s\\
		$t_{\text{p}}$ & Observational sample time & 2.4244$\times10^{-4}$ & s \\
		$\texttt{gamdec}$ & Decoherence rate as a multiple of bin width & 1.2122$\times10^{-3}$ & \\
		$\texttt{gausslor}$ & Ratio of gaussian and Lorentzian widths & 2.0089$\times10^{4}$ & \\
		$\texttt{ndopw}$ & Multiplier of Doppler width across full spectral profile & 3 & \\
		\hline
		Simulation parameters & & & \\
		$\texttt{accst}$ & The number of allowed re-starts in the iterative solver
		& 10 & \\
		$\texttt{avwait}$ & The maximum number of orthomin iterations allowed & & \\
		& within a single re-start & 50 & \\
		$\texttt{nprev}$ & Previous solutions for extrapolated guess & 3 -- 9 & \\
		$\texttt{raypow}$ & Control the number of rays per tile in nodal solution & 5 & \\
		$\texttt{rayform}$ & Control the number of rays per tile in formal solution & 10-200 & \\
		$\texttt{epsilonK}$ & Target accuracy for orthomin(K) solution & 10$^{-8}$& \\
		$\texttt{shtol}$ & Sum-Test tolerance for shape functions
		& 7.5$\times10^{-13}$ & \\
		\hline
	\end{tabular}
\end{table*}

\subsection{Solution Algorithm}
\label{ss:solalg}
The four forms of 
equation~(\ref{eq:finalmatco}) and equation~(\ref{eq:inversion}) are used to calculate the inversion at node 
position $\boldsymbol{r}$ and channel $n$. 
As saturation increases, the primary effect on the population inversions
is to reduce them from an initial value of 1 towards 0. This is in accord with the expectations for most astrophysical masers \citep{Lankhaar2019, Richards2011A&A}. The number of equations to be
solved is equal to the number of nodes multiplied by the number of transitions,
since electric field amplitudes may be eliminated from equation~(\ref{eq:inversion}) via
equation~(\ref{eq:rtsoln}). For the pointy137 domain, consisting of 137 nodes and 3 transitions, there 
are a total of 411 unknown inversions.

The electric field background introduced in equation~(\ref{eq:rtsoln}) originates from a celestial sphere 
source with a usual radius of 5 domain units, rendering it entirely external to the domain. This source is 
divided into ray-based tiles by subdividing the faces of a regular icosahedron inscribed on the
sphere (see \citealt{Satoh2014} for details). The number of rays is determined by a parameter 
known as $\texttt{raypow}$, which has been set to 5, resulting in 1002 rays directed towards each target node. 
Each ray interacts with the domain if it traverses at least one element on its path to the target. 
Once a solution has been obtained at a certain depth, the depth multiplier, $\tau_M$ in
equation~(\ref{eq:finalmatco}), can be incremented by a small value and another solution sought. 
A number of previous solutions, obtained at lower depths, specified by the parameter \texttt{nprev}, can be used
to compute starting estimates for this process.

The background electric field amplitudes that are used in both nodal and formal solutions have 
randomly chosen phase and amplitude with a seed number to control the list of random numbers.
Formal solutions are described later in this subsection.
The random number generator is essentially the {\sc ran2} algorithm from
\citet{Press1996numerical}. Unless otherwise stated, the same seed was used
for all domain and simulation calculations. In the nodal solutions the $x_\xi$ axis of ray $\xi$,
propagating in the $z_\xi$ direction, is also randomly chosen within the plane perpendicular to $z_\xi$.
Axes perpendicular to ray directions are selected differently in formal 
solutions (see Section~\ref{subsec:form_results}). Our background structure is similar in
style to that used by \citet{Dinh-v-Trung2009MNRAS}, who used the same generator to produce
random phases, but employed constant amplitudes.

Expressions involving molecular dipoles, such as the dot-products of dipoles and field amplitudes
in equation~(\ref{eq:inversion}) have been kept in a general form for as long as possible: here we
describe how they are evaluated during computations.
Since we know the unit vectors for each ray in the domain coordinates (from the background calculation
just described), and we know the magnetic field, and therefore the direction of
the dipole of the $\pi$ transition, in the same coordinates from its specification at
every node, the dot products are straightforwardly evaluated in the $(x,y,z)$ system of the domain.
The only difficulty is that the $\hv{x}'$ and $\hv{y}'$ unit vectors needed for the $\sigma$
dipoles are only constrained to be in the plane perpendicular to the magnetic field. We fix
$\hv{y}'$ through the cross-product expression $\hv{y}' = \hv{z} \times \vec{B}/ (B \sin \theta_B)$,
except in the case where $\hv{z}$ and $\vec{B}$ are parallel, when $\hv{y}' = \hv{y}$. The remaining
unit vector $\hv{x}'$ then follows from a cross product of $\hv{y}'$ with $\vec{B}$.

At each iteration, we use the non-linear Orthomin(K) algorithm, as outlined in \cite{Chen2001}, to minimize the residuals from all equations. The Orthomin algorithm is iterative and allows for multiple restarts (usually 10-15) if convergence is excessively slow. We set the number of allowed-restarts, \texttt{accst}, and the number of iterations allowed within a single-restart, \texttt{avwait}, equal to 10 and 50, respectively. Our target convergence, \texttt{epsilonK}, is typically set at 10$^{-8}$, but under conditions of strong saturation, it has been necessary to accept convergence levels up to a few times 10$^{-6}$. The most challenging convergence issues typically arise at the onset of saturation. The result of this iterative process is referred to as a nodal solution.

After obtaining a nodal solution for the inversions, subsidiary calculations, known as formal solutions, were computed for the electric field components passing through the saturated domain from a source disc to a distant observer, typically located at a distance of 1000 domain units. Ray positions on the source disc were determined using an algorithm proposed by \cite{Beckers2012}, capable of generating rays with an equal-area distribution. Similarly to the nodal solution process, the number of rays can be specified by a parameter called $\texttt{rayform}$, which was set to 200, resulting in 4107 rays directed towards the observer's position. 
This calculation utilizes equation~(\ref{eq:RT_eq}) as the primary equation to compute the output rays in terms of electric field components within the observer's sky plane.
Finally, the electric fields are converted into the observer's Stokes parameters to complete the formal solution.

Referring to Table~\ref{tab:parms}, the simulation parameters are configured to facilitate the rapid attainment of a converged solution. If convergence is not achieved, adjustments can be made by increasing the parameters $\texttt{accst}$ or $\texttt{nprev}$ as necessary. The main calculation is implemented in the Fortran programming language using the GNU compiler. The depth multiplier ranges from 0.1 to 20.0, or 40.0 in
the case of the tube domain, with a standard step of 0.1, to give a
well sampled set of simulations at different degrees of saturation.

\section{Simulation Results}\label{sec:Results}
\subsection{Nodal Solution}\label{ss:nodsols}
\begin{figure*}
	\centering
	\includegraphics[width=0.96\textwidth]{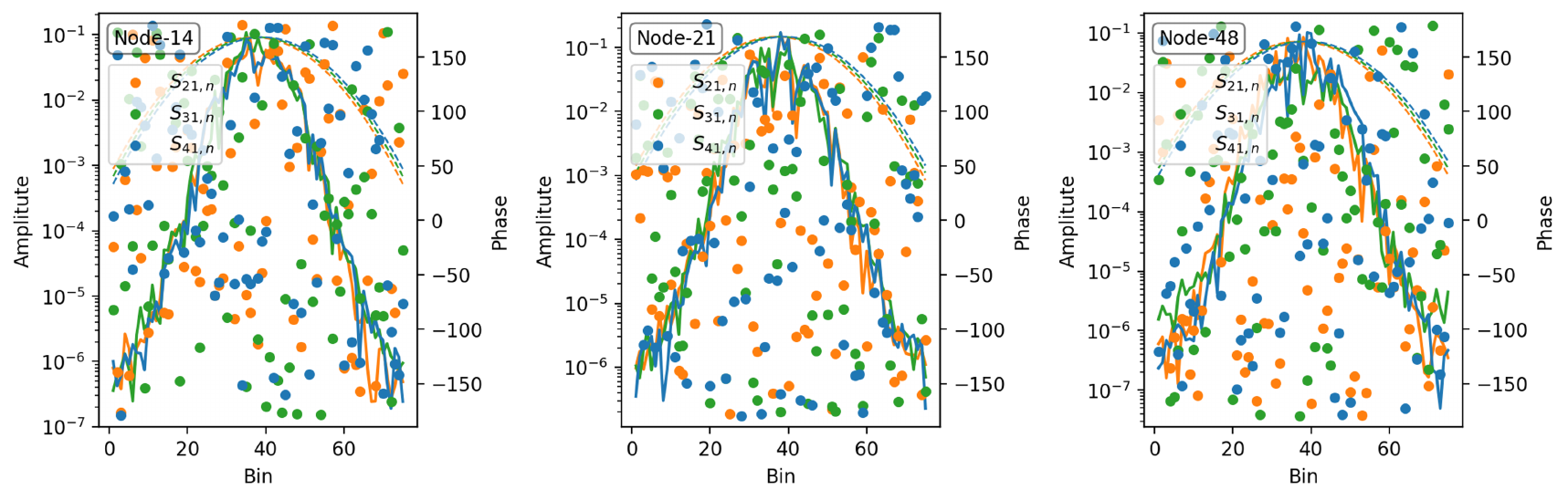}
	\caption{Spectra of ray-summed off-diagonal elements at depth 20.0: from left to right the panels display the 
    results for the most saturated node from the oblate, pseudo-spherical, and prolate shapes, respectively.
    Amplitudes are represented by solid lines, while phases are denoted by dots. Colours indicate $S_{21}$ (orange), $S_{31}$ (green), and $S_{41}$ (blue). Dashed lines represent an envelope function from
    equation~(\ref{eq:approxvoigt}). }
	\label{fig:spec_pop_p137}
\end{figure*}

Example results for a ray-summed off-diagonal DM element, in all three transitions at a depth of 20.0, are depicted in Fig.~\ref{fig:spec_pop_p137}. As an off-diagonal DM solution is a 
complex number, the figure illustrates both the amplitude and phase of this number across frequency. The selected node for each panel (or domain type) is the most saturated (smallest remaining inversion) in the $21$ transition.   
We note that the node chosen on the above basis is different for each domain type, and the node
number is shown in the top left-most box of each panel in Fig.~\ref{fig:spec_pop_p137}. The positions
of these nodes, 14, 21 and 48 for the oblate, pseudo-spherical and prolate domains respectively, are
marked in Fig.~\ref{fig:3d_p137}.

The phase, while calculated from the real and imaginary components of off-diagonal DM elements, does not significantly impact the DM, as it remains random, even under conditions of strong saturation, in line
with expectations for a Doppler-broadened model with a large `mode separation', or $\delta \omega > \Gamma$, \citep{1978PhRvA..17..701M}. The amplitude results should 
follow a scaled version of equation~(\ref{eq:offred}), where the profile function is a factor
for all rays, $\xi$. We see in Fig.~\ref{fig:spec_pop_p137} that the amplitude 
indeed peaks close to the line-centre channel for all three transitions in all three shapes. 
The largest amplitudes exceed $10^{-2}$, representing a gain over typical background levels 
of order $10^5$, and therefore an intensity gain of order $10^{10}$. 
The solution amplitudes of all transitions have a spectrum that is considerably narrowed
compared with the Voigt envelope in all three panels of Fig.~\ref{fig:spec_pop_p137}. This narrowing is
expected during amplification, but our assumption of CVR (see Section~\ref{ss:cvr_approx})
prevents possible re-broadening on saturation.

\begin{figure*}
	\centering
	\includegraphics[width=0.75\textwidth]{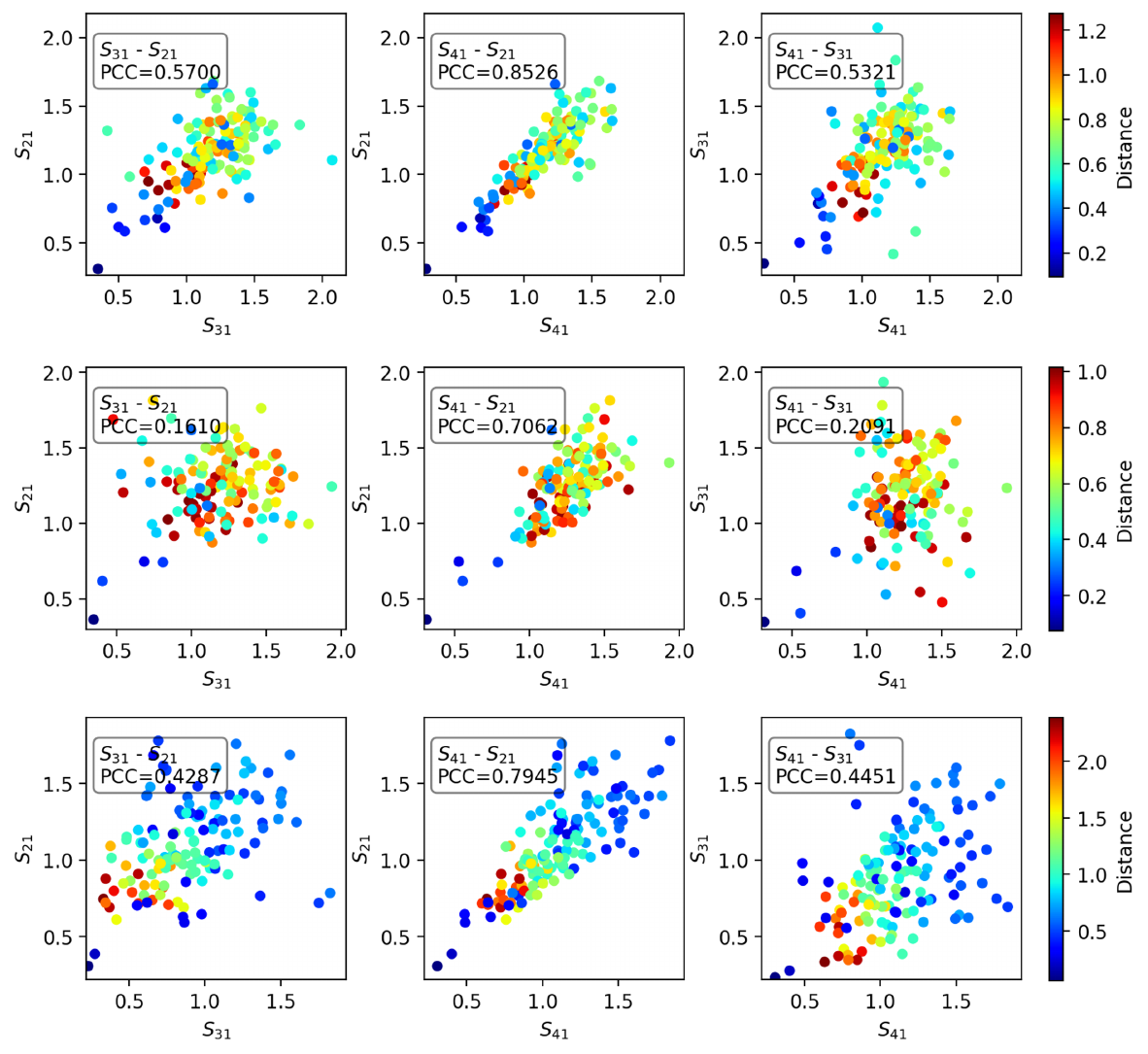}
	\caption{Correlation of the the integrated amplitudes of the three transitions at depth = 20.0. From left to right, panels display correlation plots in the $x$-$y$ axes for $S_{31}$-$S_{21}$, $S_{41}$-$S_{21}$, and $S_{41}$-$S_{31}$, along with their corresponding Pearson correlation coefficients (PCC). The graphs from top to bottom represent the correlation analysis for the oblate, pseudo-spherical, and prolate shapes, respectively. The colours in the plots indicate the distance of each node from the origin.}
	\label{fig:PCC_p137}
\end{figure*}

We computed Pearson correlation coefficients (PCC) to assess the degree of correlation 
between the integrated amplitudes of the three Zeeman transitions. The PCC value ranges 
from -1 to 1, where -1 indicates a strong negative relationship, 1 denotes a strong positive 
relationship, and 0 implies no relationship. 
We compute the integrated amplitude by taking the magnitude of the off-diagonal DM element in each channel, and
summing these magnitudes across all frequency channels.
We found that the integrated amplitudes of the $\sigma$ transitions ($S_{21}$ and $S_{41}$) exhibit the
expected strong positive correlation in all three domain shapes, while both $\sigma$ transitions 
are less well correlated with the $\pi$ transition. 
The correlation across all nodes from the three shapes is depicted in Fig.~\ref{fig:PCC_p137}.

We might also expect integrated amplitude to correlate with saturation. From the distribution
of the most saturated nodes (smallest remaining inversions) displayed in Fig.~\ref{fig:3d_p137}, this
would suggest that the nodes with the largest integrated amplitudes should be found predominantly
at large distances from the origin. In the oblate and pseudo-spherical domains, such a distance-amplitude
relation does appear to hold at least approximately: nodes closest to the domain origin (darkest
blue colours in Fig.~\ref{fig:PCC_p137}) tend to have smaller integrated amplitudes than the
more distant nodes (dark red colours). However, in most panels of Fig.~\ref{fig:PCC_p137}, the
nodes with the highest integrated amplitudes (farthest from their {\it graph} origin) have only
moderate distances from their domain origin. The distance-amplitude relation is not replicated in the prolate shape:
there is a significant population of high-amplitude nodes in this case that are quite close
to the domain origin (blue colours). Overall, while inversion solutions follow the general expectation
of masers with an unsaturated core and a saturated exterior, in agreement with unpolarized 3D models
\citep{Gray2018} and 1D spherical masers (for example, \citealt{elitzurbook}), the spatial distribution
of integrated amplitudes requires further explanation, which we give, at least partially, in
relation to Fig.~\ref{fig:evo_p137} below.

We will focus on $S_{21}$, the integrated amplitude of the $\sigma^{+}$ transition for the remainder of this subsection.
Subsequent findings are applicable to $S_{41}$ as well, given its strong correlation with $S_{21}$. Results
could be more different for $S_{31}$, the integrated amplitude of the $\pi$ transition.
Fig.~\ref{fig:evo_p137} showcases the evolution of nodal solutions,
with the depth multiplier ranging from 0.1, in steps of 0.1, to 20.0. This evolution is considered
in terms of the following parameters:
the node-averaged stimulated emission rate, the integrated amplitude, $S_{21}$, and the inversions at 
the three most saturated nodes and the three least saturated nodes. 
We calculate the stimulated emission rate in transition $p1$ at nodal position $\boldsymbol{r}$ by, 
\begin{equation}
	R_{p1}(\boldsymbol{r}) = B_{p1}\Bar{J}_{p1}(\boldsymbol{r}),
\end{equation}
where $B_{p1}$ is the Einstein B coefficient for stimulated emission, and $\Bar{J}_{p1}(\boldsymbol{r})$ is the mean intensity of radiation arriving at node position $\boldsymbol{r}$ which is approximately related to the inversion by $1/\Delta_{p1}(\boldsymbol{r}) = 1+(\Bar{J}_{p1}(\boldsymbol{r})/I_{\text{sat}})$ \citep{Gray2018}. According to the definition of $I_{\text{sat}}$ from \cite{Tobin2023}, we obtain the following simple form for the nodal stimulated emission rate,
\begin{equation}
	R_{p1}(\boldsymbol{r}) = \frac{8\Gamma}{3}\Bigl(\frac{1}{\Delta_{p1}(\boldsymbol{r})}-1\Bigr) .
\end{equation}

\begin{figure*}
	\centering
	\includegraphics[width=0.96\textwidth]{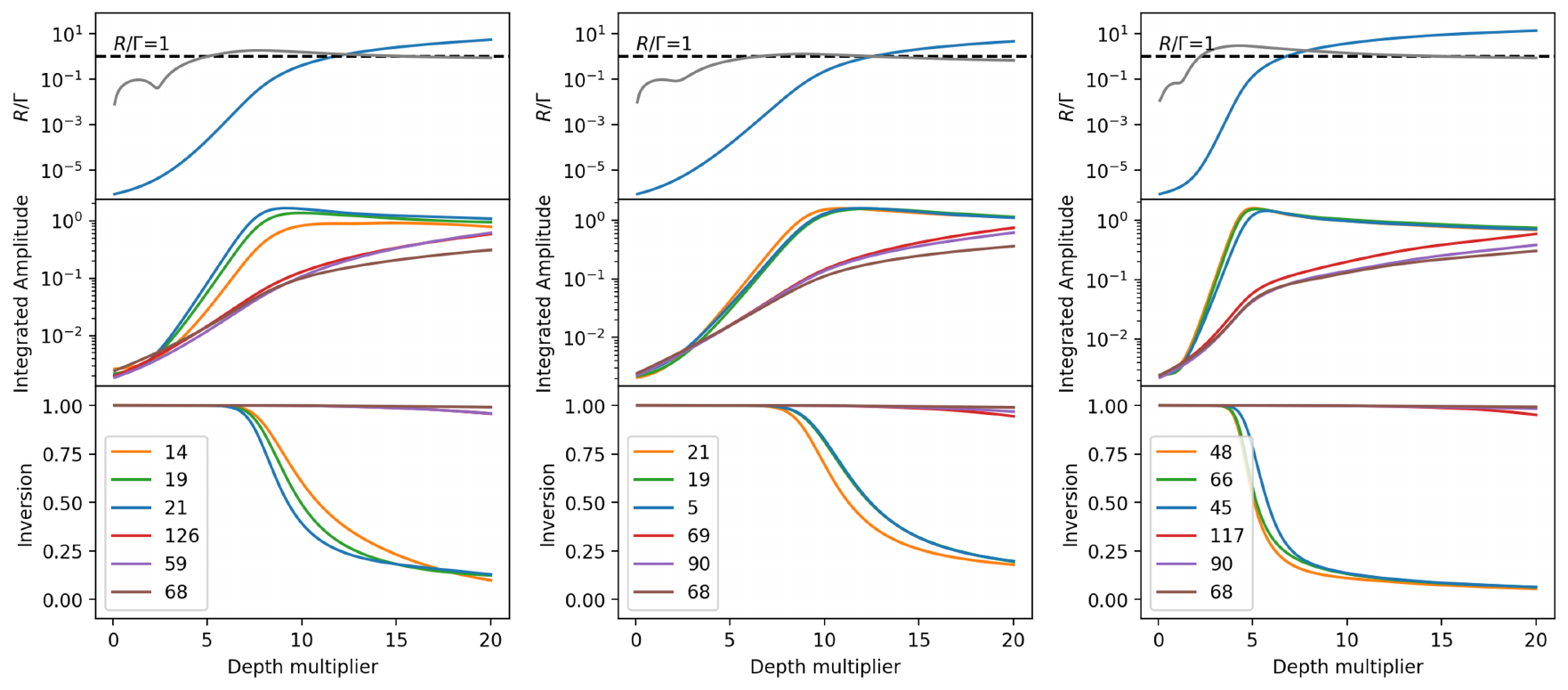}
	\caption{From top to bottom, as a function of depth $\tau_M$, 
    nodal averages of the stimulated emission to loss rate ratio, $R/\Gamma$, (solid blue line) 
    with normalized standard 
    deviations (S.D./average, solid gray line), ray and integrated amplitudes of $S_{21}$ 
    (second row) and inversions ($\Delta_{21}$) (bottom row) for the three most saturated 
    nodes (3 upper numbers in box), and the three least saturated nodes (3 lower numbers in box). 
    In the top row, a black dashed line is drawn at $R/\Gamma =1$. The left, middle, and right panels 
    refer to oblate, pseudo-spherical, and prolate shapes, respectively.}
	\label{fig:evo_p137}
\end{figure*}

Inversions, which are the fundamental quantities solved for, are limited to the range 0-1 in
their scaled form. In Fig.~\ref{fig:evo_p137}, we observe saturation in all the nodes plotted, with
the numbers falling towards zero. At the onset of saturation, inversions vary most rapidly with
depth, and the onset depth differs considerably between the domain shapes. From Fig.~\ref{fig:evo_p137}, the
onset of saturation occurs at depth multipliers of approximately 7, 8, and 4 for the oblate, 
pseudo-spherical, and prolate shapes, respectively.
Integrated amplitudes of the off-diagonal DM elements start at around $10^{-3}$ and rise
towards a final value of approximately $1$. The nodal average of $R/\Gamma$, the
saturation measure, starts below 10$^{-5}$, exceeds the threshold value of $R/\Gamma = 1$ near the onset 
of saturation, and then increases smoothly, reaching a final value of about $10$. These
smooth curves also display the expected form of exponential growth in the unsaturated regime, changing
towards linear growth when saturated.
We also plot in Fig.~\ref{fig:evo_p137}, on the same axes as $R/\Gamma$, the standard deviation of this 
quantity divided by the nodal average. This normalized standard deviation rises to values of 
order 1.0 at, or somewhat before, the onset of saturation, indicating a strongly non-uniform distribution of stimulated
emission rates over the depth range where $R/\Gamma$ changes most rapidly. This effect is
particularly strong in the prolate domain.
The node-averaged stimulated emission rate, $R$, at the final depth is about 50 s$^{-1}$ while the Zeeman 
precession rate, $g\Omega$, is about 4000 s$^{-1}$. These figures confirm that the situation at the 
final depth is still consistent with the case $g\Omega \gg R$, where the magnetic field axis 
remains a good quantization axis. 
If our value of $R/\Gamma \simeq 10$ at the highest depth multiplier is taken as our measure
of saturation, we can use it to estimate the fidelity of the extreme spectral limiting
approximation \citep{Wyenberg2021MNRAS}: if we take $I/I_0$ as their saturation measure, and
$z=0.6L$ as the onset of saturation (see their Figure~4), then a value of $I/I_0$ that is
10 times the value corresponding to $z=0.6L$ will still be $<10^9$. At this level, effects
of strong spectral limiting appear to be modest, although the models are considerably different. 

In each domain, nodes closer to the origin (examples are the lower 3 nodes in the 
keys of Fig.~\ref{fig:evo_p137}) exhibit the very earliest signs of saturation before 
those further from the origin, for example the upper 3 nodes in the keys of Fig.~\ref{fig:evo_p137}.
This is possibly because the interior nodes are intercepted by typically greater
numbers of competing rays.
However, by the onset depth, this situation has changed to the expected situation of an
unsaturated core, and saturated extremities, of the domain. 
At the onset depth, the most heavily saturated nodes are those located close to the longest 
axis, such as nodes along the circumference in the oblate shape and nodes close to the
points of the prolate shape. Fig.~\ref{fig:3d_p137} clearly illustrates that, by the
final value of the depth multiplier, nodes closer to the origin experience weaker 
saturation, while those farther from the origin exhibit stronger saturation that
is changing only slowly.
We note the gentle decay of the integrated amplitudes of the most saturated nodes in Fig.~\ref{fig:evo_p137} (middle row of panels). This is the result of more general saturation, and 
rising integrated amplitudes at some nodes elsewhere in the domain, explaining in part how
a significant population of high-amplitude nodes can be close to the domain origin in Fig.~\ref{fig:PCC_p137}.

\subsection{Formal Solution} \label{subsec:form_results}
\begin{figure}
	\centering
	\includegraphics[width=0.4\textwidth]{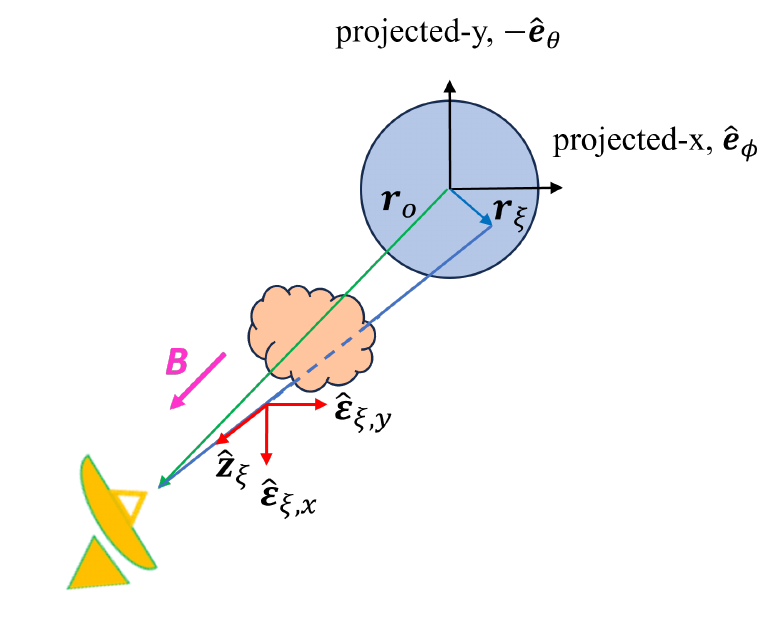}
	\caption{Geometry of a formal solution for an observer at $\boldsymbol{r}_o$ (green arrow 
    aligned with the magnetic field (magenta arrow)). The diagram shows a plane source with 
    projected-x and -y coordinates (unit vectors $\hat{\boldsymbol{e}}_{\phi}$ and $-\hat{\boldsymbol{e}}_{\theta}$, respectively). For ray $\xi$ with position $\boldsymbol{r}_\xi$ (blue arrow) on the source plane, red arrows 
    refer to the directions of ray propagation, $\hat{\boldsymbol{z}}_\xi$, and electric field components $\tilde{\boldsymbol{\varepsilon}}_{\xi,x}$ and $\tilde{\boldsymbol{\varepsilon}}_{\xi,y}$.}
	\label{fig:form_sol_diagram}
\end{figure}

Unless otherwise stated, inversion solutions at the final depth, 20, are used to compute
formal solutions, and we set the \texttt{raypow} parameter to 200, the largest value in
Table~\ref{tab:parms}, which generates 4107 rays towards the observer.
In a formal solution, the observer can be positioned at any location in 
spherical polar coordinates, allowing background rays originating behind the domain to travel
towards the observer, who perceives the domain projected onto their sky $(x,y)$ plane. To ensure a 
safety margin, the radius of the background plane is determined by the longest sky-projected 
inter-node distance in the domain, multiplied by 1.2. The output electric field of each ray,
in each frequency channel, is observed in the projected $x$-$y$ plane. These projected $x$ and $y$ coordinates 
are computed from the observer's position in spherical coordinates ($r, \theta, \phi$) using 
the following equations:
\begin{subequations}
	\begin{align}
		\text{projected-x} & = \hat{\boldsymbol{e}}_{\phi},\\
		\text{projected-y} & = -\hat{\boldsymbol{e}}_{\theta},
	\end{align}
\end{subequations}
where $\hat{\boldsymbol{e}}_{\phi}$ and $\hat{\boldsymbol{e}}_{\theta}$ are the spherical polar 
unit vectors of the observer. In this subsection, we will consider a formal solution with the observer 
placed on the positive z-axis, or in spherical 
coordinates $\boldsymbol{r}_o$ = ($r,\theta,\phi$) = (1000, 0, 0), which aligns with 
the direction of the magnetic field as shown in Fig.~\ref{fig:form_sol_diagram}. In this setup, the 
projected $x$ and $y$ axes correspond to $+y$ and $-x$ of the original (domain) system, respectively. A 
source position of ray $\xi$ is on the background plane with a 
vector of $\boldsymbol{r}_\xi$, so the unit vector in the direction of propagation of this
ray is $\hat{\boldsymbol{z}}_\xi = (\boldsymbol{r}_o-\boldsymbol{r}_\xi)/|\boldsymbol{r}_o-\boldsymbol{r}_\xi|$. 
We define the unit vectors along the $x$ and $y$ components of the electric field of ray $\xi$ by 
\begin{subequations}
	\begin{align}
            \hv{y}_\xi &= \frac{-\hat{\boldsymbol{e}}_\theta \times \hat{\boldsymbol{z}}_\xi}{\sin{\theta_\xi}},\\
		\hv{x}_\xi &=\hv{y}_\xi \times \hat{\boldsymbol{z}}_\xi,
	\end{align}
\end{subequations}
where $\theta_\xi$ is the angle between vector $-\hat{\boldsymbol{e}}_{\theta}$ and $\hat{\boldsymbol{z}}_\xi$, which is close to $\upi/2$ for all rays.

\begin{figure*}
	\centering
	\includegraphics[width=0.96\textwidth]{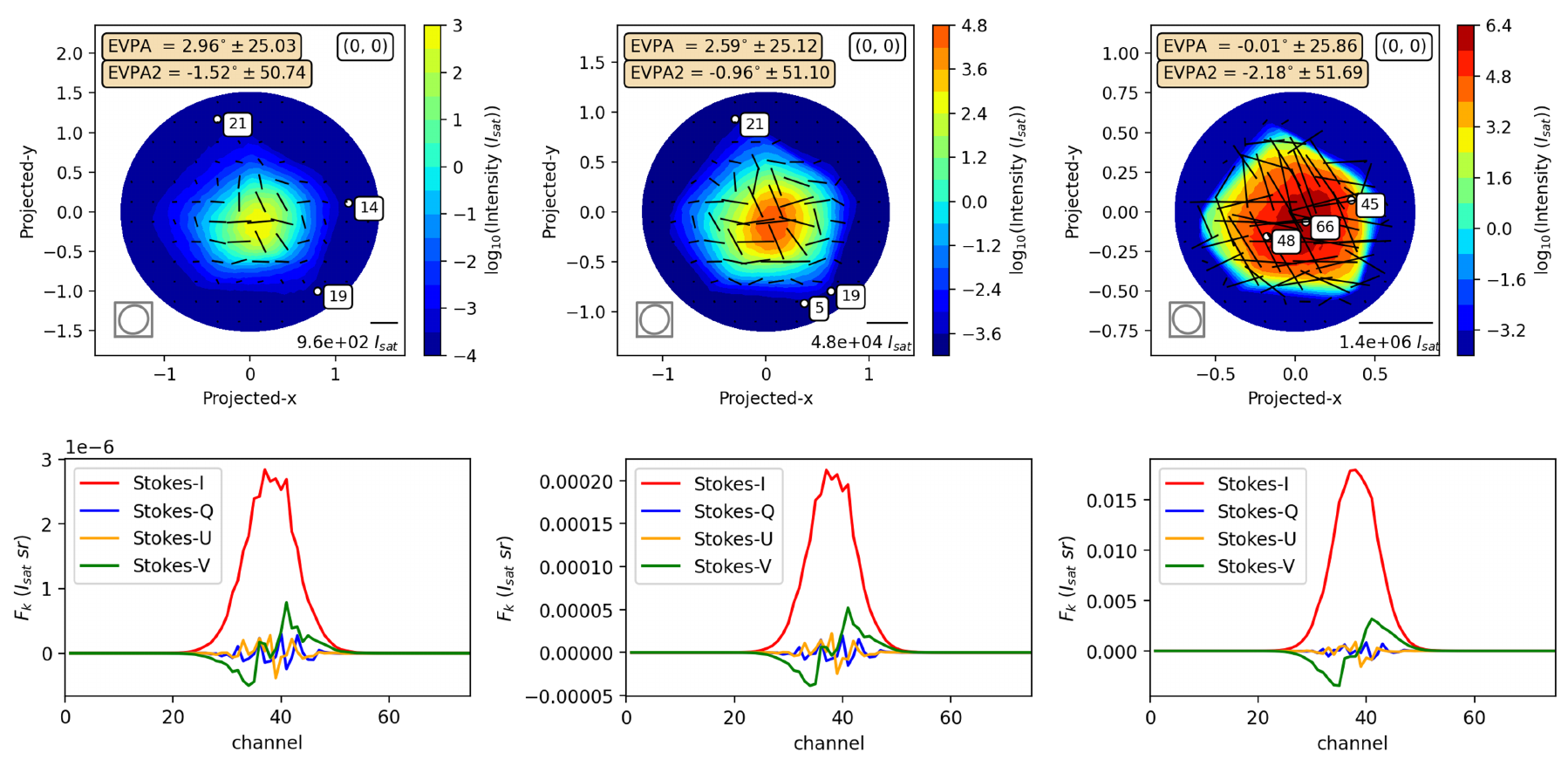}
	\caption{Images and spectra as seen by an observer on the positive $z$ axis (the +z view); the 
    magnetic field of 35\,G points directly at the observer. Top row: Contours of frequency-integrated Stokes-$I$ (intensity) with colour scales to the  right-hand side of each panel. The average EVPA2 of the rays from each cell of a 12x12 grid is shown as a 
    black solid line, with a scale to intensity in the bottom-right corner of each panel. The small box 
    at the bottom-left of each panel shows the polarization ellipse of a Poincar\'e 
    sphere \citep{Deschamps1973poincare}. The average and standard deviation of EVPA and EVPA2 are shown in the top-left boxes, while the top-right box shows the observer's position as ($\theta$, $\phi$). The small, white 
    circles with black-edges mark the projected positions of the three most saturated nodes, with their numbers
    in boxes to the top-right of each node. Bottom row: Stokes spectra with their colour codes shown in the top-left 
    of each panel. Left, middle, and right panels refer to the oblate, pseudo-spherical, and prolate shapes, respectively.}
	\label{fig:contour_p137}
\end{figure*}

The observer's Stokes parameters of the ray $\xi$ at frequency $n$ are given by:
\begin{subequations}\label{eq:stk_calc}
	\begin{align}
		I_{\xi,n} &=|\mathfrak{E}_{\xi x,n}|^2 + |\mathfrak{E}_{\xi y,n}|^2,\\
		Q_{\xi,n} &=|\mathfrak{E}_{\xi x,n}|^2 - |\mathfrak{E}_{\xi y,n}|^2,\\
		U_{\xi,n}&=2\Re(\mathfrak{E}_{\xi x,n} \mathfrak{E}_{\xi y,n}^{\ast}),\\ 
            V_{\xi,n}&=-2\Im(\mathfrak{E}_{\xi x,n} \mathfrak{E}_{\xi y,n}^{\ast}),
	\end{align}
\end{subequations}
where $\Re$ and $\Im$ refer to real and imaginary parts of a complex number. 
The Stokes parameters defined in equation~(\ref{eq:stk_calc}) represent a complete
description of the polarization of ray $\xi$ as seen by the observer.
The results of formal solutions for all 3 cloud shapes are shown as contour maps of 
channel-integrated Stokes-$I$ in units of $I_{\text{sat}}$, with boxes containg EVPA and EVPA2 
values, in Fig.~\ref{fig:contour_p137}. The EVPA is calculated 
from $0.5\arctan(U/Q)$ (range -45 to +45 degrees) to compare with the results 
of 1D models while EVPA2 is calculated from $0.5${\sc arctan2}$(U,Q)$, using the
the {\sc fortran} intrinsic function {\sc arctan2}, which gives a
position angle in the range -90 to +90 degrees. 
The maximum intensity in a map is related to the ray path perpendicular to the sky
plane, and found to be approximately $10^{3}$, $10^{4.8}$, and $10^{6.4}$ for the 
oblate, pseudo-spherical, and prolate shapes, respectively. We note that the prolate (oblate)
domain has its long (short) axis pointing approximately at the observer.

For comparison with spectra, it is also useful to convert ray intensities to a flux density:
\begin{equation}\label{eq:flux_calc}
	\begin{split}
		F_{k,n}&=\oint I_{k,n} \cos\theta d\Omega \\
		&\simeq \sum_{\xi} I_{k,\xi,n} \cos\theta_{\xi} \delta\Omega_\xi,
	\end{split}
\end{equation}
where $k$ refers to a type of Stokes parameter (for $k=0\dots3$ these are $I$, $Q$, $U$, and $V$, respectively). $I_{k,\xi,n}$, $\theta_{\xi}$, and $\delta\Omega_{\xi}$ are the Stokes intensity, offset angle from $\boldsymbol{r}_o$, and solid angle of ray $\xi$, respectively. The bottom panel of Fig.~\ref{fig:contour_p137} shows the Stokes spectra as flux densities. We note that, with the random background electric field amplitudes described in Section~\ref{ss:solalg},
flux densities for the $k\neq 0$ Stokes parameters are $\lesssim 1/1000$ of the Stokes-$I$ flux density in a
background channel.
As with the maps, we also see the effect of long axis orientation in the spectra, with flux 
densities of 3$\times10^{-6}$, 2$\times10^{-4}$, and 1.7$\times10^{-2}$ for the oblate, pseudo-spherical and prolate
shapes, respectively. Flux density values are smaller than contour map intensities owing to the
total solid angle (of all formal-solution rays) of about 3.14 to 3.16 times $10^{-6}$ steradian, dependent on the shapes,
when calculated using equation~(\ref{eq:flux_calc}). 

The spectra in this formal solution are dominated by the two $\sigma$ transitions because the 
magnetic field is pointing towards the observer, leaving the dipole of the $\pi$ transition with no projection on the sky, whilst the $\sigma$ components have left and right-handed dipoles that can generate significant Stokes-$V$
from a coupling to Stokes-$I$. Linear polarization can appear in Fig.~\ref{fig:contour_p137} because of the
form of the electric field background, introduced in Section~\ref{ss:solalg}, and the fact that all four
Stokes parameters can amplify themselves with the same gain coefficient ($\gamma_I$ in equation~(17) of
\citet{Tobin2023}), which is non-zero for lines of sight parallel to the magnetic field and depends on
the $\sigma$ dipoles. Our formal solutions are therefore somewhat analogous to the
single realizations in \citet{Dinh-v-Trung2009MNRAS}. However, the present work employs many more
rays, so that, from the observer's view in Fig~\ref{fig:contour_p137}, we see a rather random arrangement
of polarization vectors in all three domains. The EVPA2 values
of all three domains are dominated by standard deviations exceeding 45\,degrees, reinforcing
the impression of random orientation.
Absolute values of the flux density in the oblate
and pseudo-spherical domains are significantly lower than in the prolate case, and we note
that the most saturated nodes in the former two domains, for example node 21, are located near the map
edges. By contrast the two most saturated nodes in the prolate domain, nodes 48 and 66, are located towards the centre of the right-hand figure, close to the most intense rays. This leads to the idea that
the formal solution rays in the oblate and pseudo-spherical domains are amplifying through
a distribution of inversions controlled, through saturation, by other, higher intensity rays.
Consequently, in the spectrum of the prolate domain where the formal solution rays are
similar to the set controlling saturation, we see features that most closely resemble the expectations from
1D models: a smooth `S-curve' spectrum in Stokes-$V$, and weak linear polarization (especially Stokes-$U$)
when averaged across the spectrum. Larger departures from these expectations occur in the
other two domains, where amplification of the formal solution rays is affected by saturation
controlled by a different set - a situation that is impossible with a single ray in 1D.
See Appendix~\ref{app:ref_view} for alternative formal solutions, including examples where the 
observer's polar angle is 90 degrees. 

The Standard Deviation (S.D.) value in EVPA for a particular ray is calculated from
\begin{equation}
	\begin{split}
		\sigma_{\text{EVPA}} &= \sqrt{\sum_{I_k=Q,U} \bigg(\frac{\partial\text{EVPA}}{\partial I_k} \bigg)^2\sigma_{I_k}^2}\\
		&=\frac{1}{2(Q^2+U^2)}\sqrt{U^2\sigma_Q ^2 + Q^2\sigma_U ^2},
	\end{split}
\end{equation}
where $\sigma$ with a subscript denotes the S.D. of that variable and is also shown in Fig.~\ref{fig:contour_p137}.

\begin{figure*}
	\centering
    \includegraphics[width=0.96\textwidth]{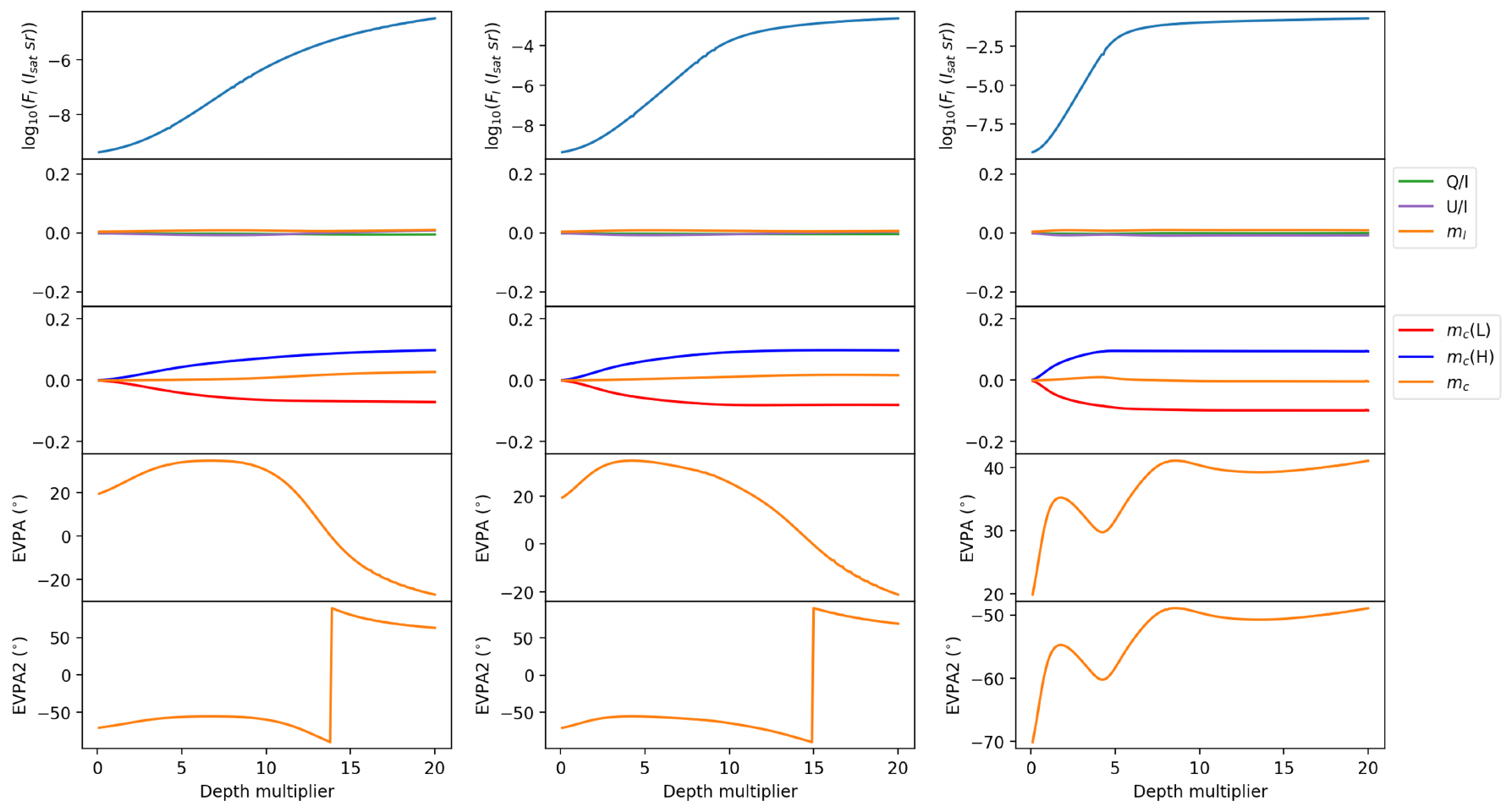}
	\caption{Stokes parameter evolution as a function of depth multiplier for an observer viewing from
    ($\theta$, $\phi$)=(0,0), as in Fig.~\ref{fig:contour_p137}, so the magnetic field points directly 
    at the observer. From top to bottom, rows
    show log$_{10}$(Flux density of $I$), linear polarization ($Q/I$, $U/I$, and $m_l$), circular polarization ($m_c(L)$, $m_c(H)$, and $m_c$), EVPA, and EVPA2, respectively. Columns, from left to right, show the evolution of the 
    oblate, pseudo-spherical, and prolate shapes. For circular polarization, the $L$ (low) and $H$ (high) modifiers
    indicate data generated from channel numbers below and above the line centre, respectively.}
	\label{fig:Evo_stk}
\end{figure*}

We consider in more detail the effect of saturation on polarization by calculating formal solutions at all 
steps of the depth multiplier. Results are shown for all three domain shapes in Fig.~\ref{fig:Evo_stk} for
a view from the $+z$ observer's position. Fractional polarization in a particular channel is calculated from
\begin{subequations}
	\begin{align}
		m_l &= \sqrt{Q^2 + U^2}/I\text{,} \\
		m_c &= V/I\text{,}
	\end{align}
\end{subequations}
where $I,Q,U,V$ are flux densities.
These quantities are channel-averaged, with a Stokes-$I$ weighting, in Fig.~\ref{fig:Evo_stk}.

Initial values of Stokes-$I$ are about $10^{-10}$, from frequency integration of the channel flux densities 
that are themselves of order $10^{-12}$ in the background. Both linear (at other view angles) and
circular polarization develop before the onset of saturation, with circular polarization
appearing at lower depth. The pre-saturation EVPA2 
is in the range -50 to -70 degrees for all three shapes due to the use of the same set of random background rays. 
At the onset of saturation, $V/I$ values become almost constant, whilst $Q/I$ continues
to grow smoothly through the saturated regime at other observer's views, though in the $+z$ view 
shown in Fig.~\ref{fig:Evo_stk} it is never more than 1 per cent at any depth in all three domains.

Circular polarization develops largely before the onset of saturation. Frequency channels
below the line centre, numbered 1-37 were averaged at each depth to form the
low set curve in the middle row panels of Fig.~\ref{fig:Evo_stk}, marked $m_c(L)$, that 
shows left-handed polarization at a level
of 8-10 per cent, whilst the channels above the line centre (numbered 39-75) were used to
generate the high-set curve, denoted by $m_c(H)$, that exhibits right-handed circular
polarization at a similar absolute level. However, the
overall channel-averaged circular polarization is close to zero (curve $m_c$), which is the
result expected for an antisymmetric spectral profile.

The general values of $m_l$, of under 1 per cent in all the domains
are consistent with a viewing direction aligned 
parallel to both the electric dipole of the pi-transition and the magnetic field, given
that the background used in the present work has zero polarization only on average, and
background properties remain important in the unsaturated regime. Departures from zero
in $Q/I$ and $U/I$ are largest in the oblate domain, where the most saturating rays are
perpendicular to the line of sight, of unequal length in general, and the absolute values of the flux density
in Fig.~\ref{fig:Evo_stk} are very small. 

In the oblate and pseudo-spherical domains, the EVPA2 flips through 180 degrees after the onset of saturation,
settling to a final value of approximately 60 degrees. By contrast, the EVPA2 in the prolate
domain exhibits a smooth curve similar to the EVPA in the same domain, but with angles offset by about -90 degrees.
The EVPA shows only
smooth variation, eventually rising by about 20 degrees in the prolate domain. In the oblate
and pseudo-spherical domains, the EVPA rises slowly from 20\,degrees to a peak of about
35\,degrees before declining through the saturated regime to a final value of
approximately -20\,degrees. The EVPA2 flips coincide with sign changes in EVPA.

\subsection{View rotation}\label{subsec:view_rot}
\begin{figure}
	\centering
	\includegraphics[width=0.42\textwidth]{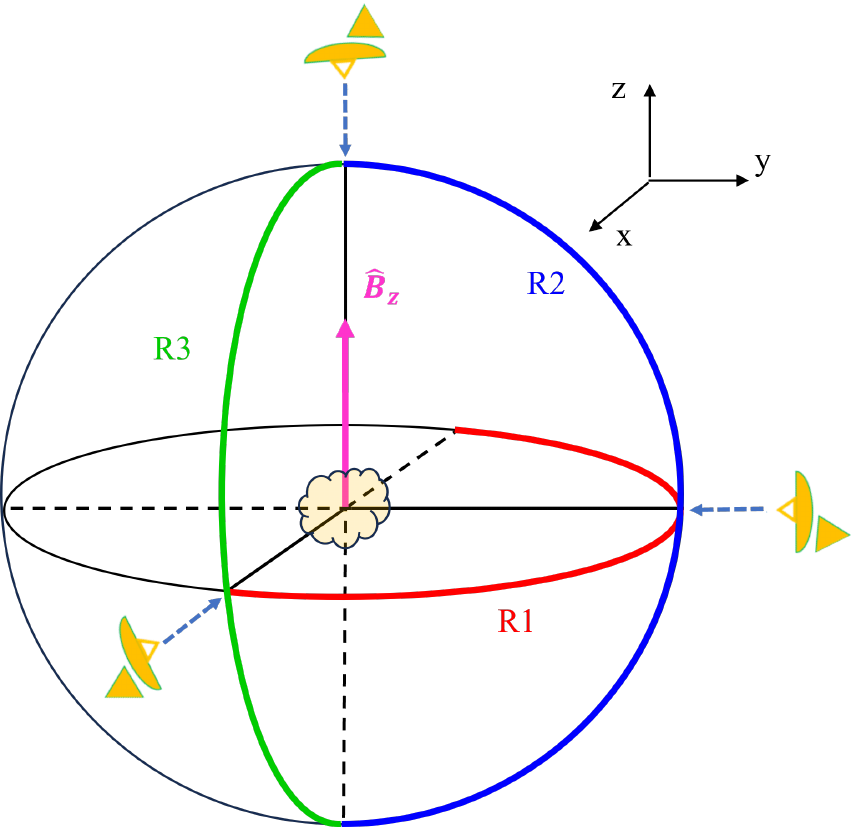}
	\caption{The diagram illustrates the rotation of the observer's view. The solid red, blue, and green lines depict the R1, R2, and R3 rotation patterns, respectively. The cartoon of the antennas represents the reference views considered. A solid magenta arrow indicates the direction of the magnetic field, pointing towards the +z direction.}
	\label{fig:view_rotate}
\end{figure}

In this subsection, we investigate the effect of the observer's viewpoint relative to the
magnetic field (and to the electric dipoles of the three transitions). The direction of the magnetic field is 
again fixed to the $+z$ axis of the domain. As mentioned previously, the observer can be set to 
view the domain from any position. We initially consider three reference views: $+z$, $+x$, and $+y$, represented 
by the yellow antennas in Fig.~\ref{fig:view_rotate}, illustrating the rotation of 
the observer's view. All three shapes are studied in this subsection to compare the effect of cloud shape 
in conjunction with rotated views. The results of the reference views are 
presented in Appendix \ref{app:ref_view}.

Next, we rotate the observer's view smoothly through three arcs labeled R1, R2, and R3, respectively in Fig.~\ref{fig:view_rotate}, in order to observe the resulting changes in polarization. Arc R1 represents 
the rotation at a polar angle of 90 degrees, across the azimuth direction from 0 degrees ($+x$ direction) to 
180 degrees ($-x$ direction). R2 and R3 are configured for fixed azimuth angles of 90 and 0 degrees, respectively, 
with rotation of the polar angle from 0 ($+z$ direction) to 180 degrees ($-z$ direction).

\begin{figure*}
	\centering
    \includegraphics[width=0.89\textwidth]{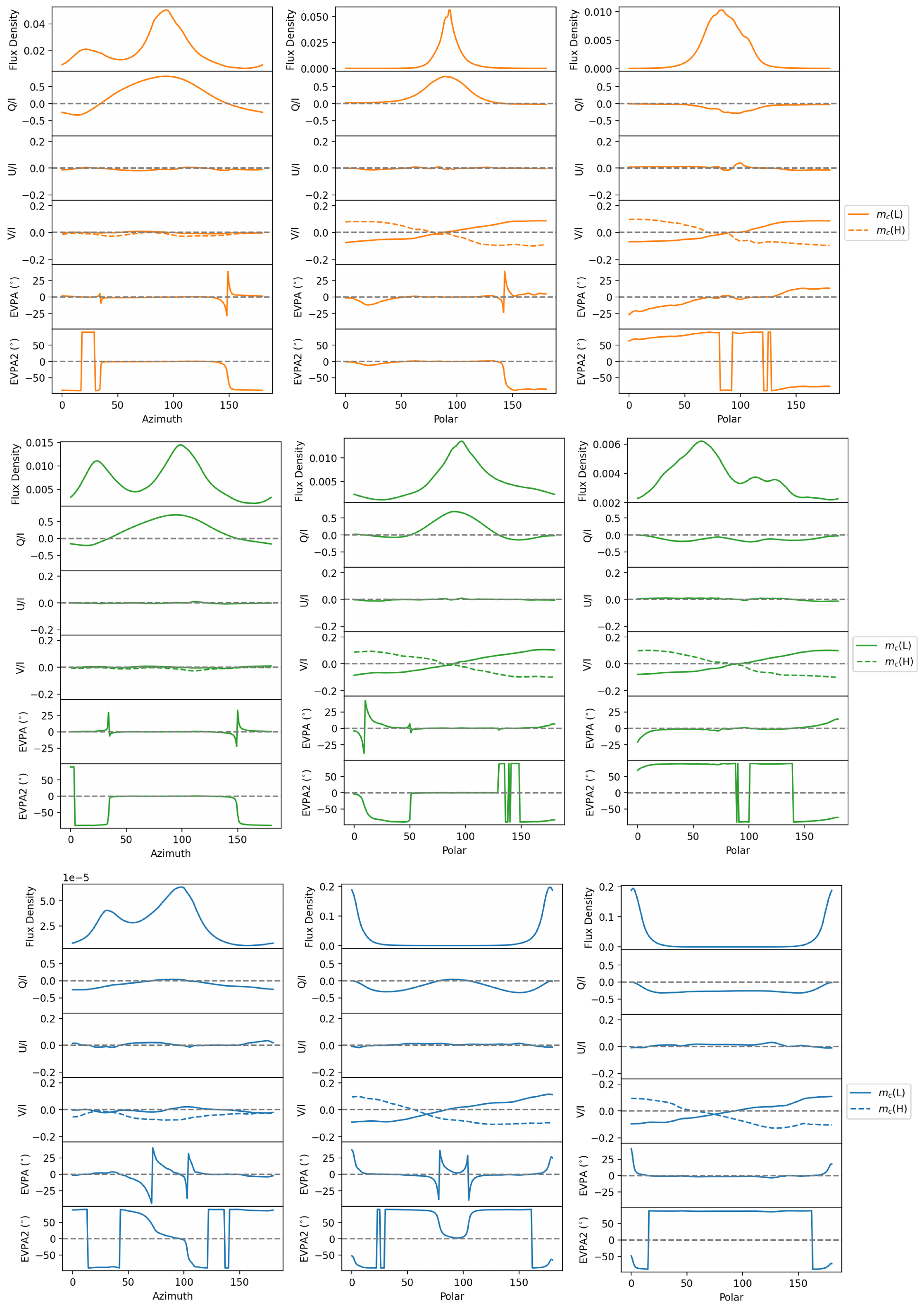}
	\caption{Stokes results for the three rotations shown in the diagram of Fig.~\ref{fig:view_rotate}. Top-to-bottom
    blocks show results for the oblate domain (orange lines), pseudo-spherical domain (green lines), and
    prolate domain (blue lines). Within each block, rows from top to bottom
    plot the flux density in Stokes $I$, fractions $Q/I$, $U/I$, $V/I$ which is separated into $m_c(L)$ (solid line) and $m_c(H)$ (dashed line), EVPA and EVPA2, respectively. Left to right graphs refer to R1, R2, and R3 rotation. Along each arc, there is one formal solution per degree.}
	\label{fig:form_rot_view}
\end{figure*}

Arcs R2 and R3 are used to examine the polarization outcomes as the observer is shifted
from being parallel to perpendicular, and then to anti-parallel, with respect to the magnetic field.
Arc R1 is employed to investigate the polarization outcomes with only a perpendicular magnetic field, and
therefore we expect little or no change in the polarization properties along this path if the domain
is accurately axisymmetric.
At each new observer's position, a maximum extent of the domain is 
calculated, as projected on the observer's sky plane, and a new source disc is constructed, based on 
this extent, in a position further from the observer than the domain. Background amplitudes and phases
of rays originating on the source disc are re-randomized.
Subsequently, we calculate the observer's results using formal solutions similar to those
computed in the previous subsection. Results of traversing R1, R2 and R3, in terms of the 
observer's Stokes parameters, are depicted in Fig.~\ref{fig:form_rot_view} for all three
shapes. EVPA and EVPA2 are displayed within the ranges of [-45, +45] and [-90, +90] degrees, respectively.
Channel-averaged fractional Stokes parameters are calculated in the same way as in
Section~\ref{ss:nodsols}.

Following the R1 arc, the observer's view is always perpendicular to the magnetic field, and
little or no Stokes-$V$ is expected at any azimuthal angle. Calculated maximum absolute
values are indeed significantly smaller than those from the R2 and R3 arcs and everywhere
smaller than 3 per cent in the oblate and prolate domains. Stokes-$V$ reaches 7.5 per cent
for the prolate domain, in the high-channel part near $\phi = 90$ degrees. The presence of
non-zero $V$ can also be seen in the related diagram, Fig.~\ref{fig:form_ref_view_pz}, centre
panel, where the $V$ spectrum has a negative bias, and the Poincar\'{e} sphere projection
is significantly open in the image. This is not expected, but we note that the prolate 
domain is geometrically and optically thin in this direction, with a Stokes-$I$ level of 
only about 1/3000 of that in a polar view.
In the R2 and R3 arcs, we see the expected sign-swap as the polar angle passes through
90 degrees. Maximum absolute values (of about 10 per cent) occur as expected near the extremes of the polar angle.
We also see the expected `mirror' behaviour between the Stokes-$V$ values from the high
and low sets of channels, depicted as solid and dashed lines in Fig.~\ref{fig:form_rot_view}.

The easiest Stokes-$Q$ result to understand in comparison with 1D models is the R2 rotation. This
view, at an azimuth of 90 degrees, passes through a high flux-density zone at polar angles near 90\,degrees
in both the oblate and pseudo-spherical domains. $Q/I$ reaches its greatest positive value in all 3 domains at a polar
angle close to 90 degrees. At this point, the model looks somewhat like the classic situation of
propagation perpendicular to the magnetic field. However, in the 3D model, $Q/I$ approaches
100 per cent in the oblate domain and $>$50 per cent in the pseudo-spherical domain. As the polar 
angle is varied, the $Q/I$ curves for all three domains display two sign changes, placed roughly symmetrically about the 
90-degree point (in the oblate domain, the lower angle is marked by changes in EVPA and EVPA2, even
if $Q$ does not actually change sign). These sign changes are somewhat analogous to the Van Vleck crossing in a 1D
system, but we note that in the present work the absolute $Q$ flux density is very small in the prolate domain
near the polar angle of 90 degrees, whilst in the other two domains the flux density is
very small near both zero and 180 degrees.
 
Both of the other rotations have a less expected Stokes-$Q$ behaviour: in R3, no sign changes result
along the arc, although $Q/I$ does become small at the extreme values of the
polar angle as expected, and the curves are approximately symmetrical about a polar angle of
90 degrees. In R1, we would expect a constant value of $Q/I$ along the arc from symmetry
considerations in a perfectly axisymmetric domain. In our domains that are only approximately
axisymmetric, what we see is a pair of sign changes and a curve somewhat simlar to R2.
 We suggest that competition
for population between saturating rays is responsible for the unexpected results (see
Section~\ref{sec:Dis_and_Con}). Sign changes in $Q/I$ were not found when an R1 rotation
was performed around the tube domain (see Section~\ref{ss:tubedom} and Appendix~\ref{app:small_domain})
with $\vec{B}$ aligned with the $z$-axis.
Stokes-$I$ variations along R1 have a contrast of order 5-10 in the present work, compared to about 3 in an
equatorial scan of an oblate domain in \citet{Gray2019}. Since the latter domain had more nodes,
agreement is probably reasonable.

Stokes-$U$ is generally close to zero in all 3 domains as expected for the $z$-axis alignment of
the magnetic field. We note that the largest values (about 4 per cent) occur in the prolate domain for intermediate
polar angles, where amplification is weak, and effects of the random background electric field
remain strongest.

The EVPA is calculated from the single-argument {\sc atan} function. Therefore, 90-degree
flips occur only when $U/Q$ changes between $-\infty$ and $+\infty$ (or $Q$ transits across zero).
Therefore, we may see flips of the EVPA at angles where EVPA2 changes smoothly.
In contrast, strong change in EVPA2, shown in
Fig.~\ref{fig:form_rot_view}, is caused both by sign changes in $Q$ when $U=0$ that produce
90-degree flips in angle, and by $U$ passing through zero when $Q$ is negative, which results
in 180-degree flips. In the R1 and R2 rotations, the oblate and pseudo-spherical domains both
experience rapid 90-degree changes in EVPA2 that correspond to sign changes in $Q$ with $U \simeq 0$.
However, on the R1 arc, the EVPA2 changes smoothly in the prolate domain when $Q$ changes sign. There
are also several examples in R1 and R2 of sharp 180 degree flips in EVPA2, where $U$ changes sign
in zones of negative $Q$. In the R3 rotation, $Q$ is always negative for all 3 domains, and the
only flips are due to sign changes in $U$, which is anyway close to zero. The $Q/I$, $U/I$ and
$V/I$ rows in Fig.~\ref{fig:form_rot_view} demonstrate that the Stokes parameters themselves 
change smoothly and without discontinuities. Rapid changes in EVPA and EVPA2 therefore result
only from sign changes in $Q$, $U$ or both of these Stokes parameters. For further information 
about rapid changes in EVPA2, see more results related to shifted-view rotation in Appendix~\ref{app:shifted-view}.

\subsection{Tube Domains}
\label{ss:tubedom}
Since the present work contains some results that are not simply explained in terms
of 1D models, particularly the behaviour of Stokes-$Q$ in the R1 and R3 rotations, we have
also considered results from a tube domain that should be a much closer analogue of a
1D model than any of the domains considered above. The tube domain has a length to radius
ratio of 10, with its long axis $z$-aligned. Its 24 nodes are placed on the cylinder
surface. In this respect it is different to the other domains described above, where
the boundary nodes were not constrained to coincide with an enclosing spherical or
spheroidal surface.
The magnetic field (strength 5\,G) was oriented in the $xz$ plane, and its
angle to the $z$-axis could be varied between 0 and 90\,degrees, as in many 1D models.
In other respects, the tube-domain results were computed like the results for the other
shapes above, with an inversion solution for each magnetic field angle, followed by
formal solutions to calculate the Stokes parameters seen by an observer. Formal solutions
used \texttt{rayform}=10 (243 rays) as the end area of the tube is much smaller than
the projected areas of the domains considered previously.

\begin{figure}
   \centering
   \includegraphics[width=0.45\textwidth]{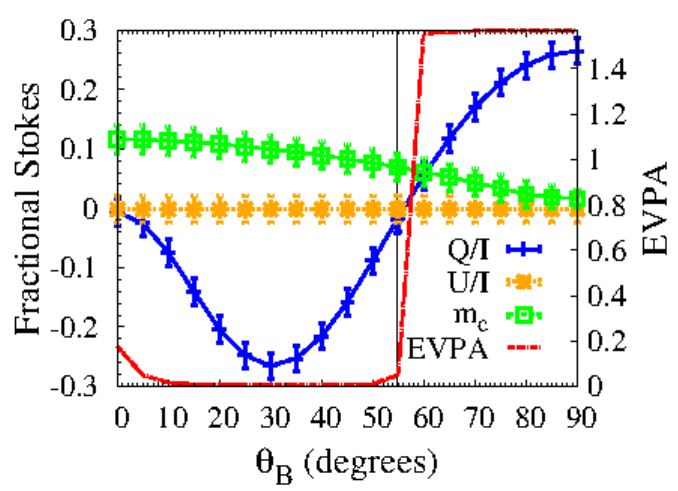}
   \caption{Variation of the observer's fractional Stokes parameters and the EVPA as a
   function of $\theta_B$: At $\theta_B = 0$, the magnetic field points towards the observer;
   at $\theta_B = 90$\,degrees, it is in the plane of the sky. We show $Q/I$ (blue) and $U/I$
   (orange), averaged over the central 0.5 Doppler widths, and weighted by Stokes-$I$: for
   example $Q/I = \sum_i Q_i /\sum_i I_i$ for channels $i$. Error 
   bars show the standard deviation. Circular polarization, $m_c$, is the average of the 
   absolute value of $V/I$ over the two sub-ranges of 5 channels where $V/I$ peaks. The EVPA
   is shown in red. A thin vertical black line is placed at the Van~Vleck angle, where
   $Q$ is expected to pass through zero.
   }
   \label{fig:vanvleck}
\end{figure}

Unless otherwise stated, the present
work used formal solutions that were computed with the observer placed
at $(r,\theta,\phi) = (1000,10^{-5},10^{-4})$, where the angles are in radians. In this
observer's view, the observer's sky-projected $x$-axis corresponds very closely to
the $y$-axis of the domain, and the sky-projected $y$-axis is the negative $x$-axis 
of the domain. Among the most important results from such formal solutions is to
follow Stokes-$Q$ as a function of the angle, $\theta_B$, between the field direction
and the $z$-axis. We plot this function, with additional information about the
behaviour of the other Stokes parameters in Fig.~\ref{fig:vanvleck}.

Stokes-$Q$ changes sign as expected at the Van-Vleck angle. It is close to zero, as 
expected at $\theta_B =0$, and $0.28$ at $\theta_B = 90$ degrees, where a value of
up to 1/3 is expected. Stokes-$U$ is zero at all angles within the uncertainty of
the standard deviation. The fractional circular polarization is, as expected, a
slowly declining function of angle, falling from approximately 0.12 at $\theta_B =0$
to $\sim 0$ at $\theta_B=90$\,degrees. The EVPA flips through $\pi/2$ at close to
the Van~Vleck angle (the resolution of the curves in angle is 5\,degrees). Additional
results from the tube domain (images and spectra) are presented in
Appendix~\ref{app:small_domain}.

\subsection{Overlapping Domains}

\begin{figure*}
	\centering
	\includegraphics[width=0.96\textwidth]{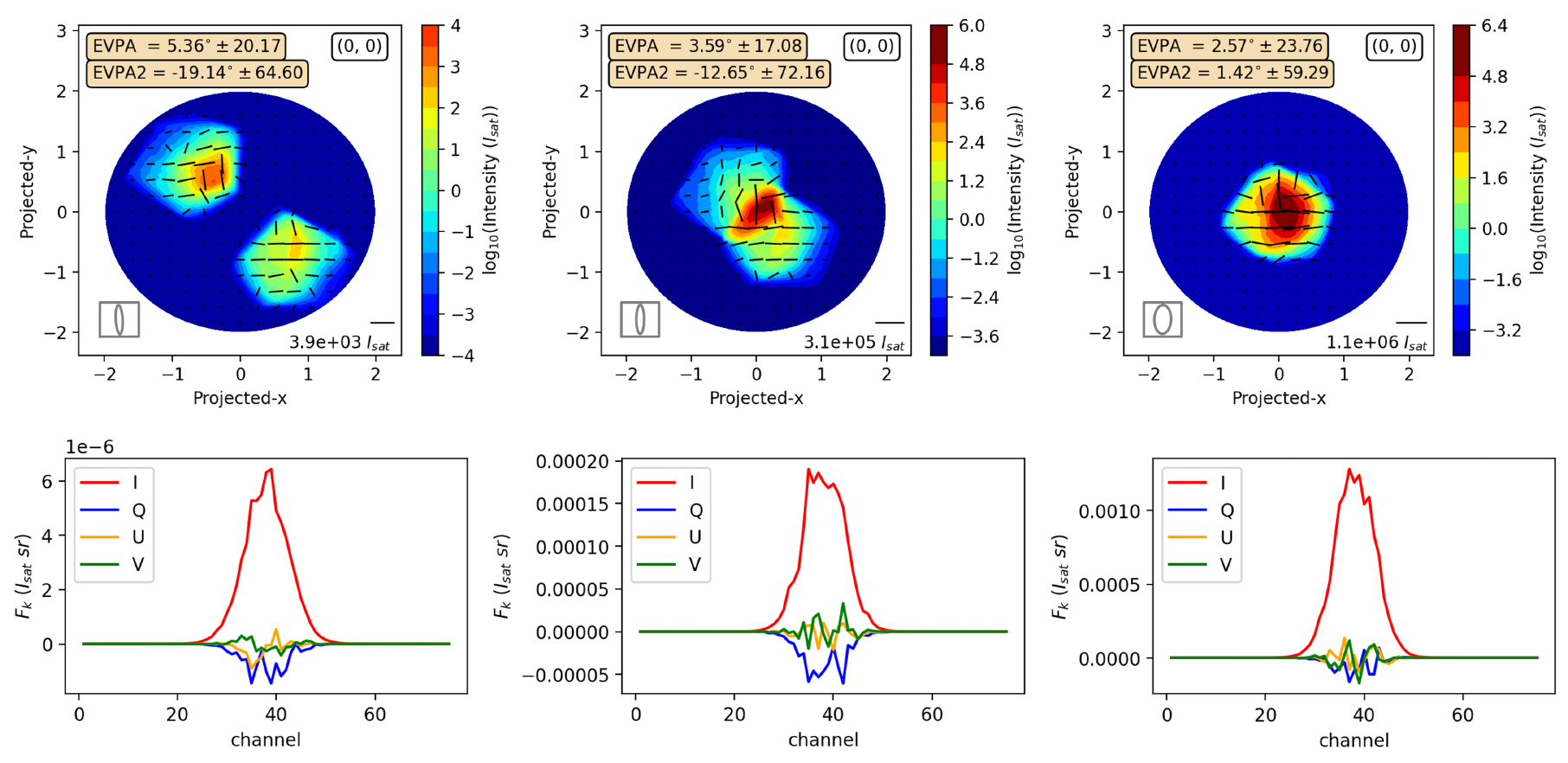}
	\caption{Saturated-cloud (depth = 20.0) overlapping with offsets of 0.8, 0.4, and 0.0 from left to right, respectively. The top panel shows intensity contours with EVPA and EVPA2, while the bottom panel displays the Stokes spectrum. The formal solutions used \texttt{rayform}=200.}
	\label{fig:overlap_p137}
\end{figure*}

In this subsection we investigate the idea that the field reversals, apparently within a
single cloud, see Section~\ref{s:intro}, arise from a line-of-sight overlap of two clouds
that have magnetic fields with different directions. This mechanism for generating field
reversals can only be studied realistically with a 3D model, and is a possible alternative to
the EVPA flip through the Van~Vleck angle due to variation of the line-of-sight component of
the magnetic field. We do not claim that overlap is better than, or a replacement for, the
Van~Vleck mechanism.
We consider a domain formed from two clouds. Both have magnetic fields perpendicular to the
line of sight to induce strong linear polarization, but with different projected orientations in
the sky plane. These clouds are modelled as pseudo-spherical sub-domains with duplicated node
distributions. The clouds were rotated using Euler rotations (axis, angle) of ($\hat{\boldsymbol{e}}_y$,$\pi/2$) and (-$\hat{\boldsymbol{e}}_x$,$\pi/2$), and then placed at 
offsets ($\Delta x$,$\Delta y$,$\Delta z$) of (0, 0.8, 1) and (0, -0.8, -1) for the first 
and second clouds, respectively. The offset $\Delta z$ is fixed to avoid errors from 
mixing nodes and elements between the two domains, while the offset $\Delta y$ could be reduced to 
0.4, and then to 0.0, to examine the polarization results from cloud overlapping. The magnetic field
followed the rotation of the clouds, resulting in $\boldsymbol{B}_1=B\hv{e}_x$ and 
$\boldsymbol{B}_2=B\hv{e}_y$ for clouds $1$ and $2$. The field strength was 35 G as for
the pointy137 models. Inversion solutions 
for the combined domain were computed, for each offset $\Delta y$, up to a depth multiplier 
of $\tau_M = 20$, which was found to be significantly saturated. Formal solutions were all
calculated at this final depth. We note that
the two separated clouds were fully coupled radiatively as a compound domain for both the population and 
formal solutions. The results of formal solutions with cloud overlapping are depicted in Fig.~\ref{fig:overlap_p137}.
In the central panel ($\Delta y = 0.4$) we see an apparent field rotation from horizontal to
vertical polarization across what appears as a single cloud image.

\section{Discussion and Conclusion}\label{sec:Dis_and_Con}

We have used a 3D maser polarization simulation to investigate the effect of magnetic field 
orientation and domain shape on the polarization properties of SiO maser radiation from 
the CSE of an AGB star. The polarization in this simulation is based on the Zeeman 
effect that splits the $J=1$ rotational level into three constituent sub-levels, and the $J=1-0$
transition into three sub-transitions. The simulation starts with a 3D molecular cloud domain 
with initial physical conditions specified at each of its nodes. Variables computed at each node 
are diagonal elements of the DM (inversions), and the model solves the combined statistical balance
and radiative transfer problem to simulate the DM up to the state of significant saturation
(remaining inversion $< 10$ per cent in the most saturated nodes) with associated 
polarization. We then used a background plane source to propagate formal solutions through 
the domain with known inversions towards observers at various positions, and analyzed the results 
in terms of observer-defined Stokes parameters. A simple simulation, based on a long cylinder
(see Appendix \ref{app:small_domain}) was 
used to test against a 1D model, shown in \cite{MPhetra2024}, and we found that Stokes $Q$ changes sign
as expected at the Van~Vleck angle while $U$ is very small at all angles. A flip of approximately
$\pi/2$ in EVPA (with the $0.5 \arctan(U/Q)$ function) occurs close to the Van Vleck angle. 
The long tubular domain with a line of sight parallel to its long axis is probably not a realistic 
example to compare with VLBI observations, so for the present work, we studied a larger cloud 
that is probably a more realistic representation of the condensations in the CSE seen in many 
observations. 

In this larger simulation, we considered three cloud shapes: oblate, pseudo-spherical, and prolate.
The results of the inversion solution show that the strongly saturated nodes are initially 
(in the very unsaturated limit) located well inside the domain. As saturation increases, nodes near the 
boundary become more strongly saturated, especially close to the longest axis, which depends on the domain shape. 
Moreover, we find that the integrated amplitude of the $\sigma$ components exhibits a strong positive correlation. Correlation between the $\pi$-component amplitude, and that of both $\sigma$ components is significantly weaker.
In the oblate and pseudo-spherical domains, we find nodes with low integrated amplitude close to the origin,
following the pattern of saturation. However, in the prolate domain, many low-amplitude nodes are far
from the origin. This behaviour is interesting, and results from the slow decay in amplitude of the
most saturated (lowest remaining inversion) nodes at high depth. Amplitudes in less saturated nodes continue
to grow (see Fig.~\ref{fig:evo_p137}, second row). A population of high amplitude, but not the most saturated, nodes then
appears close to the domain origin.

Formal solutions demonstrate that the general pattern of Stokes parameters agrees reasonably 
with theoretical expectations, particularly in the prolate domain that is closest to the 1D ideal
in having a dominant long axis. We see approximate `S-curve' circular polarization 
with a parallel magnetic field to the observer's direction 
\citep{Green2014MNRAS, Tobin2023} and see Fig.~\ref{fig:contour_p137}, right-hand panel. 
Linear polarization is weak, on average, and disorganized from this viewpoint. The sign of the `S-curve'
is reversed as expected when the view is shifted to anti-parallel to the observer's
direction (Fig.~\ref{fig:form_ref_view_pz}), right-hand panel. We have tested different values
of the parameter \texttt{rayform} that governs the number of rays in formal solutions. While
a value of 10 is suitable for end views of the tube domain (radius 0.1), we found that 100-200
was required for projections of the more complicated domains. We briefly show the effect
of changing \texttt{rayform} on spectra from the pseudo-spherical domain in
Appendix~\ref{app:rsfs}.

A new, and less expected, result here is the sign-reversal of Stokes-$Q$ when the
observer is placed in the $xy$ plane, where Stokes-$V$ is weak. The effect can be seen in 
all three domains, but is
particularly evident in the oblate and pseudo-spherical forms. In Fig.~\ref{fig:form_ref_view_s}
and Fig.~\ref{fig:form_ref_view_oz} we see that $Q$ changes sign from negative to positive
between the left (+$x$ view) and central (+$y$ view) panels. Changes in the channel-integrated 
Stokes parameters can be followed as the angle changes along the R1 arc in the left-hand
column of Fig.~\ref{fig:form_rot_view}. Stokes-$I$ peaks as the observer's direction passes
close to the $y$-axis. This exceptional flux density, at azimuthal angles near 90\,degrees,
in the oblate and pseudo-spherical domains does not correspond to the long-axis of either
domain, but to a set of rays with high gain, as viewed from these angles. For reference, the
long axis of the oblate domain has direction $(\theta,\phi)=(19\degr.6,20\degr.0)$, and
is partly responsible for the smaller flux density peak to the left of the main feature in
the top left panel of Fig.~\ref{fig:form_rot_view}.
We ignore the prolate domain here, as Stokes-$I$ along the R1 arc is very small everywhere.

\begin{figure}
	\centering
	\includegraphics[width=0.44\textwidth]{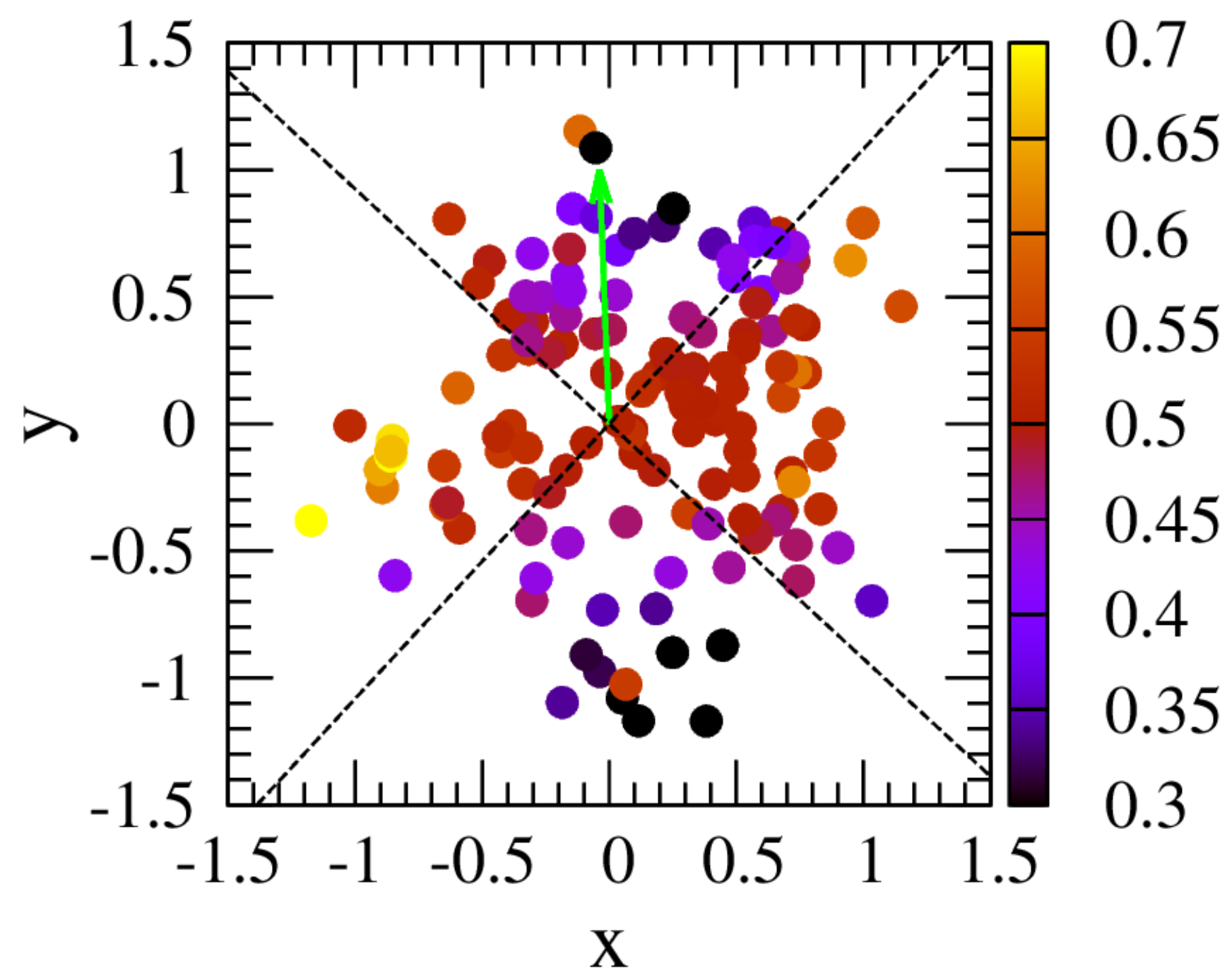}
	\caption{Points are node positions for the oblate domain, projected onto the $xy$ plane: their
    colours follow the ratio $\Delta_{31}/(\Delta_{21}+\Delta_{41})$, that is the ratio of the inversion in
    the $\pi$ transition to the combined inversion in the $\sigma$ transitions. The depth multiplier for
    this diagram is 20.0, as in Fig.~\ref{fig:form_rot_view}. The green arrow points in the direction
    corresponding to the Stokes-$I$ peak in the R1 rotation (upper block, left-hand column, top row in
    Fig.~\ref{fig:form_rot_view}), which is $\phi =$92\degr.25. Black dotted lines divide the figure 
    into quadrants offset by 45 and 135\,degrees from the green arrow.}
	\label{fig:fracpops}
\end{figure}

Stokes-$Q$ shows positive peaks in the oblate and pseudo-spherical domains close to the
Stokes-$I$ peaks, and these correspond approximately to the expectations of the 1D case.
However, an interesting difference is that Stokes-$Q$ near these peaks approaches
100 per cent, whereas a limit of $33.3$ per cent is expected, when
observing perpendicular to $\vec{B}$ in the 1D case. Obviously, the
axial symmetry about the $z$-axis of the domain, imposed by the magnetic field, is broken
by the orientation of the long axis. We suggest that this has the following consequences when
there are many competing ray paths of similar length to the long axis, as in the oblate and
pseudo-spherical domains.
Observed polarization is related to the location of the most strongly saturated nodes, which
are located close to the domain surface. If the most saturating set of rays consume most of the available
inversion in, say, the $\sigma$-transitions, then other rays must amplify at the expense
of $\pi$-inversions, leading to the observed sign change in $Q$ following the R1 arc.
This suggestion is in accord with equation~(\ref{eq:RT_eq}), in which amplification of
electric field components depends on available inversion in the various transitions.
We have explored the $\pi$ to (combined) $\sigma$ inversion ratio in the oblate domain, and
have found that this quantity is indeed fractionated in agreement with the above suggestion,
with two $\sigma$-dominated, and two $\pi$-dominated, azimuthal quadrants. Results are
plotted in Fig.~\ref{fig:fracpops}.

Our investigation of the overlap mechanism for generating 90-degree EVPA rotations shows
that this scenario is plausible. We do not claim that overlap is preferable to the Van~Vleck
mechanism of generating EVPA rotations: in fact overlap has some difficulties. The first is that we have
arranged the sky-projected magnetic field directions of the two clouds to be orthogonal.
Although this seems very artificial, there is observational evidence of a rather bimodal
distribution of projected polarization vectors, at least in some stars
\citep{Assaf2013}. A bimodal distribution of polarization vectors probably (but not certainly)
requires a similar underlying distribution of magnetic fields. Within single VLBI `clouds', then
in TX Cam \citep{Kemball2011} and R Cas \citep{Assaf2013}, we also
only see cases of EVPA rotation through approximately 90 degrees, although \citet{Kemball2011} did
detect a cloud with an EVPA flip that varies in orientation through several LSR velocity channels.
A second difficulty, or perhaps test, is that
of overall intensity: at the first offset of $\Delta y = 0.8$ the clouds are separated on the sky,
and Stokes-$I$ is only of the order of a few times $10^{-6}$, so probably unobservable.
At the second offset, ($\Delta y = 0.4$), we can see the EVPA rotation in Fig.~\ref{fig:overlap_p137},
and the Stokes-$I$ intensity has increased by a factor of about 80. However, the largest Stokes-$I$
flux densities only appear near $\Delta y = 0.0$, by which point the EVPA rotation effect has
disappeared. Therefore, we would expect all such overlaps to follow a pattern of significant
brightening as the clouds converge and the EVPA rotation appears. If the overlap is close to perfect,
and $\Delta y$ is close to zero, the EVPA rotation would disappear in the brightest, and most
compact, images.

\begin{figure}
	\centering
	\includegraphics[width=0.32\textwidth]{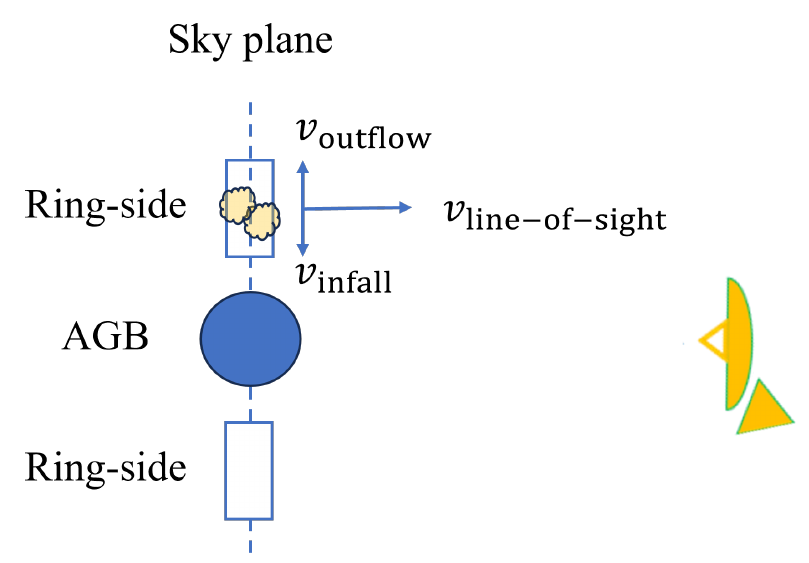}
	\caption{A diagram of maser cloud in ring-side for the observer with infall-outflow velocity, $v_{\text{infall}}$, $v_{\text{outflow}}$, and radial velocity, $v_{\text{line-of-sight}}$.}
	\label{fig:ring-side}
\end{figure}

If weak overall flux density allows, maser variability and the proper motion of maser features 
could provide evidence that two clouds have different EVPAs until they move to the overlap position. \cite{Assaf2018ApJ} demonstrated that the proper motions of SiO maser features in the CSE of R Cas exhibit non-linear motion, with a line of sight velocity of approximately 6 km\,s$^{-1}$, which is greater than the outflow or infall velocity of about +0.4 km\,s$^{-1}$. This suggests that overlapping events are rare because most of the maser features 
we measure are located on the ring-side with respect to the observer, as shown in Fig.~\ref{fig:ring-side}. 

A magneto-hydrodynamic model for the rotating AGB star, as presented by \cite{Pascoli2010PASP,Pascoli2020PASP}, suggests that the stellar winds move outward at the equator and then fall in as a polar vortex to the star, while the magnetic 
field exhibits its maximum strength at the pole of the star. We may utilize this model to generate the entire ring structure of the maser in a future simulation.

The population evolution in Fig.~\ref{fig:evo_p137} is in the regime where $R < g\Omega$. However, it would be desirable for future versions of the code to be able to run to higher values of $\tau_M$ where the stimulated emission rate, $R$, may exceed the Zeeman splitting or $R \gg g\Omega$. In this more saturated regime, we need to consider the sub-state mixing effect as predicted in \cite{Lankhaar2019,Western1984Apj285}. Such effects would be expected to modify, and ultimately diminish, the polarization levels relative to those in our current simulations. 

We should also conduct fidelity
tests on similar domains that have different numbers of nodes and elements. Much of the literature on this subject is
related to the use of the FEM to solve differential equations, whereas we use the FEM only to convert integral equations
in the inversions to non-linear algebraic equations (see Section~\ref{subsec:dom_set}), a once-only procedure that is
carried out before the iterative stage of an inversion solution. We should therefore investigate the behaviour
of solutions as a function of element count, as it may differ considerably from the expectations for differential
equation models.

\section*{Acknowledgement}
We would like to express our gratitude to the Department of Physics and Materials Science, Faculty of Science, Chiang Mai University, the Science Achievement Scholarship of Thailand (SAST), and the National Astronomical Research Institute of Thailand (NARIT) for their support of this research. This work has also made use of results obtained with the Chalawan HPC cluster, operated and maintained by NARIT under the Ministry of Higher Education, Science, Research and Innovation, Royal Thai government.

We would like to thank NARIT and Fundamental Fund (FF) under the Thailand Science Research and Innovation (TSRI) for supporting budget No. 02-15 FY2565-FY2567 and grant No. 168595, 180551, and 198692 during the Thai fiscal year 2022–2024 respectively.

\section*{DATA AVAILABILITY}
The data underlying this article will be shared on reasonable request to the corresponding author.

\bibliographystyle{mnras}
\bibliography{main} 

\begin{thebibliography}{}
\makeatletter
\relax
\def\mn@urlcharsother{\let\do\@makeother \do\$\do\&\do\#\do\^\do\_\do\%\do\~}
\def\mn@doi{\begingroup\mn@urlcharsother \@ifnextchar [ {\mn@doi@}
  {\mn@doi@[]}}
\def\mn@doi@[#1]#2{\def\@tempa{#1}\ifx\@tempa\@empty \href
  {http://dx.doi.org/#2} {doi:#2}\else \href {http://dx.doi.org/#2} {#1}\fi
  \endgroup}
\def\mn@eprint#1#2{\mn@eprint@#1:#2::\@nil}
\def\mn@eprint@arXiv#1{\href {http://arxiv.org/abs/#1} {{\tt arXiv:#1}}}
\def\mn@eprint@dblp#1{\href {http://dblp.uni-trier.de/rec/bibtex/#1.xml}
  {dblp:#1}}
\def\mn@eprint@#1:#2:#3:#4\@nil{\def\@tempa {#1}\def\@tempb {#2}\def\@tempc
  {#3}\ifx \@tempc \@empty \let \@tempc \@tempb \let \@tempb \@tempa \fi \ifx
  \@tempb \@empty \def\@tempb {arXiv}\fi \@ifundefined
  {mn@eprint@\@tempb}{\@tempb:\@tempc}{\expandafter \expandafter \csname
  mn@eprint@\@tempb\endcsname \expandafter{\@tempc}}}

\bibitem[\protect\citeauthoryear{{Assaf}}{{Assaf}}{2018}]{Assaf2018ApJ}
{Assaf} K.~A.,  2018, \mn@doi [\apj] {10.3847/1538-4357/aaea65}, \href
  {https://ui.adsabs.harvard.edu/abs/2018ApJ...869...80A} {869, 80}

\bibitem[\protect\citeauthoryear{{Assaf}, {Diamond}, {Richards}  \&
  {Gray}}{{Assaf} et~al.}{2013}]{Assaf2013}
{Assaf} K.~A.,  {Diamond} P.~J.,  {Richards} A.~M.~S.,   {Gray} M.~D.,  2013,
  \mn@doi [\mnras] {10.1093/mnras/stt242}, \href
  {https://ui.adsabs.harvard.edu/abs/2013MNRAS.431.1077A} {431, 1077}

\bibitem[\protect\citeauthoryear{{Beckers} \& {Beckers}}{{Beckers} \&
  {Beckers}}{2012}]{Beckers2012}
{Beckers} B.,  {Beckers} P.,  2012, \mn@doi [Computational Geometry]
  {https://doi.org/10.1016/j.comgeo.2012.01.011}, 45, 275

\bibitem[\protect\citeauthoryear{{Cernicharo}, {Alcolea}, {Baudry}  \&
  {Gonzalez-Alfonso}}{{Cernicharo} et~al.}{1997}]{Cernicharo1997}
{Cernicharo} J.,  {Alcolea} J.,  {Baudry} A.,   {Gonzalez-Alfonso} E.,  1997,
  \aap, \href {https://ui.adsabs.harvard.edu/abs/1997A&A...319..607C} {319,
  607}

\bibitem[\protect\citeauthoryear{{Chen} \& {Cai}}{{Chen} \&
  {Cai}}{2001}]{Chen2001}
{Chen} Y.,  {Cai} D.,  2001, \mn@doi [Applied Mathematics and Computation]
  {https://doi.org/10.1016/S0096-3003(00)00101-6}, 124, 351

\bibitem[\protect\citeauthoryear{{Cotton} et~al.,}{{Cotton}
  et~al.}{2006}]{2006A&A...456..339C}
{Cotton} W.~D.,  et~al., 2006, \mn@doi [\aap] {10.1051/0004-6361:20065134},
  \href {http://ukads.nottingham.ac.uk/abs/2006A%26A...456..339C} {456, 339}

\bibitem[\protect\citeauthoryear{{Cotton}, {Ragland}, {Pluzhnik}, {Danchi},
  {Traub}, {Willson}  \& {Lacasse}}{{Cotton}
  et~al.}{2009}]{2009ApJS..185..574C}
{Cotton} W.~D.,  {Ragland} S.,  {Pluzhnik} E.~A.,  {Danchi} W.~C.,  {Traub}
  W.~A.,  {Willson} L.~A.,   {Lacasse} M.~G.,  2009, \mn@doi [\apjs]
  {10.1088/0067-0049/185/2/574}, \href
  {http://ukads.nottingham.ac.uk/abs/2009ApJS..185..574C} {185, 574}

\bibitem[\protect\citeauthoryear{{Deguchi} \& {Watson}}{{Deguchi} \&
  {Watson}}{1990}]{Deguchi1990ApJ354}
{Deguchi} S.,  {Watson} W.~D.,  1990, \mn@doi [\apj] {10.1086/168722}, \href
  {https://ui.adsabs.harvard.edu/abs/1990ApJ...354..649D} {354, 649}

\bibitem[\protect\citeauthoryear{Deschamps \& Mast}{Deschamps \&
  Mast}{1973}]{Deschamps1973poincare}
Deschamps G.,  Mast P.,  1973, IEEE Transactions on Antennas and Propagation,
  21, 474

\bibitem[\protect\citeauthoryear{{Diamond} \& {Kemball}}{{Diamond} \&
  {Kemball}}{2003}]{2003ApJ...599.1372D}
{Diamond} P.~J.,  {Kemball} A.~J.,  2003, \mn@doi [\apj] {10.1086/379347},
  \href {http://ukads.nottingham.ac.uk/abs/2003ApJ...599.1372D} {599, 1372}

\bibitem[\protect\citeauthoryear{{Diamond}, {Kemball}, {Junor}, {Zensus},
  {Benson}  \& {Dhawan}}{{Diamond} et~al.}{1994}]{1994ApJ...430L..61D}
{Diamond} P.~J.,  {Kemball} A.~J.,  {Junor} W.,  {Zensus} A.,  {Benson} J.,
  {Dhawan} V.,  1994, \mn@doi [\apjl] {10.1086/187438}, \href
  {http://ukads.nottingham.ac.uk/abs/1994ApJ...430L..61D} {430, L61}

\bibitem[\protect\citeauthoryear{{Dinh-v-Trung}}{{Dinh-v-Trung}}{2009}]{Dinh-v-Trung2009MNRAS}
{Dinh-v-Trung} 2009, \mn@doi [\mnras] {10.1111/j.1365-2966.2009.15369.x}, \href
  {https://ui.adsabs.harvard.edu/abs/2009MNRAS.399.1495D} {399, 1495}

\bibitem[\protect\citeauthoryear{{Elitzur}}{{Elitzur}}{1992}]{elitzurbook}
{Elitzur} M.,  1992, {Astronomical masers}.
Kluwer, Dordrecht

\bibitem[\protect\citeauthoryear{{Elitzur}}{{Elitzur}}{1993}]{Elitzur1993ApJ}
{Elitzur} M.,  1993, \mn@doi [\apj] {10.1086/173232}, \href
  {https://ui.adsabs.harvard.edu/abs/1993ApJ...416..256E} {416, 256}

\bibitem[\protect\citeauthoryear{{Elitzur}}{{Elitzur}}{1996}]{Elitzur1996ApJ}
{Elitzur} M.,  1996, \mn@doi [\apj] {10.1086/176741}, \href
  {https://ui.adsabs.harvard.edu/abs/1996ApJ...457..415E} {457, 415}

\bibitem[\protect\citeauthoryear{{Goldreich} \& {Kwan}}{{Goldreich} \&
  {Kwan}}{1974}]{1974ApJ...190...27G}
{Goldreich} P.,  {Kwan} J.,  1974, \mn@doi [\apj] {10.1086/152843}, \href
  {https://ui.adsabs.harvard.edu/abs/1974ApJ...190...27G} {190, 27}

\bibitem[\protect\citeauthoryear{{Goldreich}, {Keeley}  \& {Kwan}}{{Goldreich}
  et~al.}{1973}]{Goldreich1973}
{Goldreich} P.,  {Keeley} D.~A.,   {Kwan} J.~Y.,  1973, \mn@doi [\apj]
  {10.1086/151852}, \href
  {https://ui.adsabs.harvard.edu/abs/1973ApJ...179..111G} {179, 111}

\bibitem[\protect\citeauthoryear{{Gray}}{{Gray}}{2003}]{Gray2003MNRAS}
{Gray} M.~D.,  2003, \mn@doi [\mnras] {10.1046/j.1365-8711.2003.06836.x}, \href
  {https://ui.adsabs.harvard.edu/abs/2003MNRAS.343L..33G} {343, L33}

\bibitem[\protect\citeauthoryear{{Gray}, {Mason}  \& {Etoka}}{{Gray}
  et~al.}{2018}]{Gray2018}
{Gray} M.~D.,  {Mason} L.,   {Etoka} S.,  2018, \mn@doi [\mnras]
  {10.1093/mnras/sty576}, \href
  {https://ui.adsabs.harvard.edu/abs/2018MNRAS.477.2628G} {477, 2628}

\bibitem[\protect\citeauthoryear{{Gray}, {Baggott}, {Westlake}  \&
  {Etoka}}{{Gray} et~al.}{2019}]{Gray2019}
{Gray} M.~D.,  {Baggott} J.,  {Westlake} J.,   {Etoka} S.,  2019, \mn@doi
  [\mnras] {10.1093/mnras/stz1137}, \href
  {https://ui.adsabs.harvard.edu/abs/2019MNRAS.486.4216G} {486, 4216}

\bibitem[\protect\citeauthoryear{{Green}, {Gray}, {Robishaw}, {Caswell}  \&
  {McClure-Griffiths}}{{Green} et~al.}{2014}]{Green2014MNRAS}
{Green} J.~A.,  {Gray} M.~D.,  {Robishaw} T.,  {Caswell} J.~L.,
  {McClure-Griffiths} N.~M.,  2014, \mn@doi [\mnras] {10.1093/mnras/stu429},
  \href {https://ui.adsabs.harvard.edu/abs/2014MNRAS.440.2988G} {440, 2988}

\bibitem[\protect\citeauthoryear{{Hamaker} \& {Bregman}}{{Hamaker} \&
  {Bregman}}{1996}]{1996A&AS..117..161H}
{Hamaker} J.~P.,  {Bregman} J.~D.,  1996, \aaps, \href
  {https://ui.adsabs.harvard.edu/abs/1996A&AS..117..161H} {117, 161}

\bibitem[\protect\citeauthoryear{{Herpin}, {Baudry}, {Thum}, {Morris}  \&
  {Wiesemeyer}}{{Herpin} et~al.}{2006}]{2006A&A...450..667H}
{Herpin} F.,  {Baudry} A.,  {Thum} C.,  {Morris} D.,   {Wiesemeyer} H.,  2006,
  \mn@doi [\aap] {10.1051/0004-6361:20054255}, \href
  {http://ukads.nottingham.ac.uk/abs/2006A%26A...450..667H} {450, 667}

\bibitem[\protect\citeauthoryear{{Houde}, {Lankhaar}, {Rajabi}  \&
  {Chamma}}{{Houde} et~al.}{2022}]{2022MNRAS.511..295H}
{Houde} M.,  {Lankhaar} B.,  {Rajabi} F.,   {Chamma} M.~A.,  2022, \mn@doi
  [\mnras] {10.1093/mnras/stab3806}, \href
  {https://ui.adsabs.harvard.edu/abs/2022MNRAS.511..295H} {511, 295}

\bibitem[\protect\citeauthoryear{{Humphreys}}{{Humphreys}}{2018}]{2018iss..confE..13H}
{Humphreys} E.,  2018, in Imaging of Stellar Surfaces. p.~13,
  \mn@doi{10.5281/zenodo.1220753}

\bibitem[\protect\citeauthoryear{{Kemball} \& {Diamond}}{{Kemball} \&
  {Diamond}}{1997}]{Kemball1997}
{Kemball} A.~J.,  {Diamond} P.~J.,  1997, \mn@doi [\apjl] {10.1086/310664},
  \href {https://ui.adsabs.harvard.edu/abs/1997ApJ...481L.111K} {481, L111}

\bibitem[\protect\citeauthoryear{{Kemball}, {Diamond}, {Gonidakis}, {Mitra},
  {Yim}, {Pan}  \& {Chiang}}{{Kemball} et~al.}{2009}]{Kemball2009}
{Kemball} A.~J.,  {Diamond} P.~J.,  {Gonidakis} I.,  {Mitra} M.,  {Yim} K.,
  {Pan} K.-C.,   {Chiang} H.-F.,  2009, \mn@doi [\apj]
  {10.1088/0004-637X/698/2/1721}, \href
  {https://ui.adsabs.harvard.edu/abs/2009ApJ...698.1721K} {698, 1721}

\bibitem[\protect\citeauthoryear{{Kemball}, {Diamond}, {Richter}, {Gonidakis}
  \& {Xue}}{{Kemball} et~al.}{2011}]{Kemball2011}
{Kemball} A.~J.,  {Diamond} P.~J.,  {Richter} L.,  {Gonidakis} I.,   {Xue} R.,
  2011, \mn@doi [\apj] {10.1088/0004-637X/743/1/69}, \href
  {https://ui.adsabs.harvard.edu/abs/2011ApJ...743...69K} {743, 69}

\bibitem[\protect\citeauthoryear{{Landi Degl'Innocenti}}{{Landi
  Degl'Innocenti}}{1987}]{1987nrt..book..265L}
{Landi Degl'Innocenti} E.,  1987, {Transfer of Polarized Radiation, using 4 x 4
  Matrices}.
p.~265

\bibitem[\protect\citeauthoryear{{Lankhaar} \& {Vlemmings}}{{Lankhaar} \&
  {Vlemmings}}{2019}]{Lankhaar2019}
{Lankhaar} B.,  {Vlemmings} W.,  2019, \mn@doi [\aap]
  {10.1051/0004-6361/201935064}, \href
  {https://ui.adsabs.harvard.edu/abs/2019A&A...628A..14L} {628, A14}

\bibitem[\protect\citeauthoryear{{Lankhaar}, {Surcis}, {Vlemmings}  \&
  {Impellizzeri}}{{Lankhaar} et~al.}{2024}]{Lankhaar2024A&A}
{Lankhaar} B.,  {Surcis} G.,  {Vlemmings} W.,   {Impellizzeri} V.,  2024,
  \mn@doi [\aap] {10.1051/0004-6361/202348420}, \href
  {https://ui.adsabs.harvard.edu/abs/2024A&A...683A.117L} {683, A117}

\bibitem[\protect\citeauthoryear{{Lewis}, {Pihlstr{\"o}m}  \&
  {Sjouwerman}}{{Lewis} et~al.}{2024}]{2024IAUS..380..314L}
{Lewis} M.~O.,  {Pihlstr{\"o}m} Y.~M.,   {Sjouwerman} L.~O.,  2024, in {Hirota}
  T.,  {Imai} H.,  {Menten} K.,   {Pihlstr{\"o}m} Y.,  eds,  IAU Symposium Vol.
  380, Cosmic Masers: Proper Motion Toward the Next-Generation Large Projects.
  pp 314--318, \mn@doi{10.1017/S1743921323002375}

\bibitem[\protect\citeauthoryear{{Liu} \& {Quek}}{{Liu} \&
  {Quek}}{2014}]{Liu2014}
{Liu} G.,  {Quek} S.,  2014, The Finite Element Method: A Practical Course,
  second edition edn.
Butterworth-Heinemann, \mn@doi{https://doi.org/10.1016/C2012-0-00779-X}

\bibitem[\protect\citeauthoryear{{Menegozzi} \& {Lamb}}{{Menegozzi} \&
  {Lamb}}{1978}]{1978PhRvA..17..701M}
{Menegozzi} L.~N.,  {Lamb} Jr. W.~E.,  1978, \mn@doi [\pra]
  {10.1103/PhysRevA.17.701}, \href
  {http://ukads.nottingham.ac.uk/abs/1978PhRvA..17..701M} {17, 701}

\bibitem[\protect\citeauthoryear{{Nedoluha} \& {Watson}}{{Nedoluha} \&
  {Watson}}{1990}]{Nedoluha1990ApJ354}
{Nedoluha} G.~E.,  {Watson} W.~D.,  1990, \mn@doi [\apj] {10.1086/168723},
  \href {https://ui.adsabs.harvard.edu/abs/1990ApJ...354..660N} {354, 660}

\bibitem[\protect\citeauthoryear{Olver}{Olver}{2010}]{Olver2010nist}
Olver F.~W.,  2010, NIST handbook of mathematical functions hardback and
  CD-ROM.
Cambridge university press

\bibitem[\protect\citeauthoryear{{Pardo}, {Alcolea}, {Bujarrabal}, {Colomer},
  {del Romero}  \& {de Vicente}}{{Pardo} et~al.}{2004}]{Pardo2004}
{Pardo} J.~R.,  {Alcolea} J.,  {Bujarrabal} V.,  {Colomer} F.,  {del Romero}
  A.,   {de Vicente} P.,  2004, \mn@doi [\aap] {10.1051/0004-6361:20040309},
  \href {https://ui.adsabs.harvard.edu/abs/2004A&A...424..145P} {424, 145}

\bibitem[\protect\citeauthoryear{{Pascoli}}{{Pascoli}}{2020}]{Pascoli2020PASP}
{Pascoli} G.,  2020, \mn@doi [\pasp] {10.1088/1538-3873/ab54a2}, \href
  {https://ui.adsabs.harvard.edu/abs/2020PASP..132c4203P} {132, 034203}

\bibitem[\protect\citeauthoryear{{Pascoli} \& {Lahoche}}{{Pascoli} \&
  {Lahoche}}{2010}]{Pascoli2010PASP}
{Pascoli} G.,  {Lahoche} L.,  2010, \mn@doi [\pasp] {10.1086/657508}, \href
  {https://ui.adsabs.harvard.edu/abs/2010PASP..122.1334P} {122, 1334}

\bibitem[\protect\citeauthoryear{{P{\'e}rez-S{\'a}nchez} \&
  {Vlemmings}}{{P{\'e}rez-S{\'a}nchez} \& {Vlemmings}}{2013}]{P-S&V2013}
{P{\'e}rez-S{\'a}nchez} A.~F.,  {Vlemmings} W.~H.~T.,  2013, \mn@doi [\aap]
  {10.1051/0004-6361/201220735}, \href
  {https://ui.adsabs.harvard.edu/abs/2013A&A...551A..15P} {551, A15}

\bibitem[\protect\citeauthoryear{{Perrin}, {Cotton}, {Millan-Gabet}  \&
  {Mennesson}}{{Perrin} et~al.}{2015}]{2015A&A...576A..70P}
{Perrin} G.,  {Cotton} W.~D.,  {Millan-Gabet} R.,   {Mennesson} B.,  2015,
  \mn@doi [\aap] {10.1051/0004-6361/201425110}, \href
  {https://ui.adsabs.harvard.edu/abs/2015A&A...576A..70P} {576, A70}

\bibitem[\protect\citeauthoryear{{Phetra}, {Gray}, {Asanok}, {Kramer},
  {Sugiyama}, {Etoka}  \& {Nuntiyakul}}{{Phetra} et~al.}{2024}]{MPhetra2024}
{Phetra} M.,  {Gray} M.~D.,  {Asanok} K.,  {Kramer} B.~H.,  {Sugiyama} K.,
  {Etoka} S.,   {Nuntiyakul} W.,  2024, in {Hirota} T.,  {Imai} H.,  {Menten}
  K.,   {Pihlstr{\"o}m} Y.,  eds,  IAU Symposium Vol. 380, Cosmic Masers:
  Proper Motion Toward the Next-Generation Large Projects. pp 435--439,
  \mn@doi{10.1017/S1743921323003551}

\bibitem[\protect\citeauthoryear{Press}{Press}{1996}]{Press1996numerical}
Press W.~H.,  1996, Numerical recipes in Fortran 90: Volume 2, volume 2 of
  Fortran numerical recipes: The art of parallel scientific computing.
~ Vol. 2, Cambridge university press

\bibitem[\protect\citeauthoryear{{Richards}, {Elitzur}  \& {Yates}}{{Richards}
  et~al.}{2011}]{Richards2011A&A}
{Richards} A.~M.~S.,  {Elitzur} M.,   {Yates} J.~A.,  2011, \mn@doi [\aap]
  {10.1051/0004-6361/201015397}, \href
  {https://ui.adsabs.harvard.edu/abs/2011A&A...525A..56R} {525, A56}

\bibitem[\protect\citeauthoryear{{Rizzo}, {Cernicharo}  \&
  {Garc{\'\i}a-Mir{\'o}}}{{Rizzo} et~al.}{2021}]{2021ApJS..253...44R}
{Rizzo} J.~R.,  {Cernicharo} J.,   {Garc{\'\i}a-Mir{\'o}} C.,  2021, \mn@doi
  [\apjs] {10.3847/1538-4365/abe469}, \href
  {https://ui.adsabs.harvard.edu/abs/2021ApJS..253...44R} {253, 44}

\bibitem[\protect\citeauthoryear{{Satoh}}{{Satoh}}{2014}]{Satoh2014}
{Satoh} M.,  2014, Icosahedral grids.
Springer Berlin Heidelberg, Berlin, Heidelberg, pp 636--660,
  \mn@doi{10.1007/978-3-642-13574-3_25}

\bibitem[\protect\citeauthoryear{{Sugano} \& {Koizumi}}{{Sugano} \&
  {Koizumi}}{1998}]{Sugano1998}
{Sugano} S.,  {Koizumi} H.,  1998, {Microcluster Physics}.
Springer Berlin, Heidelberg, \mn@doi{10.1007/978-3-642-58926-3}

\bibitem[\protect\citeauthoryear{{Tobin}, {Gray}  \& {Kemball}}{{Tobin}
  et~al.}{2023}]{Tobin2023}
{Tobin} T.~L.,  {Gray} M.~D.,   {Kemball} A.~J.,  2023, \mn@doi [\apj]
  {10.3847/1538-4357/aca595}, \href
  {https://ui.adsabs.harvard.edu/abs/2023ApJ...943..123T} {943, 123}

\bibitem[\protect\citeauthoryear{{Vlemmings}}{{Vlemmings}}{2019}]{Vlemmings2019}
{Vlemmings} W.,  2019, \mn@doi [IAU Symposium] {10.1017/S1743921318005367},
  \href {https://ui.adsabs.harvard.edu/abs/2019IAUS..343...19V} {343, 19}

\bibitem[\protect\citeauthoryear{{Watson} \& {Wyld}}{{Watson} \&
  {Wyld}}{2001}]{Watson2001}
{Watson} W.~D.,  {Wyld} H.~W.,  2001, \mn@doi [\apjl] {10.1086/323513}, \href
  {https://ui.adsabs.harvard.edu/abs/2001ApJ...558L..55W} {558, L55}

\bibitem[\protect\citeauthoryear{{Western} \& {Watson}}{{Western} \&
  {Watson}}{1983a}]{Western1983Apj268}
{Western} L.~R.,  {Watson} W.~D.,  1983a, \mn@doi [\apj] {10.1086/161007},
  \href {https://ui.adsabs.harvard.edu/abs/1983ApJ...268..849W} {268, 849}

\bibitem[\protect\citeauthoryear{{Western} \& {Watson}}{{Western} \&
  {Watson}}{1983b}]{Western1983ApJ275}
{Western} L.~R.,  {Watson} W.~D.,  1983b, \mn@doi [\apj] {10.1086/161525},
  \href {https://ui.adsabs.harvard.edu/abs/1983ApJ...275..195W} {275, 195}

\bibitem[\protect\citeauthoryear{{Western} \& {Watson}}{{Western} \&
  {Watson}}{1984}]{Western1984Apj285}
{Western} L.,  {Watson} W.,  1984, \mn@doi [\apj] {10.1086/162487}, \href
  {https://ui.adsabs.harvard.edu/abs/1984ApJ...285..158W} {285, 158}

\bibitem[\protect\citeauthoryear{{Wiebe} \& {Watson}}{{Wiebe} \&
  {Watson}}{1998}]{1998ApJ...503L..71W}
{Wiebe} D.~S.,  {Watson} W.~D.,  1998, \mn@doi [\apjl] {10.1086/311526}, \href
  {http://ukads.nottingham.ac.uk/abs/1998ApJ...503L..71W} {503, L71+}

\bibitem[\protect\citeauthoryear{{Wyenberg}, {Lankhaar}, {Rajabi}, {Chamma}  \&
  {Houde}}{{Wyenberg} et~al.}{2021}]{Wyenberg2021MNRAS}
{Wyenberg} C.~M.,  {Lankhaar} B.,  {Rajabi} F.,  {Chamma} M.~A.,   {Houde} M.,
  2021, \mn@doi [\mnras] {10.1093/mnras/stab2222}, \href
  {https://ui.adsabs.harvard.edu/abs/2021MNRAS.507.4464W} {507, 4464}

\bibitem[\protect\citeauthoryear{{Yun} et~al.,}{{Yun}
  et~al.}{2016}]{2016ApJ...822....3Y}
{Yun} Y.,  et~al., 2016, \mn@doi [\apj] {10.3847/0004-637X/822/1/3}, \href
  {https://ui.adsabs.harvard.edu/abs/2016ApJ...822....3Y} {822, 3}

\bibitem[\protect\citeauthoryear{{van Straten}, {Manchester}, {Johnston}  \&
  {Reynolds}}{{van Straten} et~al.}{2010}]{2010PASA...27..104V}
{van Straten} W.,  {Manchester} R.~N.,  {Johnston} S.,   {Reynolds} J.~E.,
  2010, \mn@doi [\pasa] {10.1071/AS09084}, \href
  {https://ui.adsabs.harvard.edu/abs/2010PASA...27..104V} {27, 104}

\makeatother
\end{thebibliography}

\appendix

\section{Derivation of Matter-Radiation Equations}
\label{a:deriv}
Some elements of the theory in the present work are well-known, and we deal briefly
with these, whilst explaining those parts that are new, or less familiar, in more
detail. The analysis may be divided into the following main blocks: (1) the development
of evolution equations for elements of the DM, (2) derivation of a Fourier-based representation 
of the broad-band electric
field\footnote{That is broad in comparison to the homogeneous profile, but still of
spectral line breadth}, (3) application of a rotating wave approximation to remove all
terms that oscillate at the wave frequency or higher, (4) grouping of the molecular
levels with transition rules that classify specific elements of the DM, (5) derivation
of transfer equations for the complex amplitudes of the radiation in the time
domain, (6) Fourier transformation of the transfer and DM equations into the frequency
domain, (7) formal solution of the transfer equations for the complex field amplitudes
and (8) analytic elimination of the electric field complex amplitudes and off-diagonal DM
elements to leave the problem as a set of non-linear algebraic equations in the inversions alone.

\subsection{Evolution equations of the DM}
\label{ss:evolDM}

For Part (1), we use essentially equation~(1) and equation~(2) of \citet{Tobin2023}, which we note
are in 3D. We therefore write down the pair of
equations,
\begin{align}
\left( \!
  \frac{\pd}{\pd t} + \vec{v} \cdot \vec{\nabla}
\! \right) \! \rho_{pq}
 & \! = \!
\frac{i}{\hbar} \! \sum_{j\neq p,q}^N \!\! \left(
                                 \rho_{pj} \ham_{qj}^* - \rho_{jq} \ham_{pj}
                                \right) \nonumber \\
& \! + \!
\frac{i\ham_{pq}}{\hbar} \left(
                          \rho_{p} - \rho_{q}
                        \right)
-
\left(
   \gamma_{pq} + i\omega_{pq}) 
\right) \rho_{pq} ,
\label{eq:rho_off}
\end{align}
representing the evolution of a general off-diagonal element of the DM, coupling the energy
levels $p$ (upper) and $q$, and its counterpart for a diagonal DM element,
\begin{align}
\left( \!
  \frac{\pd}{\pd t} + \vec{v} \cdot \vec{\nabla}
\! \right) \! \rho_{q}
& = -\frac{2}{\hbar} \Im \left\{
    \sum_{j=1}^{q-1} \rho_{qj} \ham_{qj}^* + \sum_{j=q+1}^N \rho_{jq}^* \ham_{jq}
                       \right\} \nonumber \\
& + \sum_{j \neq q}^N \left(
     k_{jq} \rho_{j} - k_{qj} \rho_{q}
                 \right) \nonumber \\
& + k_{Rq} \! \left( \!
  \phi ( \vec{v} ) \!\! \int_{\vec{v}'} \! \rho_{q} (\vec{r},\vec{v}',t) d^3 v' \! - \rho_{q}
       \!  \right) .
\label{eq:rho_diag}
\end{align}
We note that equation~(2) of \citet{Tobin2023} is formed from two versions of equation~(\ref{eq:rho_diag})
so as to represent the evolution of an inversion between a pair of levels, rather than of
an individual level.
Many symbols in equation~(\ref{eq:rho_off}) and equation~(\ref{eq:rho_diag}) need definitions at this point. Elements
of the DM are represented by $\rho_{pq}$, noting that the DM is Hermitian and of dimension
$N \times N$, where $N$ is the number of energy levels. Indices of diagonal elements are shown
without the subscript repeated, for example $\rho_{qq} = \rho_q$. Another Hermitian matrix, of the same size, is the
interaction Hamiltonian with components $\ham_{pq}$, which has zeros along its main diagonal. The velocity 
$\vec{v}$ is a molecular velocity that may have thermal and bulk contributions. 
The (angular) frequency difference
between two energy levels $p$ and $q$ is $\omega_{pq} = \omega_p - \omega_q$; the order is significant, making
$\omega_{qp} = -\omega_{pq}$. The transition-dependent constant $\gamma_{pq}$ is the rate of coherence-destroying
processes for the transition $pq$ and includes all `kinetic' proccess that transfer population into and out of
levels $p$ and $q$ with the exception of the maser radiation itself. It also includes elastic processes, such
as collisions that do not cause a transition. $\gamma_{pq}$ is also the 
width of the homogeneous line shape function of the
transition. The kinetic process rate coefficient, $k_{pq}$, that causes
transitions by all processes excluding the maser radiation field, is therefore slower than $\gamma_{pq}$, as
$k_{pq}$ contains only the inelastic subset of the kinetic processes that contribute to $\gamma_{pq}$.

Diagonal elements of the DM can be interpreted as number densities of molecules
per unit velocity range. They therefore follow the functional form $\rho_{p} = \rho_{p}(\vec{r},\vec{v},t)$,
being functions of position, velocity and time. In equation~(\ref{eq:rho_diag}), the $\Im$ symbol denotes taking the
imaginary part of an expression, and $z^*$ is the complex conjugate of a complex number $z$. The rate coefficient $k_{Rp}$
represents the redistribution rate within level $p$ via elastic processes and by 
the velocity-changing part of inelastic
processes. Such redistributive processes attempt to relax the velocity-specific population $\rho_{p}$, at
velocity $\vec{v}$ towards the 3D Gaussian $\phi (\vec{v})$, given by
\begin{equation}
\phi (\vec{v}) = \left(
                    \frac{m}
                         {2\pi k_B T(\vec{r})}
                 \right)^{3/2}
                \exp \left\{
                        \frac{-m v^2}
                             {2 k_B T(\vec{r})}
                     \right\} ,
\label{eq:3dgauss}
\end{equation}
noting that the overall number density of level $p$
at position $\vec{r}$ and time $t$ is
\begin{equation}
\varrho_p (\vec{r},t) = \int_{\vec{v}'} \rho_{p} (\vec{r},\vec{v}',t) d^3 v' ,
\label{eq:integrated_rho}
\end{equation}
and other quantities not defined previously are the molecular mass, $m$, Boltzmann's constant, $k_B$,
and the kinetic temperature $T(\vec{r})$.

\subsection{Description of the electric field}

Part (2) of the derivation is to define the electric field to be used in the model.
As this is a 3D model, the electric field at an arbitrary position within the model domain must consist of
the sum of contributions from many rays, all moving in different directions. Within each ray, there
are many frequencies, distributed over a line, or a blend of lines, so that each ray is itself a
sum of Fourier components. One Fourier component, centered on
frequency $\omega$, sampled from the ray $j$ has the analytic signal
representation,
\begin{equation}
\vec{\td}_j = \left(
              \hv{e}_{R,j} \td_{R,j} + \hv{e}_{L,j} \td_{L,j}
             \right) e^{-i\omega ( t - \hv{n}_j \cdot \vec{r}/c )} .
\label{eq:field_1ray_1compt}
\end{equation}
The field in equation~(\ref{eq:field_1ray_1compt}) obeys the same (IEEE) polarization conventions as
those used in \citet{Tobin2023}. For additional details see
\citet{1996A&AS..117..161H,2010PASA...27..104V}. Stokes $V$ is defined as the right-handed
intensity minus the left-handed intensity.

For ray $j$, equation~(\ref{eq:field_1ray_1compt}) may be generalised to a broad-band field of Fourier
strips, each centered on a frequency $\omega_n$. If $T$ is the observer's radiation sampling time, each 
strip has a finite width $2\pi /T$ in frequency. It is useful to define a local
frequency $\varpi_n = \omega_n - \omega_0$, where $\omega_0$ is the centre frequency of a
certain line or blend. The broad-band field of ray $j$ may then be written as an infinite sum, over
index $n$, of Fourier components
\begin{align}
\tilde{\vec{E}_j} & \! = \! e^{-i \omega_0 (t - \hv{n}_j \cdot \vec{r}/c ) }
                   \!\!\!\! \sum_{n=-\infty}^\infty \!\!\! \left[
                                  \hv{e}_{R,j} \td_{R,j} (\vec{r},\varpi_n)
                              \! + \! \hv{e}_{L,j} \td_{L,j} (\vec{r},\varpi_n)
                                       \right] \nonumber \\
                   & \times e^{-i \varpi_n (t - \hv{n}_j \cdot \vec{r}/c)} .
\label{eq:field_1ray_broad}
\end{align}

A pair of unit vectors in equation~(\ref{eq:field_1ray_broad})
that satisfy the IEEE/IAU definitions of right- and left-handedness for ray $j$ are,
\begin{align}
\hv{e}_{R,j} & = (\hv{x}_j + i \hv{y}_j) / \sqrt{2} \label{eq:e-right} \\
\hv{e}_{L,j} & = (\hv{x}_j - i \hv{y}_j) / \sqrt{2} .
\label{eq:e-left}
\end{align}
These definitions are private to the individual ray, $j$.
However, the DM at a particular point must interact with rays converging from sources spread across all
available solid angle. We use the unit vectors defined in equation~(\ref{eq:e-right}) and
equation~(\ref{eq:e-left}) to put the electric field from equation~(\ref{eq:field_1ray_broad}) in ray-specific
Cartesian coordinates before summing the field over a total of $J$ rays. If we define $Y_j = \omega_0 (t - \hv{n}_j \cdot \vec{r}/c)$, and let $\td_{R,jn}$ be a shorthand for the Fourier
component $\td_{R,j} (\vec{r},\varpi_n)$, then
the overall electric field at point $\vec{r}$ becomes 
\begin{equation}
\vec{E} = \frac{1}{2}  \left\{
            \sum_{j=1}^J w_{\Omega,j} e^{-iY_j} \sum_{n=-\infty}^\infty \vec{\td}_{jn} e^{-i \varpi_n t} e^{i \varpi_n \hv{n}_j \cdot \vec{r}/c}
                                 + c.c   \right\},
\label{eq:field_broad_global}
\end{equation}
where each ray has a suitable solid-angle weighting of
$w_{\Omega,j} = \sqrt{\delta \Omega_j /4\pi}$ in order to yield the standard solid angle for a ray-specific intensity. The field in equation~(\ref{eq:field_broad_global}) has been made 
real by adding the complex conjugate, denoted by $c.c.$, of
the analytic signal expression. The vector Fourier component $\vec{\td}_{jn} = \vec{\td}_{j}(\vec{r},\varpi_n)$ is
\begin{equation}
\vec{\td}_{jn} = \hv{x}_j \td_{xj,n} + \hv{y}_j \td_{yj,n} .
\label{eq:vector_fourier_j}
\end{equation}
We use the finite wavetrain Fourier transform (see below) 
to
replace these vector Fourier components, with the time-domain amplitudes of the
form $\vec{\td}_j(\vec{r},t)$.
The general electric field is now
\begin{equation}
\vec{E} = \frac{1}{2}  \left\{
            \sum_{j=1}^J w_{\Omega,j} e^{-iY_j} \vec{\td}_j(\vec{r},t) + c.c.
                               \right\} .
\label{eq:field_broad_global_time}
\end{equation}

\subsection{Interaction terms}

The electric field represented by equation~(\ref{eq:field_broad_global_time}) is coupled to the molecular density
matrix via its interaction with the dipole operator, $\hv{d}_{pq}$, of a transition $pq$, and the mathematical
form of the coupling is defined through the interaction part of the Hamiltonian operator,
\begin{equation}
\ham_{pq} = - \hv{d}_{pq} \cdot \vec{E} .
\label{eq:hampq}
\end{equation}
A number of terms similar to equation~(\ref{eq:hampq}) are evident in the DM equations~(\ref{eq:rho_diag}) and
(\ref{eq:rho_off}). It is possible, with the aid of equation~(\ref{eq:field_broad_global_time}), to write 
equation~(\ref{eq:hampq}) in terms of electric field amplitudes, for example,
\begin{equation}
\ham_{pq} = -\frac{1}{2} \left\{
             \sum_{j=1}^J w_{\Omega,j} \left[ 
                    e^{-iY_j} \hv{d}_{pq} \cdot \vec{\td}_j
                  + e^{iY_j} \hv{d}_{pq} \cdot \vec{\td}_j^*
                          \right]
                                \right\} .
\label{eq:interact}
\end{equation}

\subsection{Rotating wave approximation}
\label{ss:rwa}

Equation~(\ref{eq:interact}) contains oscillations at the carrier
frequency of the wave, $\omega_0 (pq)$. We apply a rotating wave approximation (RWA) in order to keep
only those parts of the expression where the electric field is close to resonance with $\omega_{pq}$, the
natural frequency of the transition. To this end, we expand off-diagonal DM elements in the form,
\begin{equation}
\rho_{pq} = -(i/2) \sum_{j=1}^J s_{pq}^{(j)} e^{-i \omega_0 t} e^{i \omega_0 \hv{n}_j \cdot \vec{r}/c} ,
\label{eq:defspq}
\end{equation}
where $s_{pq}^{(j)}$ is the part of the DM element specific to ray $j$ that contains only 
relatively slow frequencies of order
$|\omega_0 (pq) - \omega_{pq}|$, where $\omega_0(pq)$ is the radiation band-centre frequency in the
vicinity of the transition $pq$.
To carry out the RWA successfully, we follow \citet{Western1983Apj268} in making
the additional, very reasonable, assumption that fields from different rays cannot be coherent. That is,
terms such as $e^{i \vec{r} \cdot (\vec{k}_\xi - \vec{k}_j)}$ are fast, and therefore eliminated by the RWA, unless
$\xi = j$. Terms that result from the multiplication of an off-diagonal DM element, expanded as in
equation~(\ref{eq:defspq}), by an interaction term, corresponding 
to the complex conjugate of equation~(\ref{eq:interact}),
are reduced by the RWA to comparatively simple expressions of the form,
\begin{equation}
\rho_{qj} \ham_{qj}^* = \frac{i \hv{d}_{qj}^*}{4}  \cdot \sum_{\xi=1}^J w_{\Omega,\xi} s_{qj}^{(\xi)}
                         \vec{\td}_{\xi}^*(\vec{r},t) ,
\label{eq:rhoham}
\end{equation}
where $\vec{\td}_{\xi}^*(\vec{r},t)$ is the time-domain electric field amplitude of ray $\xi$ in the
band local to transition $qj$. 
With the RWA applied fully to equation~(\ref{eq:rho_diag}), a diagonal DM element evolves according to,
\begin{align}
\left( \!
  \frac{\pd}{\pd t} \!+\! \vec{v} \! \cdot \! \vec{\nabla}
\! \right) \! \rho_{p}
& \!=\! -\frac{1}{2 \hbar} \Re \! \left\{ \sum_{\xi=1}^J \! w_{\Omega,\xi} \left[
   \sum_{j=1}^{p-1} s_{pj}^{(\xi)} \hv{d}_{pj}^* \cdot \vec{\td}_\xi^* \!\! \right. \right. \nonumber \\   
&  - \left. \left. \! \!\!\! \sum_{j=p+1}^N \!\! s_{jp}^{*(\xi)} \hv{d}_{jp} \cdot \vec{\td}_\xi
                                                                        \right]
                       \! \right\}
  + \sum_{j \neq p}^N \left(
     k_{jp} \rho_{j} - k_{pj} \rho_{p}
                 \right) \nonumber \\
& + k_{Rp} \left(
  \phi ( \vec{v} ) \!\! \int_{\vec{v}'} \rho_p (\vec{r},\vec{v}',t) d^3 v' - \rho_{p}
        \right) .
\label{eq:rho_diag_rwa}
\end{align}
In equation~(\ref{eq:rho_diag_rwa}), the symbol $\Re$ denotes taking the real part of the expression.
Note that this equation is free of terms that oscillate at the wave frequency,
so the aim of the RWA has been achieved.

We identify two types of off-diagonal element. Type~2 elements are in transitions between
energy levels that would be degenerate in the absence of an applied magnetic field. Other transitions
are type~1, without a requirement to be
electric-dipole allowed. A type~2 element
has no fast part: a good example would be a transition in a Zeeman-split system that changes the magnetic quantum number,
say $m_J$, but preserves $J$. The frequency of a type~2 transition must be only of a similar
magnitude to the range of local frequencies, measured from the current band centre. A type~1
off-diagonal DM element evolves according to
\begin{align}
& \left( \!
  \frac{\pd}{\pd t} + \vec{v} \! \cdot \! \vec{\nabla}
\! \right) \! s_{pq}^{(\xi)} \!
=
-\frac{i w_{\Omega,\xi}}{2 \hbar} \left\{
\sum_{j=1}^{q-1} \left( \!
                      s_{pj}^{(\xi)} \hv{d}_{qj}^* \cdot \!\vec{\td}_\xi^* + s_{qj}^{*(\xi)} \hv{d}_{pj} \!\cdot
                      \vec{\td}_\xi 
                                \right) +
                            \right. \nonumber \\
& \left. \!\!\!\! \sum_{j=q+1}^{p-1} \!\!\! \left( \!
                      s_{pj}^{(\xi)} \hv{d}_{jq} \! \cdot \!\vec{\td}_\xi \! - \! s_{jq}^{(\xi)} \hv{d}_{pj} \! \cdot \!\vec{\td}_\xi
                    \! \right)
 \!\! - \!\!\!\!\! \sum_{j=p+1}^N  \!\!\!  \left( \!
                      s_{jp}^{*(\xi)} \hv{d}_{jq} \!\cdot \!\vec{\td}_\xi \! + \! s_{jq}^{(\xi)} \hv{d}_{jp}^* \!\cdot \!\vec{\td}_\xi^*
                    \! \right)
                            \!\! \right\} \nonumber \\
& + w_{\Omega,\xi}\frac{\rho_{p} \! - \! \rho_{q}}{\hbar} \hv{d}_{pq} \! \cdot \vec{\td}_\xi
 \! -
\! \left[
   \gamma_{pq} \! - i \! \left( \omega_0 - \omega_{pq} - \omega_0 \vec{v}\! \cdot \! \hv{n}_\xi / c \right)
\right] \! s_{pq}^{(\xi)} ,
\label{eq:rho_off_rwa_1}
\end{align}
where we now require $p > q$, and all paired indices are written with the
upper index first. The angular frequency 
$\omega_0$ stands for $\omega_0(pq)$, the radiation band-centre
frequency local to the $pq$ transition.  If the type~1 transition has no dipole, then the first term
on the third line of equation~(\ref{eq:rho_off_rwa_1}) is zero, and there is no direct coupling
to the inversion in the $pq$ transition itself.

The alternative to equation~(\ref{eq:rho_off_rwa_1}) that applies in the case 
of a type~2 transition and the $J=1-0$ system is
\begin{align}
& \left( \!
  \frac{\pd}{\pd t} + \vec{v} \! \cdot \! \vec{\nabla}
\! \right) \! \hat{s}_{pq}^{(\xi)}
 =
-\frac{i w_{\Omega,\xi}}{2 \hbar} \left\{
                      s_{p1}^{(\xi)} \hv{d}_{q1}^* \cdot \vec{\td}_\xi^* + 
                      s_{q1}^{*(\xi)} \hv{d}_{p1} \cdot \vec{\td}_\xi
                            \right\} \nonumber \\
& - (\gamma_{pq} + i \omega_{pq} )
\hat{s}_{pq}^{(\xi)} ,
\label{eq:rho_off_rwa_2}
\end{align}
where level 1 is the undivided $J=0$ state.
In equation~(\ref{eq:rho_off_rwa_2}), off-diagonal DM elements that are of type~2 are marked with the caret
symbol. It has been assumed, at least for the purposes of the present work, that transitions of type~2
are not dipole allowed, and that they have typical frequences of order 1-1000\,kHz.

\subsection{Frequency domain equations} \label{a:fdomeq}
As a prelude to Fourier transformation, we drop the terms containing the operator $\vec{v} \cdot \nabla$ in
equations~\ref{eq:rho_diag_rwa}-\ref{eq:rho_off_rwa_2} on the grounds that $|\vec{v}| \ll c$. We also
re-define the Fourier component of the electric field, $\vec{\td}_{jn}$, to include the spatial exponential
in $\varpi_n$ from equation~(\ref{eq:field_broad_global}).
Terms that are time-domain products transform to frequency-domain convolutions. The
\hspace*{0.3333em}Fourier transform of equation~(\ref{eq:rho_diag_rwa}) is
\begin{align}
\rho_{p,n}& \! = \! \frac{- \pi \tilde{\cal L}_{p,n}}{2 \hbar} \!\!\!\! \sum_{m=-\infty}^\infty 
                  \sum_{\xi=1}^J \! w_{\Omega,\xi} \! \left\{
      \sum_{j=1}^{p-1} \! \left[
                      \hv{d}_{pj}^* \!\cdot \! \vec{\td}_{\xi,m}^* s_{pj,m+n}^{(\xi)} 
                    + \right. \right. \nonumber \\
          & \left. \left.         
                    \hv{d}_{pj} \!\cdot \! \vec{\td}_{\xi,m} s_{pj,m-n}^{*(\xi)}
                    \right] 
                                                                         \right. \nonumber \\
         & \left. \!\! - \sum_{p+1}^N \left[
                     \hv{d}_{jp} \cdot \! \vec{\td}_{\xi,m} s_{jp,m-n}^{*(\xi)} 
                     + \hv{d}_{jp}^* \cdot \! \vec{\td}_{\xi,m}^* s_{jp,m+n}^{(\xi)}
                         \right]
                                                                         \right\} \nonumber \\
         & + 2\pi \tilde{\cal L}_{p,n} \!\! \left[ \sum_{j \neq p}^N \! k_{jp} \rho_{j,n} 
           \!+\! k_{Rp} \! \left(  \! \phi ( \vec{v} ) \!\!\! \int_{\vec{v}'} \!\! \rho_{p,n} (\vec{r},\vec{v}') d^3 v'
                               \!-\! \rho_{p,n} \right)
                                  \!\right] ,
\label{eq:rho_diag_ft} 
\end{align}
where the indices $n$ and $m$ denote frequency bins, and $\tilde{\cal L}_{p,n}$ is a complex Lorentzian
function, given by
\begin{equation}
\tilde{\cal L}_{p,n} = 1/[(2\pi)(\Gamma_p - i \varpi_n)] .
\label{eq:diaglor}
\end{equation}
The loss-rate, $\Gamma_p$, has been set equal to $ \sum_{j \neq p}^N k_{pj}$.

As in Section~\ref{ss:rwa}, two equations are used to represent the different types of off-diagonal
DM element. For the type~1 transitions, the off-diagonal DM elements are given by
\begin{align}
s_{pq,n}^{(\xi)} & \!=\!  \frac{- i \pi w_{\Omega,\xi} \tilde{L}_{pq,n}^{(\xi)}}{\hbar} \!\!\!\!\! \sum_{m=-\infty}^\infty \!\!\!\left\{ 
            2 i \left(
                  \rho_{p,n-m} - \rho_{q,n-m}
               \right) \hv{d}_{pq} \cdot \vec{\td}_{\xi,m} \right. \nonumber \\
        & \left. + \sum_{j=1}^{q-1}           
             \left(
                \hv{d}_{qj}^* \!\cdot \!\vec{\td}_{\xi,m}^* s_{pj,m+n}^{(\xi)} 
            + \hv{d}_{pj} \!\cdot \!\vec{\td}_{\xi,m} s_{qj,m-n}^{*(\xi)}
                             \right)
                                                                               \right. \nonumber \\
       & \left. + \!\! \sum_{j=q+1}^{p-1} \! \left(
               \hv{d}_{jq} \cdot \! \vec{\td}_{\xi,m} s_{pj,n-m}^{(\xi)} 
               - \hv{d}_{pj} \cdot \! \vec{\td}_{\xi,m} s_{jq,n-m}^{(\xi)}
                           \right)
                                                                                \right. \nonumber \\
       & \left. - \sum_{j=p+1}^N \left(
               \hv{d}_{jq} \cdot \! \vec{\td}_{\xi,m} s_{jp,m-n}^{*(\xi)} 
               + \hv{d}_{jp}^* \cdot \! \vec{\td}_{\xi,m}^* s_{jq,m+n}^{(\xi)}
                        \right)
                                                                           \right\} .
\label{eq:rho_off_ft_1}
\end{align}
The complex Lorentzian response profile in this case is
\begin{equation}
\tilde{L}_{pq,n}^{(\xi)}(\vec{v}) = \frac{1}{2\pi [\gamma_{pq} - i (\varpi_n - \Delta \omega_{pq} - \omega_0 \vec{v} \cdot \hv{n}_\xi/c)]} ,
\label{eq:off1lor}
\end{equation}
where $\Delta \omega_{pq} = \omega_{pq} - \omega_0 (pq)$, the difference between the natural frequency of the 
transition and the radiation band centre for that transition, or overlapping group of transitions.

The Fourier-transformed equation for the DM element of a type~2 transition is
\begin{equation}
\hat{s}_{pq,n}^{(\xi)} \!=\!\!  \frac{- i \pi w_{\Omega,\xi}\hat{L}_{pq,n}}{\hbar} \!\!\!\!\!\! \sum_{m=-\infty}^\infty 
                 \!\!\!\!\!\left\{
                \hv{d}_{q1}^* \!\cdot \!\vec{\td}_{\xi,m}^* s_{p1,m+n}^{(\xi)} \!
                \!+\! \hv{d}_{p1} \!\cdot \!\vec{\td}_{\xi,m} s_{q1,m-n}^{*(\xi)} 
                                                                           \right\} ,
\label{eq:rho_off_ft_2}
\end{equation}
with the complex Lorentzian profile,
\begin{equation}
2\pi \hat{L}_{pq,n} = [ \gamma_{pq} - i (\varpi_n - \omega_{pq})]^{-1} .
\label{eq:offlor_type2}
\end{equation}
Note that all of the equations~(\ref{eq:rho_diag_ft}-\ref{eq:offlor_type2}) are algebraic equations 
in the frequency domain; Fourier
transformation has removed their differential character. Also note that
the off-diagonal elements of the DM on the right-hand sides of equation~(\ref{eq:rho_diag_ft}) and
equation~(\ref{eq:rho_off_ft_1}) have all been written as type~1, but some
may in fact be of type~2, depending on the levels involved.

\subsection{Radiation transfer}

The electric fields considered in this work are classical fields obeying Maxwell's equations. Standard
derivations relate the field $\vec{E}$, given in the time domain by equation~(\ref{eq:field_broad_global_time}), 
to the macroscopic polarization of the medium, $\vec{P}$. This latter field is defined as
\begin{equation}
\vec{P} (\vec{r},t) = \int_{\vec{v}} Tr[ \uprho \mathrm{\hat{d}}] d^3 v ,
\label{eq:macropol}
\end{equation}
where $Tr$ denotes the trace, and $\uprho$, $\mathrm{\hat{d}}$ are, respectively, the DM and
the dipole operator in matrix form. After application of the RWA, the electric field 
of ray $\xi$ evolves according to
\begin{align}
\left(
   \frac{\pd}{\pd t} + c \hv{n}_\xi \! \cdot \! \vec{\nabla}
\right) \vec{\td}_{\xi}
& +
 \frac{\sigma}{2\epsilon_0} \vec{\td}_{\xi} \nonumber \\
& = \frac{\omega_0}{2 \epsilon_0 w_{\Omega,\xi}} \sum_{p=2}^{N_G} \sum_{j=1}^{p-1} \hv{d}_{pj}^*
\int_{\vec{v}} s_{pj}^{(\xi)}(\vec{r},\vec{v},t) d^3v ,
\label{eq:rt_time}
\end{align}
where $\hv{n}_\xi$ is a unit vector along the propagation direction of the ray, $\sigma$ is the
conductivity of the medium, and $N_G$ is the number of Zeeman-split energy levels
in a group formed from, for example, one rotational state. The presence of the term including
$\sigma$ allows the theory to include the effect of attenuation due to, for example, a
substantial population of free electrons. Fourier transformation of equation~(\ref{eq:rt_time})
leads to an equation for the spatial evolution of the amplitude-in-frequency along one axis specific to
one ray. The linearity of equation~(\ref{eq:rt_time}) makes this operation fairly straightforward. The
result is
\begin{equation}
\hv{n}_\xi \! \cdot \! \vec{\nabla} \vec{\td}_{\xi,n}
+ \left[\frac{\sigma (\vec{r})}{2 \epsilon_0 c} \!-\!  \frac{i\varpi_n}{c} \right]\!\vec{\td}_{\xi,n}
\! = \!\frac{\omega_0 }{ 2\epsilon_0 c w_{\Omega,\xi}} \!\!\sum_{p=2}^{N_G} \! \sum_{j=1}^{p-1} \! \hv{d}_{pj}^*
\! \! \! \int_{\vec{v}} \!\! s_{pj,n}^{(\xi)} d^3v .
\label{eq:rt_freq}
\end{equation}

\subsection{Velocity Integration}
\label{ss:velint}

The RT equation, equation~(\ref{eq:rt_freq}), already expresses the off-diagonal DM
elements in a velocity-integrated form, and we now formalise this relation as,
\begin{equation}
S_{pj,n}^{(\xi)}(\vec{r}) = \int_{\vec{v}} \!\! s_{pj,n}^{(\xi)}] d^3v = \int_{\vec{v}} \!\! s_{pj}^{(\xi)} (\vec{r},\vec{v},\varpi_n) d^3v.
\label{eq:offvelint}
\end{equation}
With the help of equation~(\ref{eq:offvelint}), it is straightforward to re-write equation~(\ref{eq:rt_freq})
in the style of a standard first-order linear ordinary differential equation, to be solved along the ray distance $ds_\xi$:
\begin{equation}
\frac{d \vec{\td}_{\xi,n}}{d s_\xi} + \kappa(\vec{r}) \vec{\td}_{\xi,n} = 
\frac{D}{w_{\Omega,\xi}} \sum_{p=2}^{N_G} \sum_{j=1}^{p-1} \hv{d}_{pj}^* S_{pj,n}^{(\xi)} ,
\label{eq:rt_ode}
\end{equation}
where $\kappa = [\sigma/(2 \epsilon_0 c) - i\varpi_n /c]$ is the complex
attenuation coefficient, a function of position in general, and we have defined
the new constant,
\begin{equation}
D = \omega_0 / (2\epsilon_0 c) ,
\label{eq:defdconst}
\end{equation}

Since the Lorentzian profiles
in both equation~(\ref{eq:diaglor}) and equation~(\ref{eq:offlor_type2}) are independent of velocity, it is
useful to formally integrate equations~(\ref{eq:rho_diag_ft}), (\ref{eq:rho_off_ft_1}) and 
(\ref{eq:rho_off_ft_2}) over velocity, reducing
the number of independent variables. We
use equation~(\ref{eq:integrated_rho}) to define the velocity-integrated diagonal DM element,
complementing equation~(\ref{eq:offvelint}). The velocity-integrated form of equation~(\ref{eq:rho_diag_ft}) is
\begin{align}
\varrho_{p,n}& \! = 2\pi \tilde{\cal L}_{p,n} \sum_{j \neq p}^N k_{jp} \varrho_{j,n} -
\! \frac{\pi \tilde{\cal L}_{p,n}}{2 \hbar} \!\!\! \sum_{\xi=1}^J w_{\Omega,\xi}
  \sum_{m=-\infty}^\infty \!\! \nonumber \\
   & \left\{ \!  \sum_{j=1}^{p-1} \! \left[
                       \hv{d}_{pj}^* \!\cdot \! \vec{\td}_{\xi,m}^* S_{pj,m+n}^{(\xi)} 
                     + \hv{d}_{pj} \!\cdot \! \vec{\td}_{\xi,m} S_{pj,m-n}^{*(\xi)}
                    \right] 
                                                                         \right. \nonumber \\
         & \left. - \sum_{p+1}^N \left[
                        \hv{d}_{jp} \! \cdot \! \vec{\td}_{\xi,m} S_{jp,m-n}^{*(\xi)} 
                      + \hv{d}_{jp}^* \!\cdot \! \vec{\td}_{\xi,m}^* S_{jp,m+n}^{(\xi)}
                         \right]
                                                                         \right\} ,
\label{eq:diag_vi}
\end{align}
noting the disappearance of the redistributive terms  from equation~(\ref{eq:diag_vi}).
It is also straightforward to integrate equation~(\ref{eq:rho_off_ft_2}) over velocity, and the result
is
\begin{align}
\hat{S}_{pq,n}^{(\xi)} & \!\!=\!\! \frac{- i \pi w_{\Omega,\xi}\hat{L}_{pq,n}}{\hbar} \!\!\!\!\!\! \sum_{m=-\infty}^\infty \!\!\!\!\! \left\{
              \!  \hv{d}_{q1}^* \!\cdot \!\vec{\td}_{\xi,m}^* S_{p1,m\!+\!n}^{(\xi)} 
              \!\!+\! \hv{d}_{p1} \!\cdot \!\vec{\td}_{\xi,m} S_{q1,m\!-\!n}^{*(\xi)}
                                                                           \! \right\} \!,
\label{eq:off2_vi}
\end{align}
where again,
the velocity-integrated off-diagonal elements on the right-hand side of equation~(\ref{eq:off2_vi}) are marked
as type~1, but some may in fact be of type~2.

The main difficulty in this section arises on integrating equation~(\ref{eq:rho_off_ft_1}) over
velocity, because the Lorentzian function in equation~(\ref{eq:off1lor}) depends on $\vec{v}$. To
carry out the integration, we assume that an unknown, normalized, velocity distribution function, $\psi_{pj}(\vec{v})$,
relates the velocity-dependent and velocity-integrated off-diagonal DM elements, that is
\begin{equation}
s_{pj,m}^{(\xi)} = \psi_{pj}(\vec{v}) S_{pj,m}^{(\xi)} .
\end{equation} 
When integrating equation~(\ref{eq:rho_off_ft_1}) over velocity, we now find integrals over products
of a Lorentzian and an unknown velocity distribution. for example
\begin{equation}
\int_{\vec{v}} \! s_{pj,m}^{(\xi)}(\vec{r},\vec{v}) \tilde{L}_{pq,n}^{(\xi)}(\vec{v}) d^3v 
       = \!\! S_{pj,m}^{(\xi)}(\vec{r}) \int_{\vec{v}} \! \psi_{pj}(\vec{v}) \tilde{L}_{pq,n}^{(\xi)}(\vec{v}) d^3v ,
\label{eq:convlor}
\end{equation}
noting that in general the Lorentzian and the unknown distribution belong to different transitions.
In the special case where $\psi_{pj}(\vec{v}) = \phi (\vec{v})$, a Gaussian, the integral on the right-hand side
of equation~(\ref{eq:convlor}) reduces to a combination of Voigt and Faraday-Voigt profiles. However, we will not
immediately assume this, and write the integral as
\begin{equation}
\Xi_{pj,n}^{pq} = \int_{\vec{v}} \! \psi_{pj}(\vec{v}) \tilde{L}_{pq,n}^{(\xi)}(\vec{v}) d^3v .
\label{eq:bigxi}
\end{equation}
With the aid of equation~(\ref{eq:bigxi}), we may now write down the 
velocity-integrated form of equation~(\ref{eq:rho_off_ft_1}) as
\begin{align}
S_{pq,n}^{(\xi)} & \!\!=\!  \frac{- i \pi w_{\Omega,\xi}}{\hbar} \!\!\!\!\! \sum_{m=-\infty}^\infty \!\!\!\left\{
             2 i \hv{d}_{pq} \!\cdot \vec{\td}_{\xi,m} \left(
                  \varrho_{p,n-m} \Xi_{p,n}^{pq} \!-\! \varrho_{q,n-m} \Xi_{q,n}^{pq}
               \right) \right. \nonumber \\
       & \left. \! + \!\! \sum_{j=1}^{q-1} \! \left(
                \hv{d}_{qj}^* \!\cdot \!\vec{\td}_{\xi,m}^* S_{pj,m+n}^{(\xi)} \Xi_{pj,n}^{pq} 
              + \! \hv{d}_{pj} \!\cdot \!\vec{\td}_{\xi,m} S_{qj,m-n}^{*(\xi)} \Xi_{qj,n}^{pq}
                           \!\right)
                                                                               \right. \nonumber \\
       & \left. \!+ \!\!\!\! \sum_{j=q+1}^{p-1} \!\! \left(
               \hv{d}_{jq} \!\cdot \!\vec{\td}_{\xi,m} S_{pj,n-m}^{(\xi)} \Xi_{pj,n}^{pq} \! 
               \!- \! \hv{d}_{pj} \!\cdot \!\vec{\td}_{\xi,m} S_{jq,n-m}^{(\xi)} \Xi_{jq,n}^{pq}
                           \!\right)
                                                                                 \right.       \nonumber \\
       & \left. \!- \!\!\!\! \sum_{j=p+1}^N \!\! \left(
               \hv{d}_{jq} \!\cdot \!\vec{\td}_{\xi,m} S_{jp,m-n}^{*(\xi)} \Xi_{jp,n}^{pq} \! 
               \! + \! \hv{d}_{jp}^* \!\cdot \!\vec{\td}_{\xi,m}^* S_{jq,m+n}^{(\xi)} \Xi_{jq,n}^{pq}
                        \right) 
                                                                          \! \right\} \!,
\label{eq:off1_vi}
\end{align}
where the symbols $\Xi_{p,n}^{pq}$ and $\Xi_{q,n}^{pq}$ in the 
first row of equation~(\ref{eq:off1_vi}) behave similarly to the definition in
equation~(\ref{eq:bigxi}), but use a velocity distribution $\psi_X(\vec{v})$, where $X$ is an energy level, $p$ or $q$.
Equations~(\ref{eq:rt_ode}), (\ref{eq:diag_vi}), and (\ref{eq:off1_vi}) are very general, and 
may be applied to an 
arbitrary Zeeman system of $N$ levels.

\section{Simplified Equation Sets}
\label{s:simplo}

Any solution of the equations developed in Appendix~\ref{a:deriv} involves a large
number of unknowns: typically one population for each position, energy level and frequency channel (Fourier component) of
the model, and one slow off-diagonal DM element for each position, transition (whether electric
dipole-allowed or not), frequency channel and ray.
It is therefore very advantageous to
seek some simplified equation sets, particularly for checking against earlier work, much of which
is limited to 1D and the simplest Zeeman patterns, particularly $J=1-0$.

\subsection{The J=1-0 pattern}
\label{ss:j1-0pattern}

The most widely considered system in maser polarization modelling is the $J=1-0$ Zeeman pattern,
with an unsplit $J=0$ state (level~1) and a triplet of Zeeman-split levels, $2,3$ and $4$ in
$J=1$. This system has one dipole-allowed transition of each helical type: transitions $21$
($\sigma^+$), $31$ ($\pi$) and $41$ ($\sigma^-$). Each of these transitions corresponds to an
off-diagonal DM element of type~1, but there are also three type~2 transitions amongst the
Zeeman-split levels of the $J=1$ state. The system is summarised in Fig.~\ref{fig:trans_diagram}.

In the $J=1-0$ pattern, equation~(\ref{eq:off1_vi})
reduces to the general form, for upper level $p$, of
\begin{align}
S_{p1,n}^{(\xi)} & = \! -\frac{i\pi w_{\Omega,\xi} \Xi_{p1,n}^{p1}}{\hbar} \!\!\!\! \sum_{m=-\infty}^\infty \!\!\! \left\{
            2 i \Delta_{p1,n-m} \hv{d}_{p1} \! \cdot \! \vec{\td}_{\xi,m}  \right. \nonumber \\
 & \left. \! + a_q  \hv{d}_{q1} \! \cdot \! \vec{\td}_{\xi,m} \hat{S}_{pq,n-m}^{(\xi)}
         \! + a_{q'} \hv{d}_{q'1} \! \cdot \! \vec{\td}_{\xi,m} \hat{S}_{pq',n-m}^{(\xi)}
                                                                             \right\},
\label{eq:j10_off1}
\end{align}
where $q, q' \neq p$ are the other energy levels of the $J=1$ state, and $\Delta_{p1,m} = \varrho_{p,m}-\varrho_{1,m}$
is Fourier component $m$ of the inversion between levels $p\neq 1$ and $1$. The constants
$a_q$, $a_{q'}$ are 1 if $p$ is greater than the respective $q$ or $q'$, but $-1$ otherwise.
If a type~2 off-diagonal element $\hat{S}_{pq,n-m}$ appears, or a similar expression with $q'$,
and $p<q$ or $p<q'$, then the off-diagonal element is customarily replaced by its Hermitian
conjugate $\hat{S}_{qp,m-n}^*$. In the first line of
equation~(\ref{eq:j10_off1}), we have replaced individual level populations with a
population inversion, and extracted a common integral expression, $\Xi_{p1,n}^{p1}$,
throughout. This amounts to assuming the same molecular velocity distribution in all transitions
as a prelude to Section~\ref{ss:cvr_approx}.

Off-diagonal elements of type~2 follow equation~(\ref{eq:off2_vi}), which requires
no further reduction, but note that $p$ may be 3 or 4, and $q<p$, but $q\neq 1$.
We do not yet set terms of this type to zero, on the grounds that we wish to consider possibly very
high degrees of saturation. We also note that the only surviving off-diagonal elements on the right-hand
side of equation~(\ref{eq:j10_off1}) are of type~2, and therefore may be eliminated in favour of other
type~1 off-diagonal elements with the help of equation~(\ref{eq:off2_vi}).

The electric field complex amplitude amplifies according to equation~(\ref{eq:rt_ode}), and for
the $J=1-0$ system the summations on the right-hand side reduce to a single sum over the three electric dipole-allowed
transitions. We are therefore left with,
\begin{align}
\frac{d \vec{\td}_{\xi,n}}{d s_\xi} \! + \! \kappa(\vec{r}) \vec{\td}_{\xi,n}
    \! = \! \frac{D}{w_{\Omega,\xi}}
                        \! \left[
                            \hv{d}_{21,\xi}^* S_{21,n}^{(\xi)} \!+\! \hv{d}_{31,\xi}^* S_{31,n}^{(\xi)} \!+\! \hv{d}_{41,\xi}^* S_{41,n}^{(\xi)}
                         \right] \! .
\label{eq:j10_eamp}
\end{align}

Differences of pairs of equations like equation~(\ref{eq:diag_vi}), with the second population always
being $\varrho_{1,n}$, result in expressions for inversions in the dipole-allowed transitions. This
is a slight simplification in the $J=1-0$ system, since there are three inversions, as opposed to
four energy levels. Inversions are described by the equation,
\begin{align}
\Delta_{p1,n} & \! = \! -\frac{\pi \tilde{\cal L}_n}{2 \hbar} \!\!\! \sum_{\xi=1}^J \!\! w_{\Omega,\xi}
\!\!\!\!\!\! \sum_{m=-\infty}^\infty \!\!\!\!\! \left\{
   \hv{d}_{q1}^* \! \cdot \! \vec{\td}_{\xi,m}^* S_{q1,m+n}^{(\xi)} 
   \! + \! \hv{d}_{q1} \! \cdot \! \vec{\td}_{\xi,m} S_{q1,m-n}^{*(\xi)} \right. \nonumber \\
  & \left. \!\!\!\!\! +  2 \left( 
    \hv{d}_{p1}^* \! \cdot \! \vec{\td}_{\xi,m}^* S_{p1,m+n}^{(\xi)} 
    + \hv{d}_{p1} \! \cdot \! \vec{\td}_{\xi,m} S_{p1,m-n}^{*(\xi)}
   \right) \right. \nonumber \\
  & \left. \!\!\!\!\! + \hv{d}_{q'1}^* \! \cdot \! \vec{\td}_{\xi,m}^* S_{q'1,m+n}^{(\xi)} 
    \!\! + \hv{d}_{q'1} \! \cdot \! \vec{\td}_{\xi,m} S_{q'1,m-n}^{*(\xi)} \right\}
   \! + \! 2\pi \tilde{\cal L}_n P_{p1,n} \delta_0 \! ,
\label{eq:j10_inv}
\end{align}
where the energy levels $q$ and $q'$ are those
other then $p$ in the $J=1$ rotational state, and
where $\Delta_{p1,n} = \varrho_{p,n}- \varrho_{1,n}$. $P_{p1,n}$ is an overall pumping term for
transition $p1$, Fourier component $n$. The presence of the Kronecker delta, $\delta_0$, ensures
that the pump feeds only the central Fourier component (with index $0$). The pumping
term has the formal definition,
\begin{equation}
P_{p1,n} = \sum_{j \neq p}^N k_{jp} \varrho_{j,n} - \sum_{j \neq 1}^N k_{j1} \varrho_{j,n} ,
\label{eq:pumpdef}
\end{equation}
where the $\varrho_{j,n}$ are Fourier components of the many other energy levels, $j$, of the
system that we cannot reasonably model in detail. In order to write both equation~(\ref{eq:j10_inv})
and equation~(\ref{eq:pumpdef}), we have also assumed that levels $1$ to $4$ all have the same loss
rate, $\Gamma$, and therefore the same Lorentzian, $\tilde{\cal L}_n$.
Equations~(\ref{eq:j10_off1})-(\ref{eq:j10_inv}) and (\ref{eq:off2_vi}) provide a complete 
solution, in principle, to the semi-classical maser polarization problem for the $J=1-0$ system in 3D. 
However, suggestions for further simplification are made below.

\subsection{Complete velocity redistribution}
\label{ss:cvr_approx}

A further assumption that we will generally make is that of complete velocity redistribution (CVR),
even under conditions of strong saturation. This is partly for reasons of computational
expediency, but also has a reasonable physical basis, dependent upon a network of optically-thick
IR pumping transitions \citep{1974ApJ...190...27G}. Under CVR, redistribution amongst velocity
subgroups of molecules occurs essentially at the pumping rate, and the velocity distribution function
of any energy level coupled by such an IR network remains approximately Gaussian. 
Wihtout making this approximation, we have already written down a convolution integral over velocity,
involving a complex Lorentzian function with an unknown velocity distribution in equation~(\ref{eq:bigxi}). On
making the CVR approximation, this integral becomes
\begin{equation}
\Xi_{n}^{pq} = \int_{\vec{v}} \! \phi(\vec{v}) \tilde{L}_{pq,n}^{(\xi)}(\vec{v}) d^3v ,
\label{eq:bigxidef}
\end{equation}
where $\phi (\vec{v})$ is a 3D Gaussian, defined in 
equation~(\ref{eq:3dgauss}), that is no longer specific to a particular
transition. In the case of a real Lorentzian, an integral analogous
to equation~(\ref{eq:bigxidef}) forms the standard definition of a Voigt profile. In the less familiar
case of a complex Lorentzian, a Voigt profile arises from the real part, and a Faraday-Voigt profile,
from the imaginary part. 
The result of the integration is
\begin{equation}
\Xi_{n}^{pq} = \frac{1}{2} \! \left( \!
                          \frac{mc^2}
                               {2\pi k_B T \omega_0^2}
                        \!  \right)^{\!\! 1/2} \!\! H(a,q_{pq,n})
                       + \frac{i}
                              {2\sqrt{\pi}\gamma_{pq}} \!F(a,q_{pq,n}) ,
\label{eq:bigxiresult}
\end{equation}
where the Voigt and Faraday-Voigt profiles are, respectively
\begin{equation}
H(a,q_{pq,n}) = \frac{a}{\pi} \int_{-\infty}^\infty \frac{e^{-y^2} dy}{a^2 + (y - q_{pq,n})^2 } ,
\label{eq:hvoigt}
\end{equation}
which is an even function, and
\begin{equation}
F(a,q_{pq,n}) = \frac{a}{\pi} \int_{-\infty}^\infty \frac{e^{-y^2} (y - q_{pq,n}) dy}{a^2 + (y - q_{pq,n})^2 } ,
\end{equation}
which is odd, with shared parameters,
\begin{equation}
a =  \frac{1}{\Delta v_D} \frac{c \gamma_{pq}}{\omega_0} \;\; ; 
    \;\; q_{pq,n} = \frac{1}{\Delta v_D} \frac{c (\Delta \omega_{pq} (\vec{r}) -\varpi_n)}{\omega_0} ,
\end{equation}
where $\Delta v_D = \sqrt{ 2 k_B T(\vec{r})/m }$.
In the case where the Lorentzian in equation~(\ref{eq:bigxidef}) is so 
much narrower than the Gaussian that it may
reasonably be approximated by a $\delta$-function, the integral in the imaginary part reduces
to zero, and the real part, to a Gaussian in frequency, centered on the Zeeman-shifted molecular
response centre of the $pq$ transition, so that
\begin{equation}
\Xi_{n}^{pq} \! \simeq \! \frac{1}{2} \!\! \left( \!
                          \frac{mc^2}
                               {2\pi k_B T \omega_0^2 l^2}
                        \!  \right)^{\!\! 1/2}
          \!\!\!\!  \exp \left\{
                         \frac{-m c^2 (\varpi_n \! - \! \Delta \omega_{pq}(\vec{r}) )^2}
                              {2 k_B T(\vec{r}) \omega_0^2}
                       \right\}.
\label{eq:approxvoigt}
\end{equation}
Once the velocity integration has been carried out under the CVR approximation, the
common velocity integral in equation~(\ref{eq:j10_off1}) becomes simply $\Xi_n^{p1}$, which
is independent of any particular ray.

\subsection{Classical reduction}
\label{ss:classic}

Approximations made in this section are not always made, and we attempt to quantify their effects by
comparisons with more accurate models. A set of approximations that may be grouped together as the
`classical approximation' are made for a mixture of computational expendiency and
comparison with earlier work carried out in this approximation. In the
classical approximation, there are no population pulsations significantly populated: we ignore all diagonal DM elements with
Fourier components $\neq 0$.

It is possible to simplify the equation set further, in the case of modest saturation
($\Gamma \ll R \ll g\Omega$), or a stimulated emission rate intermediate between the
loss and Zeeman precession rates. In this regime, it is possible to ignore the
type~2 off-diagonal elements, which are second-order in field-dipole products of the
form $\hv{d} \cdot \vec{\td}$. Inclusion of the type~2 off-diagonal DM elements is expected
to introduce mixing of the magnetic sublevels of the $J=1$ state 
\citep{Lankhaar2019,Tobin2023}. The simplest possible set of equations,
derived by implementing all of the above approximations consists of the reduced
form of equation~(\ref{eq:j10_inv}) for the inversion,
\begin{align}
\Delta_{p1} & =  \frac{P_{p1}}{\Gamma} -\frac{1}{2 \hbar \Gamma} 
      \!\!\! \sum_{\xi=1}^J w_{\Omega,\xi} \!\!\!\! \sum_{m=-\infty}^\infty \!\!\! \Re \left\{
  2  \hv{d}_{p1} \! \cdot \! \vec{\td}_{\xi,m} S_{p1,m}^{*(\xi)} \right. \nonumber \\
  & \left. + \hv{d}_{q'1} \! \cdot \! \vec{\td}_{\xi,m} S_{q'1,m}^{*(\xi)} 
           + \hv{d}_{q1} \! \cdot \! \vec{\td}_{\xi,m} S_{q1,m}^{*(\xi)} \right\} ,
\label{eq:diagred}
\end{align}
the type~1 off-diagonal DM element, simplified from equation~(\ref{eq:j10_off1}),
\begin{equation}
S_{p1,n}^{(\xi)} = \frac{2\pi w_{\Omega,\xi}}{\hbar} \Delta_{p1} \hv{d}_{p1} \! \cdot \! \vec{\td}_{\xi,m} \Xi_{n}^{p1} ,
\label{eq:offred}
\end{equation}
where $\Xi_{n}^{p1}$ is the complex Voigt function in frequency defined in equation~(\ref{eq:hvoigt}), and
an electric field amplitude to be derived from an unchanged equation~(\ref{eq:j10_eamp}). 
This equation set is used to demonstrate scaling to dimensionless forms in the main text of Section~\ref{ss:proxandscal}.

\subsection{Formal solution of the transfer equation}
\label{ss:formsol}

As a prelude to this operation, we assume in the present work that the medium has
zero conductivity ($\sigma = 0$), and we restore the original definition of the Fourier-transformed
electric field amplitude without the exponential in $\varpi_m$ (see Appendix~\ref{a:fdomeq}). The
combined effect is to eliminate $\kappa(\vec{r})$ from equation~(\ref{eq:j10_eamp}), while
equation~(\ref{eq:deltafinal}) does not change form.
We then substitute equation~(\ref{eq:offred}) into the
attenuation-free version of equation~(\ref{eq:j10_eamp}), allowing us to re-write
the latter expression in the vector-matrix form,
\begin{align}
   \frac{d}{ds_\xi}
   \left(
   \begin{array}{c}
   \td_{\xi x,n} \\
   \td_{\xi y,n}
   \end{array}
   \right)
=               \left[
   \begin{array}{cc}
      A(s_\xi) & B(s_\xi)  \\
      C(s_\xi) & D(s_\xi) 
   \end{array}
                \right]
                \left(
                   \begin{array}{c}
   \td_{\xi x,n} \\
   \td_{\xi y,n}
   \end{array}
                \right),
\label{eq:gainmat}
\end{align}
where the matrix element $A(s_\xi)$ is equal to
\begin{equation}
    A(s_\xi) = \frac{\pi \omega_0}{c \epsilon_0 \hbar} \sum_{p=2}^4
                \Xi_n^{p1}(s_\xi) \Delta_{p1}(s_\xi) | \hat{d}_{p1,x} |^2(s_\xi) .
\label{eq:matelA}
\end{equation}
In equations~(\ref{eq:gainmat}) and (\ref{eq:matelA}) a subscript index $x$ indicates $x_\xi$, the
x-component in the ray frame, where the ray is propagating along $z_\xi$. The matrix elements
$B$, $C$ and $D$ follow the pattern of equation~(\ref{eq:matelA}), and may be obtained by
replacing $| \hat{d}_{p1,x} |^2$ by the respective functions, $\hat{d}_{p1,x}^* \hat{d}_{p1,y}$,
$\hat{d}_{p1,x} \hat{d}_{p1,y}^*$ and $| \hat{d}_{p1,y} |^2$.

We solve equation~(\ref{eq:gainmat}) via the method of \citet{1987nrt..book..265L,Tobin2023}, so that
the electric field amplitude at ray distance $s_\xi$ starts with background $\vec{\td}_{\xi,n}(0)$, and becomes
\begin{align}
\vec{\td}_{\xi,n}(s_\xi) & = \vec{\td}_{\xi,n}(0)
                         + \! \sum_{k=1}^\infty 
                        \! \frac{1}{k!} \!\! \int_0^{s_\xi} \!\! ds_1 \! \int_0^{s_\xi} \!\! ds_2...
                            \nonumber \\
                        & \int_0^{s_\xi} \!\! ds_k
                         \upgamma_{\xi,n}(s_1) \upgamma_{\xi,n}(s_2)...\upgamma_{\xi,n}(s_k) \vec{\td}_{\xi,n}(0) ,         
\end{align}
\label{eq:rtsoln}
where $\upgamma_{\xi,n}$ is a $2\times2$ matrix of the type that appears in equation~(\ref{eq:gainmat}).

\section{The reference views}\label{app:ref_view}
\begin{figure*}
	\centering
	\includegraphics[width=0.96\textwidth]{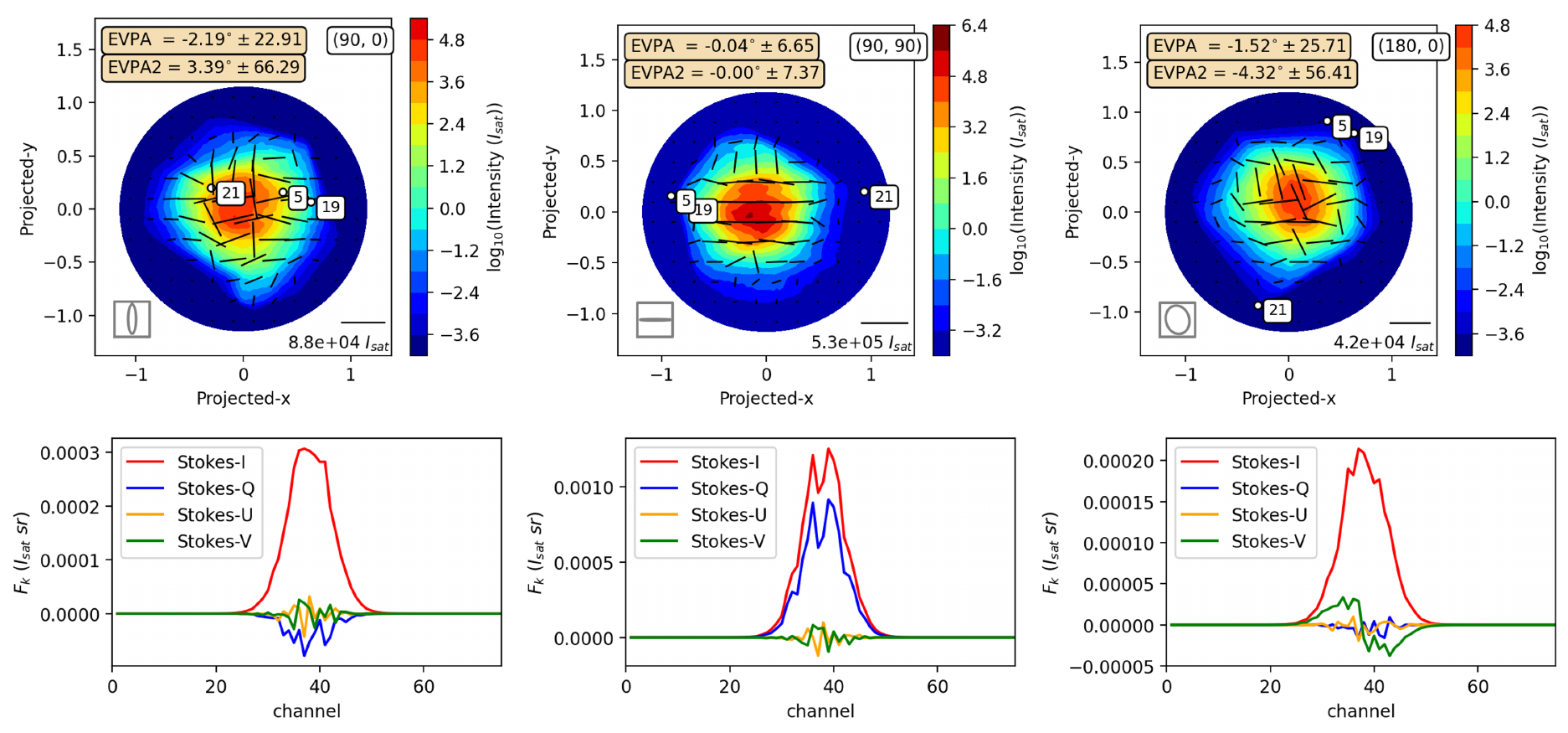}
	\caption{A formal solution of view $+x$, $+y$, and $-z$ from left to right, respectively for the pseudo-spherical shape. The top and bottom panels show a contour map and Stokes spectrum similar to Fig.~\ref{fig:contour_p137}.}
	\label{fig:form_ref_view_s}
\end{figure*}

\begin{figure*}
	\centering
	\includegraphics[width=0.96\textwidth]{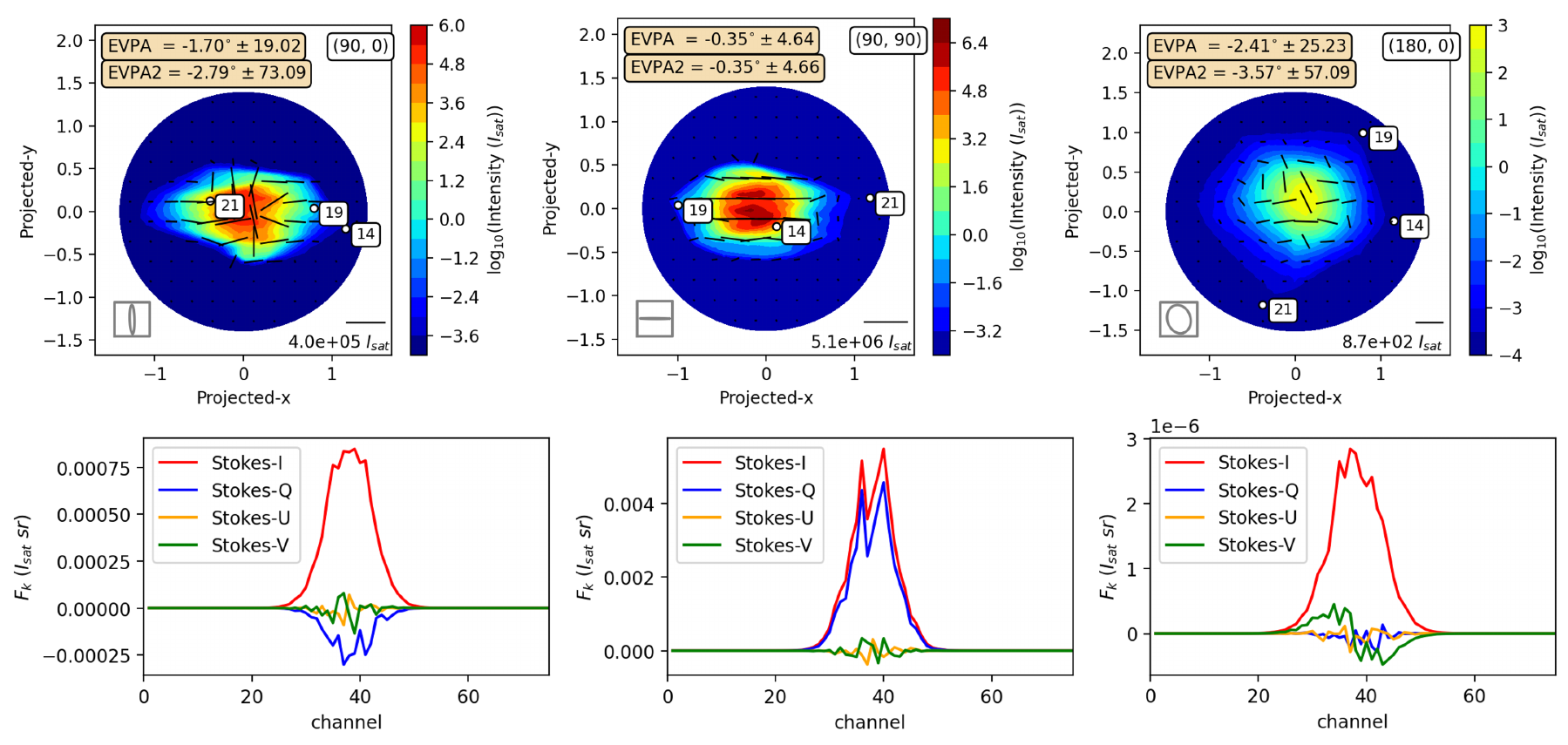}
	\caption{Similar to Fig.~\ref{fig:form_ref_view_s} but for the oblate shape.}
	\label{fig:form_ref_view_oz}
\end{figure*}

\begin{figure*}
	\centering
	\includegraphics[width=0.96\textwidth]{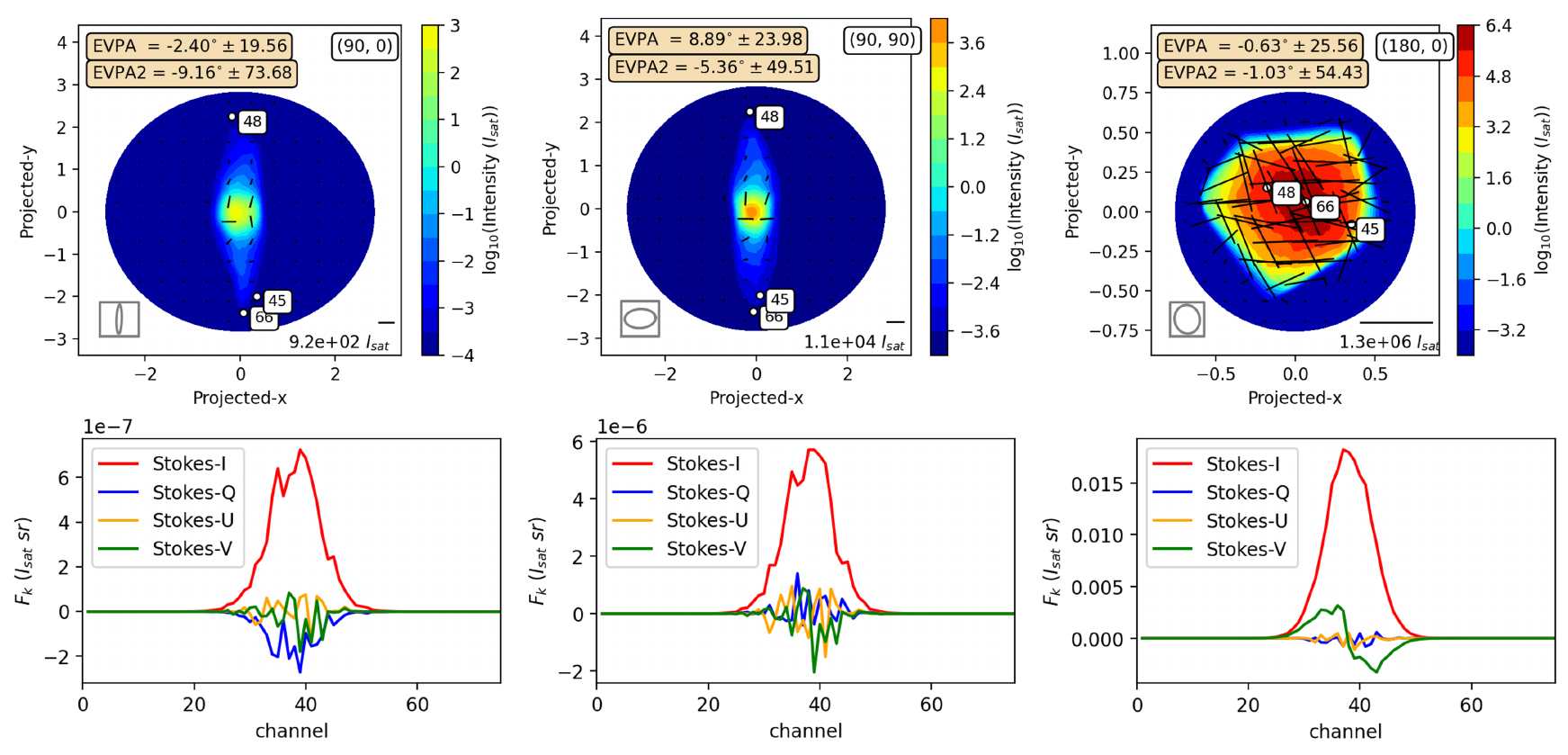}
	\caption{Similar to Fig.~\ref{fig:form_ref_view_s} but for the prolate shape.}
	\label{fig:form_ref_view_pz}
\end{figure*}

This section adds diagrams of the views $+x$, $+y$, and $-z$, corresponding to observer positions with coordinates ($r,\theta,\phi$) = (1000,90,0), (1000,90,90), and (1000,180,0) respectively. 
The projected-$x$ and projected-$y$ coordinates for the reference views $+x$, $+y$, and $-z$ are as follows: ($+y$, $+z$), ($-x$, $+z$), and ($+y$, $+x$). The $-z$ view is the opposite perspective to that discussed in sub-section \ref{subsec:form_results}, while views $+x$ and $+y$ provide side views of the domain. These three reference 
views elucidate the impact of strongly saturated nodes on polarization across the three different shapes, and the
effects of view rotation as
discussed in sub-section \ref{subsec:view_rot}. Fig.~\ref{fig:form_ref_view_s} illustrates 
the contour map with EVPA, EVPA2, and their Stokes spectra for the pseudo-spherical 
shape viewed from perspectives $+x$, $+y$, and $-z$. Figures~\ref{fig:form_ref_view_oz} and \ref{fig:form_ref_view_pz} depict the same reference views for the oblate and prolate shapes, respectively.

The $-z$ view presents a y-reflected perspective from sub-section \ref{subsec:form_results}. The
principal effect is on Stokes $V$, which flips sign in accord with the reversal of the magnetic field 
direction relative to the observer. The Stokes-$I$ intensity is contingent on the length of the domain 
segment traversed, with longer segments producing higher amplitudes compared to shorter ones. This explains why the intensity in views $+x$ and $+y$ for the oblate shape is higher than that for the prolate shape, while 
the pseudo-spherical shape exhibits similar intensities across all three reference views. 
The EVPA and EVPA2 in views $+z$ and $-z$ displays uncertainties that considerably exceed the
notional angle, reflecting the weak, random nature of linear polarization parallel or
anti-parallel to the magnetic field. The only views in which the uncertainty in the EVPA and EVPA2
angles is small (less than 10\,degrees) is the $+y$ view for the oblate and pseudo-spherical shapes.
We note that the most saturated nodes in the pseudo-spherical and oblate domains are found
mostly towards the circumference, while in the prolate shape they appear at the points of the domain
in the $+x$ and $+y$ views, but towards the centre (nodes 48 and 66) in the $-z$ view.

The Poincar\'e sphere projection in the $+x$ and $+y$ views 
shown in Fig.~\ref{fig:form_ref_view_s}-Fig.~\ref{fig:form_ref_view_pz} takes 
the form of highly eccentric ellipses, indicating almost pure linear polarization. By comparison, the 
more open ellipses in the $-z$ view (right-hand panels of 
Fig.~\ref{fig:form_ref_view_s}-Fig.~\ref{fig:form_ref_view_pz}) indicate stronger circular 
polarization, and reinforce the conclusion that linear polarization is weak in this view. The
prolate domain does show a clear rotation of the major axis of the ellipse between the $+x$ and
$+y$ views, but the ellipse is more open in the $+y$ view, showing some circular polarization
is present.

\section{The shifted-view rotation}\label{app:shifted-view}
\begin{figure}
	\centering
	\includegraphics[width=0.42\textwidth]{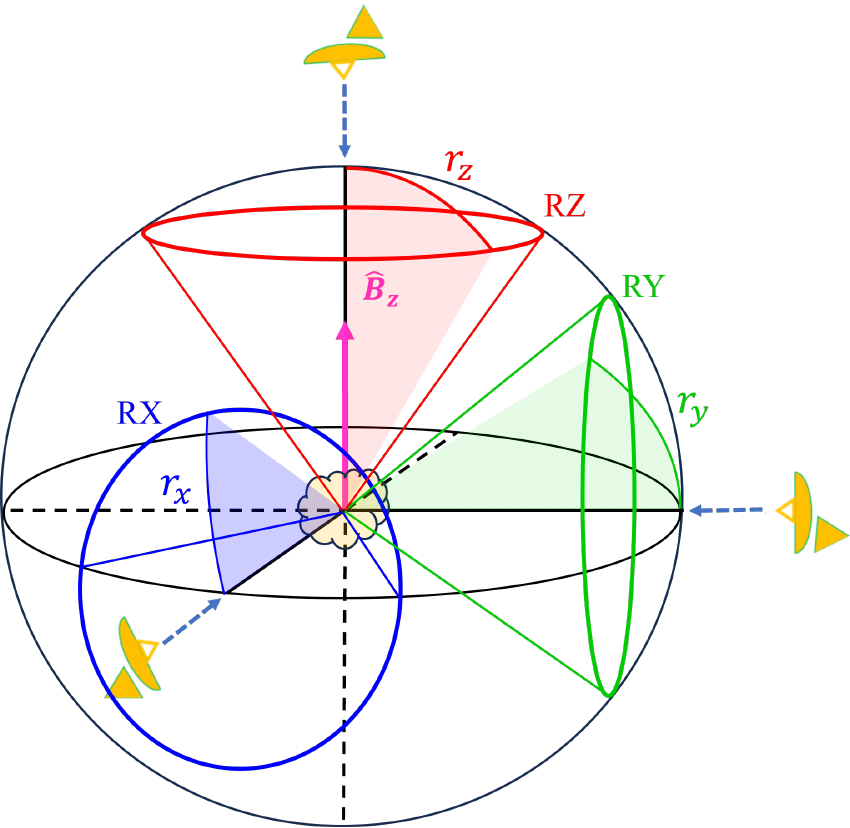}
	\caption{The diagram of shifted-view from the reference observer view. The solid red, blue, and green lines refer to RZ, RX, and RY rotation patterns, with their shifted-angle of $r_z$, $r_x$, and $r_y$ respectively. }
	\label{fig:diagram_shift_rotate}
\end{figure}

\begin{figure*}
	\centering
	\includegraphics[width=0.96\textwidth]{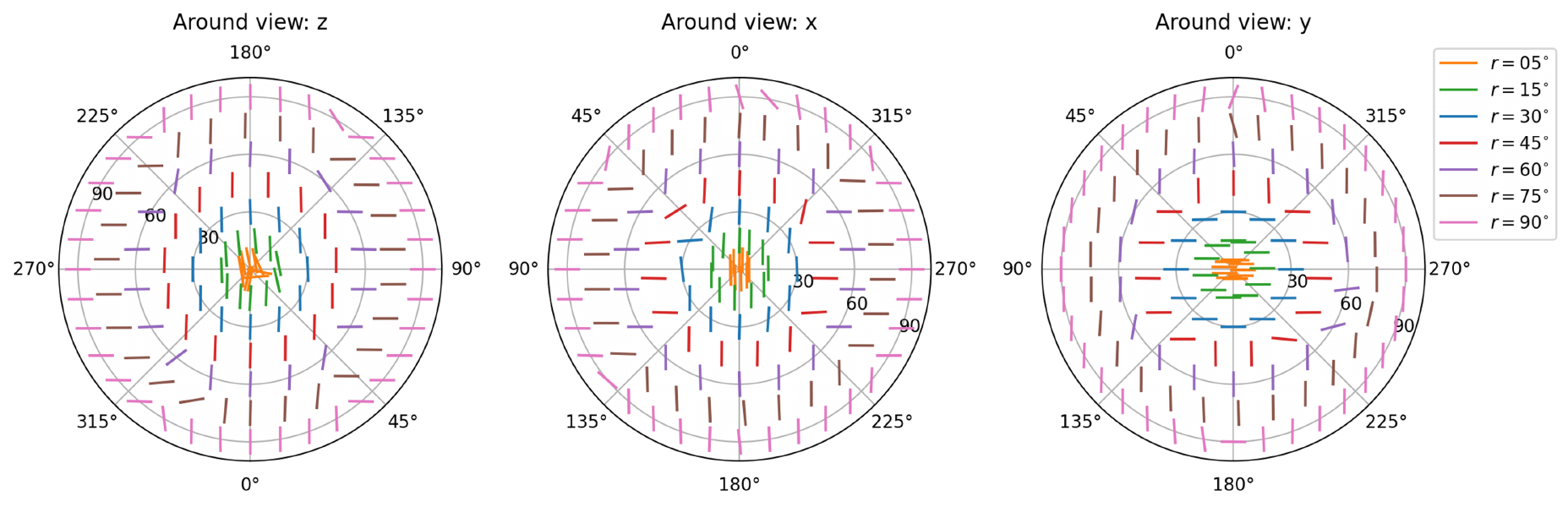}
	\caption{The results of EVPA2 for three shifted-rotations 
    shown in the diagram of Fig.~\ref{fig:diagram_shift_rotate} of the pseudo-spherical domain. Left to right graphs refer to RZ, RX, and RY rotation as a polar coordinate ($r_{\text{polar}}$, $\theta_{\text{polar}}$). $r_{\text{polar}}$ refers to the shifted angle while $\theta_{\text{polar}}$ refers to the projected azimuth angle related to the reference view. Colors orange, green, blue, red, purple, brown, and pink refer to the shifted angle as 5, 15, 30, 45, 60, 75, and 90 degrees, respectively. In all plotted positions, the vector displayed has the orientation seen
    in the observer's $xy$ plane for that view.}
	\label{fig:rot_shift_view_s}
\end{figure*}

\begin{figure*}
	\centering
	\includegraphics[width=0.96\textwidth]{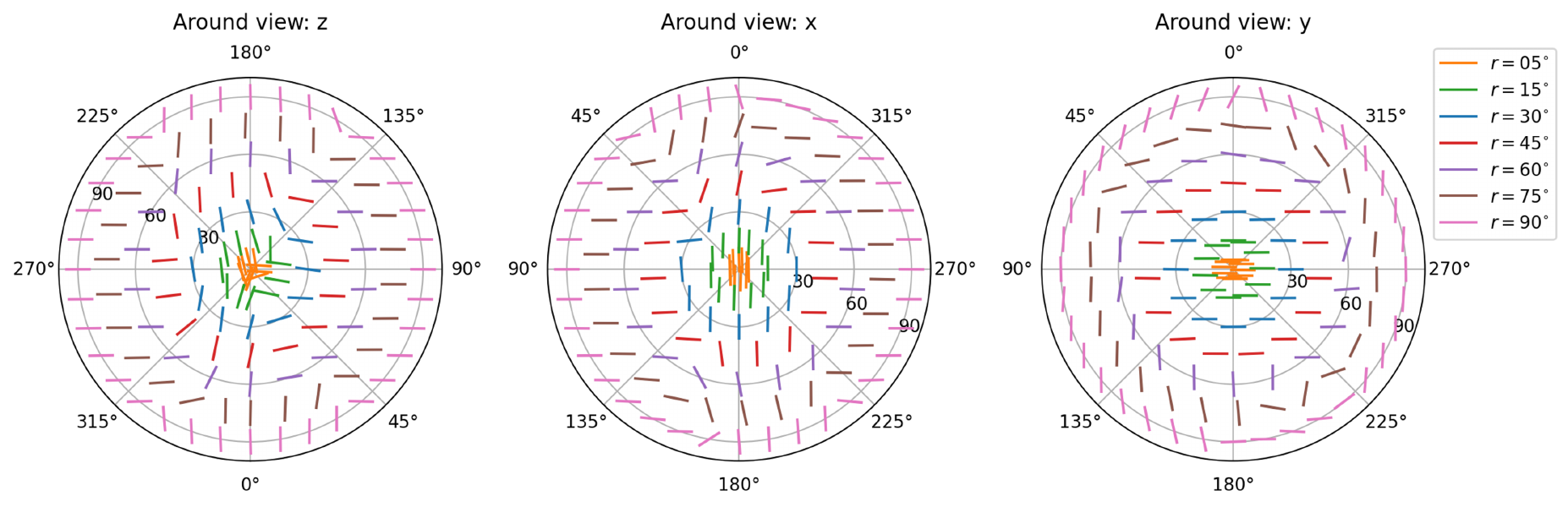}
	\caption{Similar to Fig.~\ref{fig:rot_shift_view_s} but for the oblate shape.}
	\label{fig:rot_shift_view_oz}
\end{figure*}

\begin{figure*}
	\centering
	\includegraphics[width=0.96\textwidth]{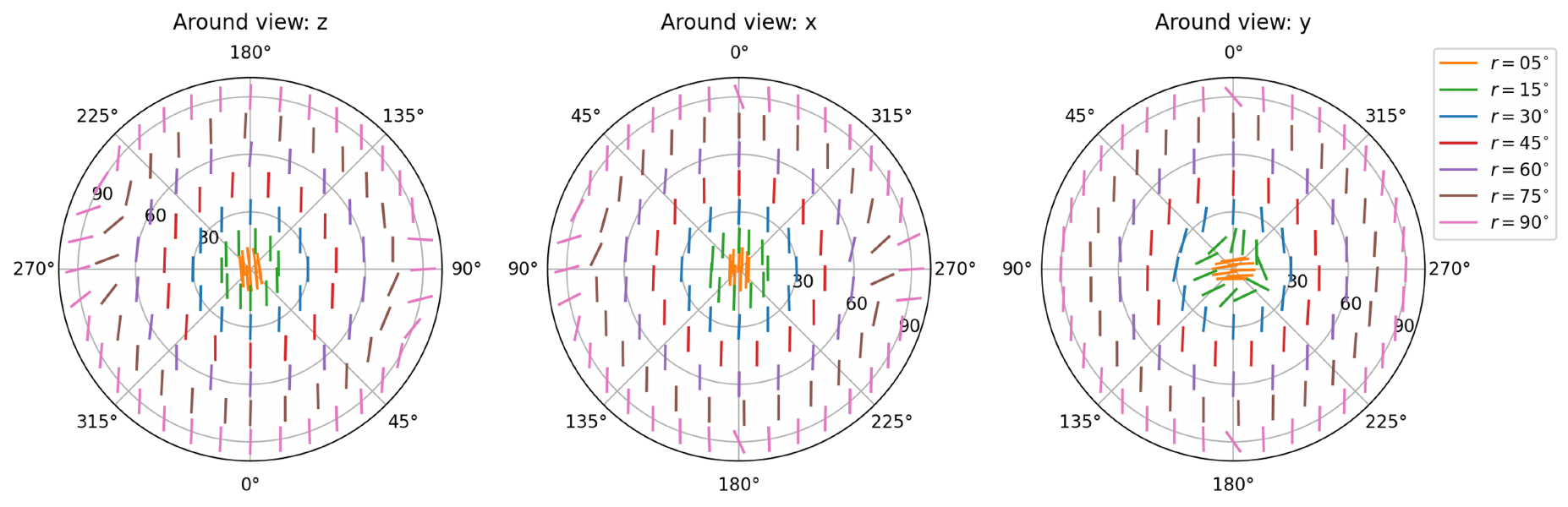}
	\caption{Similar to Fig.~\ref{fig:rot_shift_view_s} but for the prolate shape.}
	\label{fig:rot_shift_view_pz}
\end{figure*}

This appendix aims to extract exact positions that exhibit EVPA2 rotation by considering views 
based on small-circle arcs, as shown in Fig.~\ref{fig:diagram_shift_rotate}. However, as $+z$ and $-z$
views show approximately opposite polarizations in Stokes-$V$ (see appendix \ref{app:ref_view}), we consider 
only samples from half the sphere here. For easy comparison with sub-section \ref{subsec:view_rot}, we 
computed formal solutions along small-circle arcs around the reference views $+x$, $+y$, and $+z$. These
arcs have respective spherical angles $r_x$, $r_y$, and $r_z$, as shown in Fig.~\ref{fig:diagram_shift_rotate}. 
Example small-circles are shown as RX, RY and RZ. As $r_x$ tends to zero, the small-circle RX tends to a
point corresponding to the $+x$ reference view, whilst at $r_x = 90$\,degrees, RX becomes a great-circle
consisting of the R2 arc and its mirror in the $yz$ plane. Similar definitions apply to RY and RZ.
We examined rotation of the plane of linear polarization in the plane of the sky for small, moderate, and large 
angles away from the reference positions (or towards the great-circle arcs). Results are presented in Figures~\ref{fig:rot_shift_view_s}, \ref{fig:rot_shift_view_oz}, and \ref{fig:rot_shift_view_pz} for the pseudo-spherical, oblate, and prolate domains, respectively. The values of $r_x$, $r_y$ and $r_z$ in these
figures ranges from 5 to 90 degrees.

In the RZ small-circle, the projected azimuth angle corresponds to the azimuth angle of the main sphere, while the
RX and RY circles combine polar and azimuthal motion. For all three domains, and all three small-circles
we observe mostly smooth changes, both in traversing any circle and when moving from one circle to another (changing
shift angle).
However, we identify some zones of more rapid change in EVPA2 that we discuss in more detail below. Perhaps
the broadest distinction that can be made is between the RY circles (where change in the shift angle leads only
to a rotation of the plane of polarization), and the pair RX and RZ (where a single polarization plane at low
shift angle changes to a pattern with four rotations within each circle at high shifts). We attribute the
EVPA2 rotations within small circles to
the saturation effect due to the
orientation of the most intense rays, as introduced regarding the R1 view rotation in Section~\ref{subsec:view_rot}.

In the RY circles we find that the rotation in polarization angle, from approximately vertical to approximately
horizontal, occurs when the shift-angle $r_y$ is small (5 degrees or less) in the prolate domain
(right panel, Fig.~\ref{fig:rot_shift_view_pz}). A similar change occurs between 60-90\,degrees (oblate domain)
and 45-75\,degrees (pseudo-spherical).

The RZ (or RX) circles both show four rapid changes of approximately 90\, degrees in EVPA2 in the circles
at high $r_z$ (or $r_x$). One pair appears along a line joining azimuths of approximately 45 and 225\,degrees,
with the other pair aligned at right-angles to this. For RZ, where the rotation with $r_z$ = 90 degrees corresponds 
to the R1 arc, this is exactly as expected from the reference views and the
polarization rotation observed in sub-section \ref{subsec:view_rot}. In RZ, the change to this pattern, from
approximately constant horizontal polarization vectors, occurs over the shift-angle ranges of
45-60\,degrees in both the pseudo-spherical and oblate domains. It occurs later (60-75\,degrees) in the
prolate domain. In RX, the pattern change occurs between $r_x=45$ and 75\,degrees in the prolate case
and 30-45\,degrees in the other two domains.

In conclusion, the strong changes of EVPA2 observed in subsection \ref{subsec:view_rot} along the R1 arc happen
also for small-circle arcs down to about 60\,degrees from the $z$-axis (75\,degrees in the prolate case).
Closer to the $z$-axis, and away from this equatorial belt, the strong change behaviour gives way to
a pattern of approximately constant-direction polarization vectors, closer to the 1D expectation.

\section{Tube Domain}\label{app:small_domain}
\begin{figure*}
	\centering
	\includegraphics[width=0.98\textwidth]{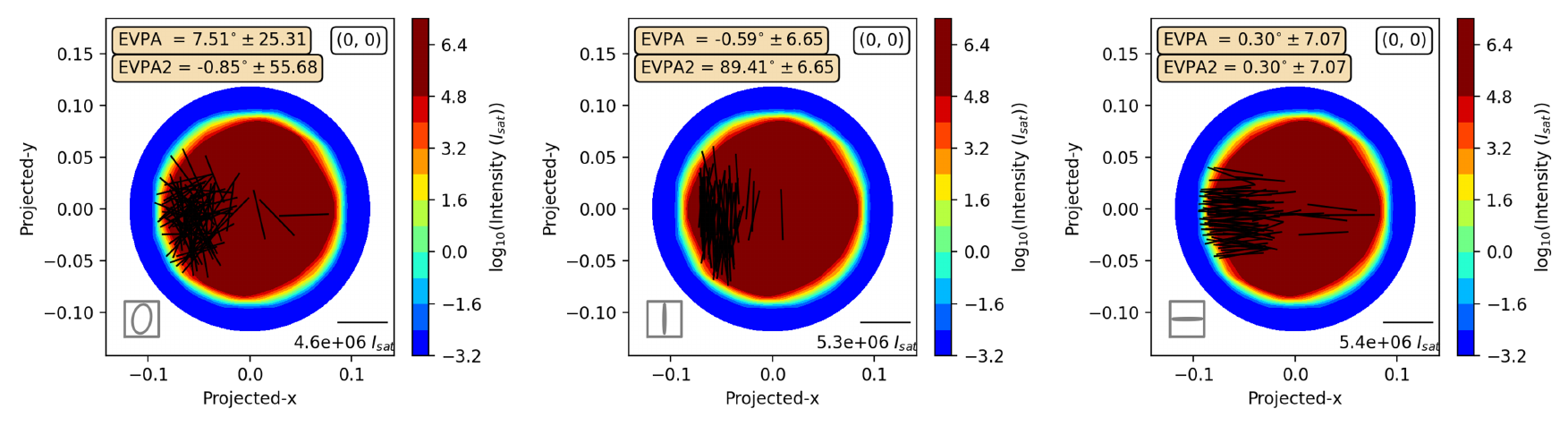}
	\caption{Images of the tube domain from view +z with overlaid linear polarization
    vectors for the 60 brightest rays: From left to right, the magnetic field is offset
    from the $z$-axis by respective angles of 0.001, 30 and 89.999\,degrees.}
	\label{fig:form_p43_view_z}
\end{figure*}

Here, we present additional results for the tube domain described in Section~\ref{ss:tubedom}.
This was done to investigate the conformity of our model with previous 1D results. For all results,
except an experimental R1 rotation,
the observer views anti-parallel to the $z$-axis (the long axis of the cylindrical domain).
For this domain, $T_K =1500$\,K, $B=5$\,G, with consequent changes to the frequency
parameters to maintain a resolution of 1 channel per Zeeman shift. The \texttt{rayform} 
parameter was 10 for formal solutions. Otherwise, we used the
same simulation parameters as listed in Table~\ref{tab:parms}. 
In Fig.~\ref{fig:form_p43_view_z} we show images of the tube domain with the magnetic field
at (from left to right) 0.001, 30 and 89.999\,degrees. At the smallest angle, with the field
pointing at the observer, there is no $\pi$-transition dipole in the plane of the sky, and
the $\sigma$-dipoles provide only circular polarization, so the orientation of the
linear polarization vectors is random. There is an approximately 90-degree rotation in the
strongly organized EVPA vectors between the images at 30 and 89.999\,degrees, as expected.

\begin{figure*}
	\centering
	\includegraphics[width=0.96\textwidth]{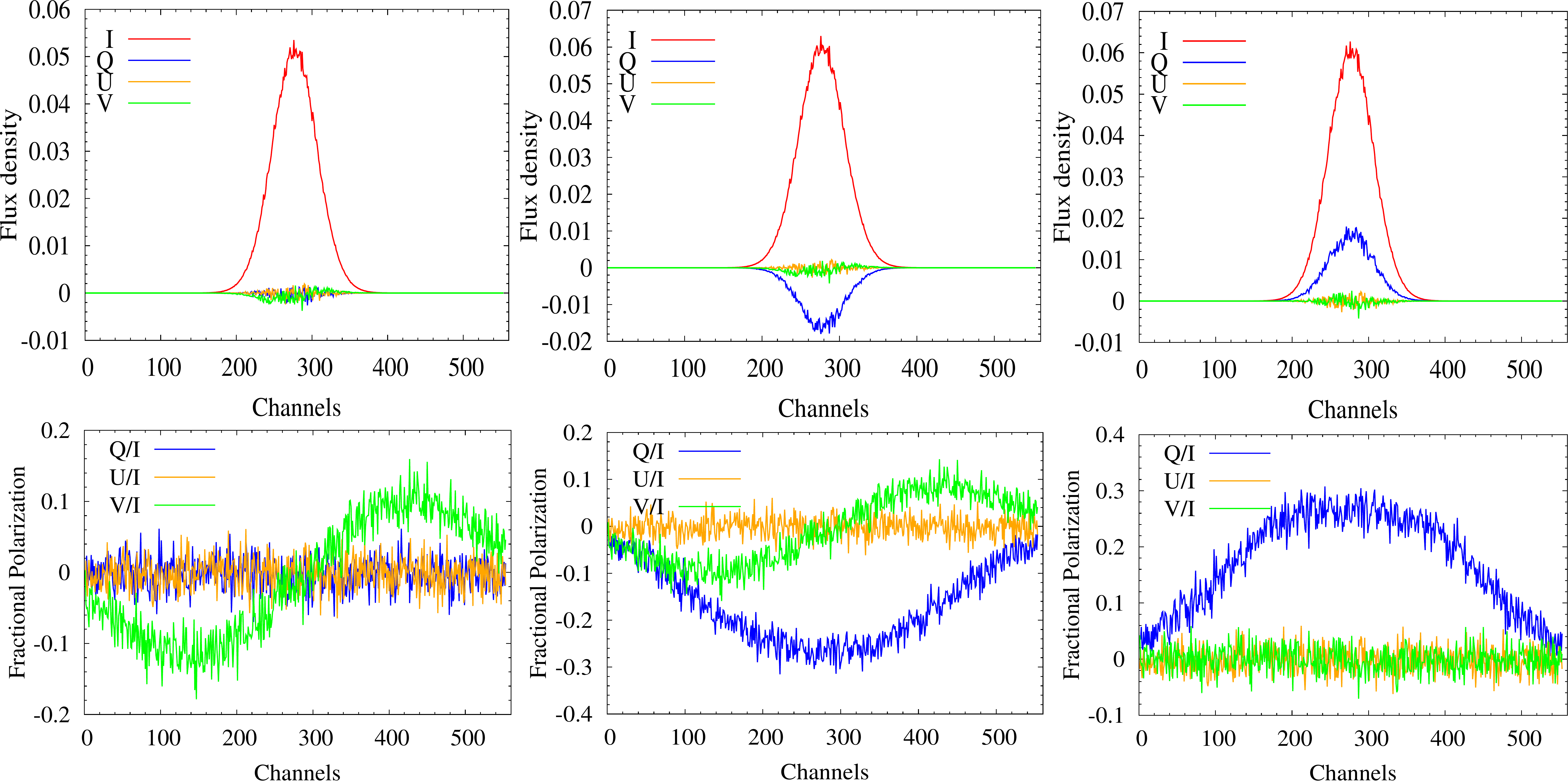}
	\caption{Stokes spectra (top row) and spectra relative to Stokes-$I$ (bottom row) for
    magnetic field angles $\theta_B = 0.001,30$ and $89.999$\,degrees (columns, left to right).}
	\label{fig:form_p76_view_z}
\end{figure*}

Fig.~\ref{fig:form_p76_view_z} shows the corresponding spectra. In the upper row, we have
plotted all four Stokes parameters, while the lower row shows fractional polarizations with
respect to Stokes-$I$. The following points are important: Stokes-$V$ is antisymmetric about the
pattern centre, and strongest at $\theta_B=0.001$\,degrees. The ratio $V/I$ has become a
`noisy zero' at $\theta_B=89.999$\,degrees. These results are in accord with expectations from
1D models, as is the $U/I \simeq 0$ at all angles. Finally, $Q/I$ is a noisy zero at $\theta_B = 0.001$\,degrees,
before taking on negative values as the angle is increased (the middle column is a good example
at 30\,degrees). Stokes-$Q$ changes sign at the Van~Vleck angle (not shown) before reaching
a final level of +(25 to 30) per cent with the magnetic field aligned with the projected $y$-axis.
For the level of saturation in this model, the maximum absolute value of $Q/I$ is similar on
both sides of the Van~Vleck angle. It is possibly useful to compare these figures with
results in \citet{Dinh-v-Trung2009MNRAS}.

When the observer's position was moved to a polar angle of 90\,degrees and an R1 (azimuthal)
view rotation carried out, in the case where $\vec{B} = B \hv{z}$, we found that there were
no sign changes in Stokes-$Q$ of the type found along the R1 rotations in Section~\ref{subsec:view_rot}.
The mean value of $Q/I$ along the R1 arc was $-0.1147 \pm 0.0132$ (the corresponding mean $U/I$
was $-0.0070 \pm 0.0130$). 

\section{Ray Sampling in Formal Solutions}\label{app:rsfs}

\begin{figure*}
	\centering
	\includegraphics[width=0.96\textwidth]{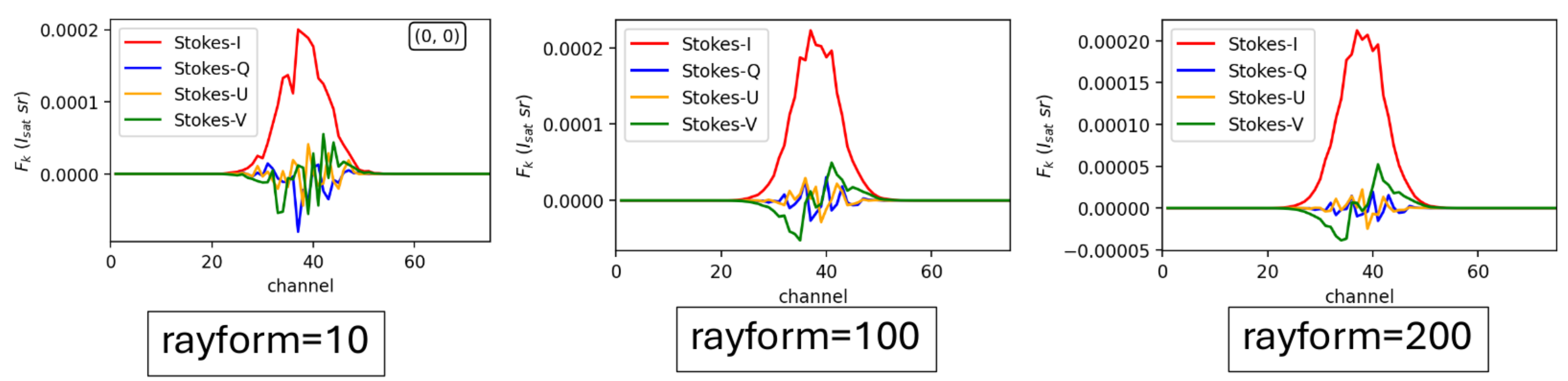}
	\caption{The effect of increased ray sampling (parameter \texttt{rayform} as marked) on Stokes spectra. From
    left to right, spectra use \texttt{rayform}=10, 100 and 200, with respective numbers of rays equal to 243, 2023
    and 4107. The domain is the pseudo-spherical version of pointy137 with an inversion solution at
    depth 20.0. The view is `+z', so the magnetic field of 35\,G points directly at the observer.}
	\label{fig:rsfs}
\end{figure*}

Increasing the number of rays in a formal solution via the parameter \texttt{rayform} has a smoothing
effect on the spectra of all Stokes parameters. We show spectra for three different values of this parameter
in Fig.~\ref{fig:rsfs}, noting that the right-hand panel is the same spectrum as in the central panel of Fig.~\ref{fig:contour_p137}. There is a `law of diminshing returns' in that smoothing improvements become
minimal after about \texttt{rayform}=100. Improving the ray sampling helps reduce noise resulting
from the amplification of stochastic background radiation, but does not remove effects resulting
from departures of the domain from axial symmetry.

\bsp	
\label{lastpage}

\end{document}